\pdfoutput=1

\documentclass[12pt, a4paper]{article}

\usepackage{notes2bib}
\usepackage{amsmath}
\usepackage{amsfonts}
\usepackage{amssymb}
\usepackage{bbm}
\usepackage{verbatim}
\usepackage{dsfont}
\usepackage{booktabs}
\usepackage{slashed}
\usepackage{bm}
\usepackage{enumitem}

\usepackage{epsfig}
\usepackage[table]{xcolor}
\usepackage{graphicx}
\usepackage[]{caption}
\usepackage[listofformat=empty,subrefformat=empty]{subfig}

\usepackage{float}
\usepackage{overpic}

\usepackage{enumerate}
\usepackage{hhline}
\usepackage{multirow}
\usepackage[hidelinks]{hyperref}

\usepackage{cite}
\usepackage{a4wide}
\usepackage{bbold}

\usepackage{geometry}
\renewcommand{\baselinestretch}{1.2}

\newcommand{\la}{\lambda}

\newcommand{\bk}{\bar{k}}
\newcommand{\bx}{\bar{x}}

\definecolor{darkgreen}{HTML}{3C8031}

\begin{document}

\thispagestyle{empty}

\begin{flushright}
DESY 24-105
\end{flushright}
\vskip .8 cm
\begin{center}
  {\Large \bf DeWitt wave functions
  for de Sitter  JT gravity}\\[12pt]

\bigskip
\bigskip 
{
{\bf{Wilfried Buchm\"uller$^\dagger$}\footnote{E-mail:
    wilfried.buchmueller@desy.de}},
{\bf{Arthur Hebecker$^\ast$}\footnote{E-mail:
    a.hebecker@thpys.uni-heidelberg.de}} and
{\bf{Alexander Westphal$^\dagger$}\footnote{E-mail:
    alexander.westphal@desy.de}}
}
\bigskip\\[0pt]
\vspace{0.23cm}
{\it $^\dagger$Deutsches Elektronen-Synchrotron DESY, Notkestr. 85, 22607 Hamburg, Germany \\
  \vspace{0.2cm}
  $^\ast$Institute for Theoretical Physics, 
  Heidelberg University,\\ Philosophenweg 19, 69120 Heidelberg, Germany
}
\\[20pt] 
\bigskip
\end{center}

\date{\today}

\begin{abstract}
\noindent
Jackiw-Teitelboim (JT) gravity in two-dimensional de Sitter space 
is an intriguing model for cosmological ``wave functions of the universe". 
Its minisuperspace version already contains all physical information. The size 
of compact slices is parametrized by a scale factor $h > 0$. The dilaton $\phi$
is chosen to have positive values and interpreted as size of an
additional compact slice in a higher-dimensional theory. At the boundaries 
$h=0$, $\phi=0$, where the volume of the universe vanishes, the curvature is
generically singular. According to a conjecture by DeWitt, solutions of the
Wheeler-DeWitt (WDW) equation should vanish at singular loci. Recently,
the behaviour of JT wave functions at large field values $h$, $\phi$ has
been obtained by means of a path integral over Schwarzian degrees of
freedom of a boundary curve. We systematically analyze solutions of the
WDW equation with Schwarzian asymptotic behaviour. We find real analytic solutions that vanish on the entire boundary, in agreement with DeWitt's
conjecture. Projection to expanding and contracting branches may lead to
singularities, which can however be avoided by an appropriate superposition of solutions. Our analysis also illustrates the limitations of
semiclassical wave functions.
\end{abstract}

\newpage 
\setcounter{page}{2}
\setcounter{footnote}{0}
{\renewcommand{\baselinestretch}{1}\tableofcontents}

\section{Introduction}
\label{sec:introduction}

Accepting the universal validity of quantum mechanics, also the universe has to be treated as a quantum system that is described by
a Wheeler-DeWitt~\cite{DeWitt:1967yk,WDWWheeler} ``wave function of the universe'' (WF). An idea that has inspired
four decades of quantum cosmology is the no-boundary proposal
of Hartle and Hawking \cite{Hartle:1983ai}, and Linde \cite{Linde:1983mx} and Vilenkin \cite{Vilenkin:1984wp}, where
de Sitter space with Lorentzian signature is analytically continued to a Euclidean geometry without boundary. The WF is a solution of the Wheeler-DeWitt (WDW) equation of quantum gravity
(for a review and references, see, for example \cite{Kiefer:2007ria}).
Although over the years much progress has been made (for a recent review, see \cite{Lehners:2023yrj}), the precise definition of the WF is still a
topic of current research \cite{Halliwell:2018ejl}, and one may worry that the no-boundary proposal has finally failed \cite{Feldbrugge:2017mbc,Maldacena:2024uhs}. 

An interesting toy model for quantum gravity is Jackiw-Teitelboim (JT) gravity in two dimensions \cite{Teitelboim:1983ux,Jackiw:1984je}. The model is exactly
solvable \cite{Henneaux:1985nw,Louis-Martinez:1993bge} and,
like all 2d dilaton-gravity theories, its minisuperspace version
already contains all physical information \cite{Louis-Martinez:1993bge}. Over the past years, JT gravity has been
studied in detail in anti-de Sitter ($\text{AdS}_2$) space, shedding new light
on the structure of Euclidean quantum gravity (for a recent
review, see \cite{Mertens:2022irh}).

An important new development is the computation of the
no-boundary wave function in de Sitter ($\text{dS})$
JT gravity at large field values 
\cite{Maldacena:2019cbz,Cotler:2019nbi}. This is achieved
by reducing the path integral for the wave function to
a path integral over the Schwarzian degrees of freedom
of a boundary curve, as in $\text{AdS}_2$ gravity
\cite{Stanford:2017thb}. Summing up an infinite series of 
extrinsic curvature terms of the boundary curve, the asymptotic
form of the wave function has been extended to the entire
field space \cite{Iliesiu:2020zld}. However, as noted
in \cite{Iliesiu:2020zld}, the obtained wave function is
singular at the de Sitter radius. This has been criticized in
\cite{Fanaras:2021awm} as unphysical since it points toward
a source not contained in the no-boundary proposal.
Further developments on $\text{dS}$ JT gravity include  
the connection to tunneling processes 
\cite{Moitra:2022glw}, semiclassical thermodynamics in an extension
with conformal matter \cite{Svesko:2022txo}, the emergence of time
\cite{Nanda:2023wne}  and the proposal of isometric time 
evolution \cite{Cotler:2023eza,Cotler:2024xzz}.
  
The goal of this paper is to clarify the status of the
no-boundary wave function in $\text{dS}$ JT gravity.  We consider this model as an effective low-energy description
of a higher-dimensional theory with compact spatial slices. For example, one may think of it as a toy model for Kantowski-Sachs cosmology \cite{Kantowski:1966te}, where the universe has the topology of $S^1\times S^2$ and the dilaton $\phi$ parametrizes the size of the $S^2$ \cite{Fanaras:2021awm, Fanaras:2022twv,Svesko:2022txo}. Hence, we consider only positive values of $\phi$. 
In addition, one has the usual compact slices of $\text{dS}_2$, the size of which we will parameterize by $h$. Clearly, the scale factor $h$ also takes only positive values.  We then analyze the boundary problem for the WDW equation, focusing on the `right-upper quadrant' $\{h>0,\phi>0\}$. We find exact solutions of
the WDW equation which vanish on  the entire boundary. The latter consists of the two rays $\{h=0,\phi>0\}$ and $\{\phi=0,h>0\}$. These rays represent characteristic curves of the WDW equation viewed as a hyperbolic partial differential equation. One may think of such characteristics as `lightlike' since they constrain the way in which perturbations of solutions propagate. Our solutions have Stokes lines separating a region with large scale factor and oscillatory behaviour from a region with small scale factor and exponential behaviour.  The general `quantum no-boundary wave function' is a superposition
of such solutions. For
sufficiently large scale factors one obtains the familiar
semiclassical behaviour. For small scale factors there is no interpretation in terms of a semiclassical geometry.

As explained, our wave function is defined in the quadrant
$\{h>0,\phi>0\}$. One of the two boundary lines is characterized by
the universe having zero spatial extension. The other may be viewed as
the locus of a spacelike end-of-world brane, defined by $\phi=0$,
similarly to the `boundary proposal' of
\cite{Friedrich:2024aad}. Interpreting $\Psi(h,\phi)$ as the
probability amplitude for finding a universe with size $h$ and dilaton
value $\phi$, it then appears reasonable that $\Psi$ should vanish at
the two boundary lines just described. One may take this as one of the
defining features of our proposed wave function. Given the emphasis
DeWitt placed on the vanishing of the wave function on singular
geometries~\cite{DeWitt:1967yk}, we hence call our solutions
`DeWitt wave functions of the universe'. The term DeWitt wave function
has recently also been used in the context of Horava-Lifshitz gravity
in~\cite{Matsui:2021yte,Martens:2022dtd}, and DeWitt boundary
conditions in quantum cosmology have been discussed
in~\cite{Esposito:2023ymk,Matsui:2023tkw}.

The paper is organized as follows. In Section~\ref{sec:HHJT}
we recall the computation of the asymptotic wave function
in terms of the Schwarzian degrees of freedom of the boundary
curve. For comparison, the semiclassical Hartle-Hawking wave
function is briefly reviewed in Appendix~\ref{sec:semiHH}, with emphasis on the asymptotic
behaviour at large and small scale factors and its singularities,
and in Appendix~\ref{sec:semiJT} the semiclassical wave function
for JT gravity is computed. Section~\ref{sec:holo} deals
with the representation of the JT wave function in terms of the
standard JT bulk amplitude. The result is compared with 
the asymptotic wave function discussed in Section~\ref{sec:HHJT}.
The characteristic initial value problem posed by the
JT WDW equation is systematically analyzed in Section~\ref{sec:charivp}, and it is shown how to obtain
a wave function free of singularities by analytic
continuation in field space across a Stokes line. We discuss probabilistic constraints on our proposed WF in Section \ref{prob} before concluding in Section \ref{sec:conclusion}. Four appendices are devoted to technical aspects.
Appendix~\ref{sec:canonical} deals with canonical quantization 
of JT gravity and with the connection between the path integral measure and factor ordering of the WDW equation. In Appendix~\ref{sec:nonSingBWF}
boundary wave functions with compact support are considered, and
exact solutions
of the WDW equation in terms of Airy functions are given in
Appendix~\ref{sec:airy}. Initial-value problems in two dimensions
and the relevant Green functions are discussed in
Appendix~\ref{sec:cauchy}.\\

\section{Nearly-de Sitter gravity and boundary modes}
\label{sec:HHJT}

We will be dealing with quantum gravity on de Sitter space in two space-time dimensions. A minimal model for a theory of quantum gravity in  $\text{dS}_2$  is JT gravity,  defined by the Lorentzian action
\begin{equation}\label{jtL}
  S[g,\phi] = \frac{1}{2} \int_{\mathcal{M}}d^2x \sqrt{g} \phi (R - 2\lambda^2) +
  \int_{\partial\mathcal{M}} d\theta \sqrt{h} \phi K \ .
\end{equation}
Here $g$, $R$, $\lambda^2$, $h$ and $K$ denote  metric tensor, 
Ricci scalar and cosmological constant, and induced metric and extrinsic curvature on the boundary $\partial\mathcal{M}$, respectively. 

Since the action \eqref{jtL} is linear in the dilaton, it can be
integrated out in a Lorentzian path integral. This fixes the bulk metric to the hyperboloid of global de Sitter \cite{Cotler:2019nbi},
\begin{equation}
  g_{\mu\nu}dx^\mu dx^\nu = -dt^2 + \alpha^2\lambda^{-2}\cosh^2(\lambda t) (d\theta + \gamma \delta(t) dt)^2\ ,
  \end{equation}
where $2\pi |\alpha|/\lambda$ is the length of the circle at $t=0$, and $\gamma$ is an additional twist parameter.\footnote{We work in minisuperspace, which in 2d is possible without loss of generality \cite{Louis-Martinez:1993bge}.} The regions $t\geq 0$ and $t\leq 0$ describe
future and past "trumpets", respectively.
For $\alpha =1$, a trumpet can be matched to a Euclidean half-sphere. This
complex Lorentzian/Euclidean geometry is the basis of the semi-classical Hartle-Hawking wave function. 

In \cite{Maldacena:2019cbz,Cotler:2019nbi} a no-boundary wave function for JT gravity has been computed starting from the Lorentzian path integral
\begin{equation}
\Psi(h,\phi) =
\int^{(h,\phi)}[Dg][D\phi'] \exp{(iS[g,\phi'])}\ .\label{dexp}
\end{equation}
In general, a Lorentzian path integral requires an initial condition. In the spirit of Hartle and Hawking, this can be avoided in the following way \cite{Maldacena:2019cbz, Cotler:2019nbi}: One first performs the $\phi'$ integration. This enforces an on-shell metric. More precisely, integrating over purely imaginary $\phi'$ forces the Euclidean metric of the round sphere with radius $h_c= \lambda^{-2}$. Additionally, performing the $\phi'$ integration for real-valued $\phi'$ forces the metric of the Lorentzian de Sitter hyperboloid. By continuity it has critical waist size $h_c= \lambda^{-2}$.
In total, this amounts to integrating out $\phi'$ on a complex path to get the glued complex Hartle-Hawking geometry. In this way, the condition $\alpha=1$ is enforced on the Lorentzian part.

All that is left is the integration over different boundary curves in the Lorentzian region with $h > \lambda^{-2} \equiv h_c$. Following \cite{Cotler:2019nbi}, this integration is conveniently performed using conformal coordinates, $dt = \sqrt{h(\tau)} d\tau = 1/(\lambda\cos{\lambda\tau}) d\tau$, such that the metric of $\text{dS}_2$ with $\alpha=1$ becomes
($0\leq \theta \leq 2\pi$, $0 \leq \tau < \frac{\pi}{2}$)
\begin{equation}
  g_{\mu\nu}dx^\mu dx^\nu = \frac{1}{\lambda^2\cos^2{(\lambda\tau)}}(-d\tau^2 + 
  d\theta^2)\ .
  \end{equation}
The boundary of de Sitter space is
specified by two functions $\tau(u)$ and $\theta(u) = f(u)$ where $u$ is
a periodic variable, $f(u+2\pi) = f(u) + 2\pi$. Demanding that the
induced metric on the boundary
is constant implies a relation between the functions $\tau(u)$ and $f(u)$,
\begin{equation}
  g_{uu} = \frac{-\tau'(u)^2 + f'(u)^2}{\lambda^2\cos^2{(\lambda\tau(u))}} \equiv h \ ,
\end{equation}
where the prime denotes differentiation with respect to $u$.
For nearly-de Sitter space, i.e., asymptotically large values 
$\lambda\sqrt{h}$, one has
\begin{equation}
  \lambda\tau(u) = \frac{\pi}{2} - \frac{1}{\lambda\sqrt{h}} f'(u) +
  \mathcal{O}\left(\frac{1}{(\lambda\sqrt{h})^3}\right)  \ ,
\end{equation}
and the extrinsic curvature is given by
\begin{equation}
\lambda^{-1}K = \frac{f'({f'}^2 - \lambda^2{\tau'}^2)\sin{\lambda\tau} +
  \lambda(f'\tau''-f''\tau')\cos{\lambda\tau}}{({f'}^2-\lambda^2{\tau'}^2)^{3/2}} \ .
\end{equation}
The sign of $K$ corresponds to the choice of an outward
pointing normal vector, $N^{\tau} > 0$.
$K$ transforms as a scalar and it is convenient to
choose coordinates such that the dilaton is constant on the boundary. 
A reparametrization $u(\tilde{u})$ corresponds to a coordinate change
from $f(u)$ and $\tau(u)$ to $\tilde{f}(\tilde{u}) = f(u(\tilde{u}))$
and $\tilde{\tau}(\tilde{u}) = \tau(u(\tilde{u}))$, and, without loss
of generality,  one
can demand that in the new coordinates $du\phi'(u) = d\tilde{u}\phi$, where
\begin{equation}
  \phi = \frac{1}{2\pi}\int_0^{2\pi}du\phi'(u) 
\end{equation}  
is the constant boundary value.
For nearly-de Sitter space the
extrinsic curvature becomes
\begin{equation}\label{dSexc}
 \lambda^{-1}K = 1  - \frac{1}{\lambda^2h} \left(\{f(u),u\} + \frac{f'(u)^2}{2}\right) +
  \mathcal{O}\left(\frac{1}{\lambda^4h^2}\right)  \ ,
\end{equation}
where
\begin{equation}
\{f(u),u\} = \frac{f'''(u)}{f'(u)} - \frac{3}{2}\frac{f''(u)^2}{f'(u)^2}
\end{equation}
is the Schwarzian derivative of $f(u)$ with respect to $u$.

The path integral now
reduces to an integral over the boundary modes with a measure $[Df]$
appropriate for a half-hyperboloid,
\begin{equation}\label{ZJT}
  \begin{split}
    Z_{\text{JT}}\left(\frac{\lambda\sqrt{h}}{\phi}\right) &=
    \int^{(h,\phi)}[Dg][D\phi'] \exp\left\{\Big(
      \frac{i}{2} \int_{\mathcal{M}}d^2x \sqrt{g} \phi' (R - 2\lambda)\right. \\
  &\left.\hspace{4cm} -i\int_{\partial\mathcal{M}} d\theta \sqrt{h} \phi' (K-1)\Big)\right\}\\
  &= N_E\int [Df] \exp{(-i\int_0^{2\pi}du\sqrt{h}\phi' (K-1))}\Theta(h -
  \lambda^{-2})\ .
  \end{split}
\end{equation}
The integral is given by fluctuations around a circle on the
$\text{dS}_2$ hyperboloid whose size has to be larger than the de Sitter
radius $\lambda^{-1}$. This leads to the theta function. 
The normalization factor includes a
contribution of the Euclidean section to the path integral, which is
independent of the boundary curve.
Analogously to $\text{AdS}_2$ JT gravity \cite{Stanford:2017thb}, one finds
for large $h$ and $\phi$ 
\cite{Maldacena:2019cbz,Cotler:2019nbi},
\begin{equation}\label{SchwarzianLimit}
  Z_{\text{JT}}\left(\frac{\lambda\sqrt{h}}{\phi}\right) = N_E
  \left(\frac{\phi}{\lambda\sqrt{h}}\right)^{3/2}
  \exp{\left( i\pi \frac{\phi}{\lambda\sqrt{h}}\right)} \ .
\end{equation}
This result can be expressed as trace of a Hamiltonian with an appropriately chosen
density of states, 
\begin{equation}
  Z_{\text{JT}}\left(\frac{\lambda\sqrt{h}}{\phi}\right) =
  i^{3/2}\int _0^\infty dE \rho(E) e^{(2\pi i\lambda\sqrt{h}E)} , \quad
\rho(E) = 2\phi N_E\sinh(2^{3/2}\pi\sqrt{\phi E})  ,
\end{equation}
which supports the idea of $\text{dS}_2$/$\text{CFT}_1$ correspondence.

\begin{figure}[t]
 \centering
 \includegraphics[width = 0.3 \textwidth]{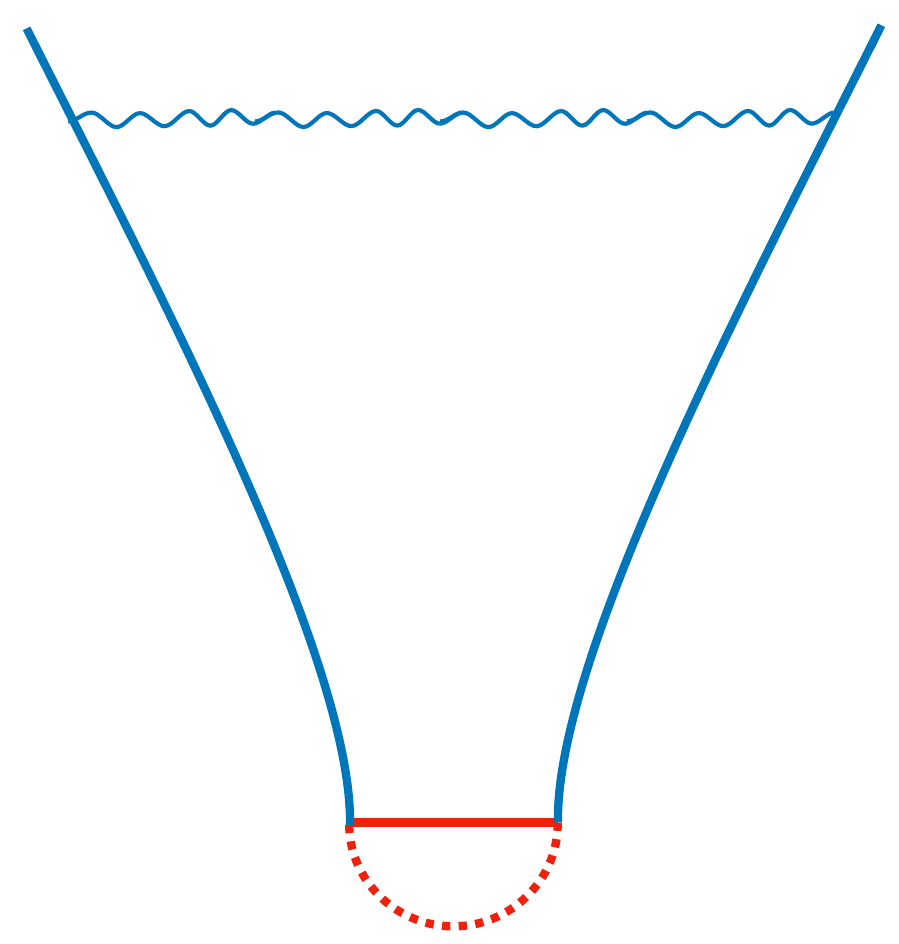} 
 \hspace{2cm}
 \includegraphics[width = 0.4 \textwidth]{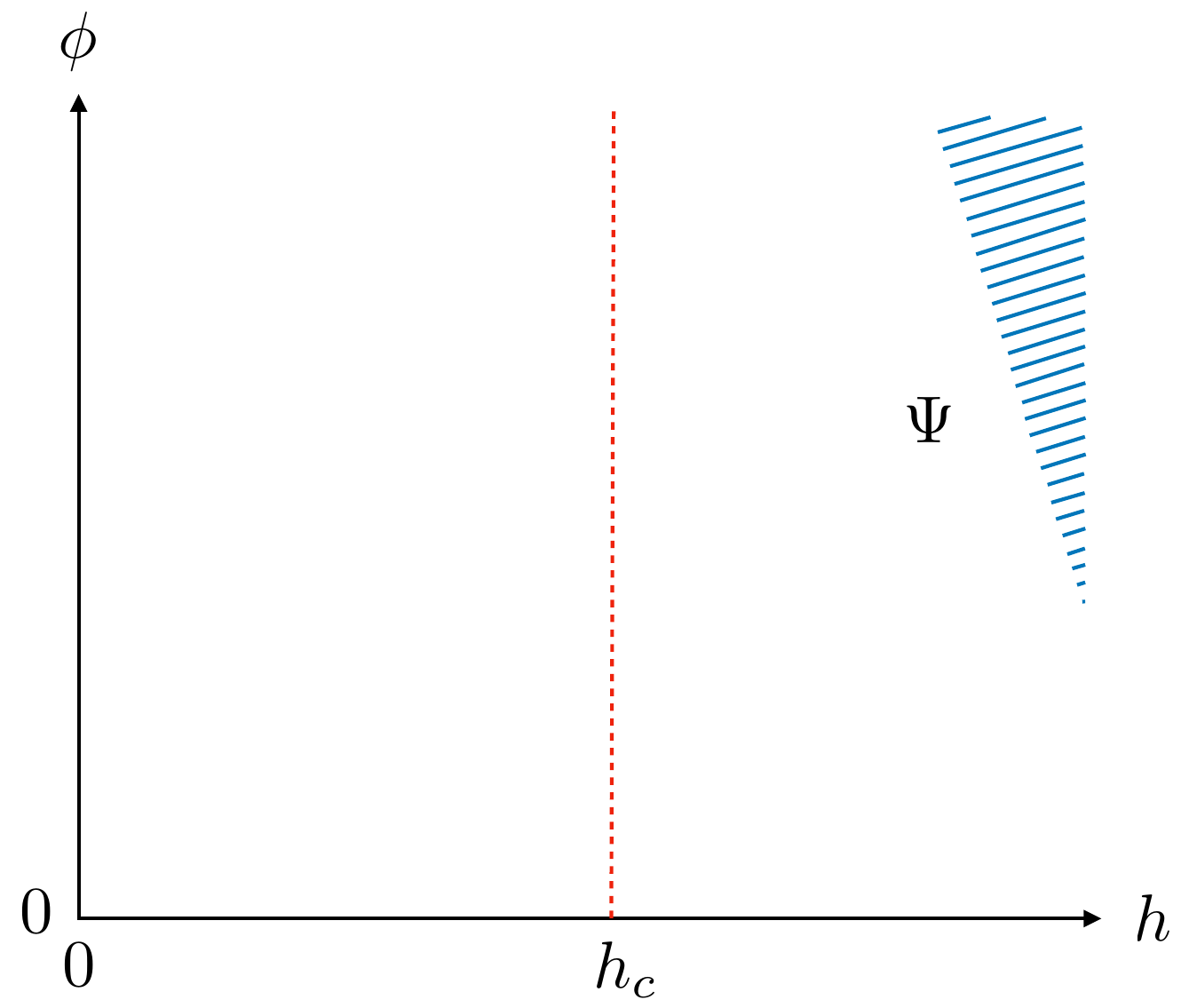} 
 \caption{Left: Complex Lorentzian/Euclidean geometry underlying no-boundary wave function. Right: Asymptotic wave function at large field values.}
\label{fig:WF1}
\end{figure}

$Z_{\text{JT}}$ is the analytic continuation of the disk partition function, but it is not a solution of a WDW equation and therefore not a wave function, as suggested in \cite{Cotler:2019nbi}. A wave function for large field values can be obtained from Eq.~\eqref{SchwarzianLimit} by multiplying with the phase factor that accounts for the leading term of the extrinsic curvature \eqref{dSexc}. Choosing the sign of $K$ corresponding to an `expanding universe', one obtains the wave function
\begin{equation}\label{CJM}
  \Psi_{\text{CJM}}(h,\phi) \sim
\left(\frac{\phi}{\lambda\sqrt{h}}\right)^{3/2} \exp{\left( -2\pi i
    \lambda \phi \sqrt{h}\left(1-\frac{1}{\lambda^2 h}\right)\right)} \ .
\end{equation}
  Alternatively, it has been argued that the connection between
  $Z_{\text{JT}}$ and wave function involves an additional factor
  $\phi$ \cite{Maldacena:2019cbz},
  \begin{equation}\label{MTY}
  \Psi_{\text{MTY}}(h,\phi) \sim \frac{1}{\phi}
\left(\frac{\phi}{\lambda\sqrt{h}}\right)^{3/2} \exp{\left( -2\pi i
    \lambda \phi \sqrt{h}\left(1-\frac{1}{\lambda^2 h}\right)\right)} \ .
\end{equation}
The two wave functions satisfy at large field values (see Fig.~\ref{fig:WF1}) two different WDW
equations. For $\Psi_{\text{MTY}}$ one has
\begin{equation}
(\partial_h\partial_\phi + 2\pi^2\lambda^2\phi)\Psi_{\text{MTY}}(h,\phi) =
0
\label{wdwmty}
\end{equation}
up to terms $\mathcal{O}(1/\chi^3)$, with $\chi = \sqrt{h},\phi$. This corresponds to `canonical factor ordering'
in canonical quantization. On the contrary, $\Psi_{\text{CJM}}$
satisfies the equation
\begin{equation}
(\sqrt{h}\partial_h\frac{1}{\sqrt{h}}\partial_\phi +
2\pi^2\lambda^2\phi)\Psi_{\text{CJM}}(h,\phi) = 0
\end{equation}
up to terms $\mathcal{O}(1/\chi^3)$, with $\chi = \sqrt{h},\phi$. As
shown in \cite{Iliesiu:2020zld}, this `Henneaux factor ordering'
arises if one solves the functional WDW equation for JT gravity by
solving the two functional momentum constraints individually
\cite{Henneaux:1985nw,Louis-Martinez:1993bge}. Factor orderings of this type have
been studied in the past (see, for example \cite{Hartle:1983ai,Kiefer:2007ria,Lehners:2023yrj}). They
correspond to different options to
quantize JT gravity, which are related to different definitions of the
path integral measure \cite{Halliwell:1988wc}, see Appendix~\ref{sec:canonical} for more details.

In the following section we shall proceed to derive a general class of exact bulk dS JT wave functions valid for a corresponding class of factor orderings. We will do so by using the bulk path integral \'a la Halliwell-Louko \cite{Halliwell:1990tu}, as well as by directly determining the bulk WDW propagator for a characteristic initial value problem. 

As we shall see, the
resulting bulk wave function has a precise asymptotic duality to the Schwarzian path integral on the boundary curve of $\text{dS}_2$, which holds for all choices of factor ordering. En passant, this explains how different results \eqref{CJM}, \eqref{MTY} for the wave function discussed above are related via a choice of factor ordering.

\section{JT wave functions from the JT bulk amplitude}\label{sec:holo}

If one is interested in the exact wave function, the situation just described is not fully satisfactory for several reasons: On the one hand, while we tried to give a path-integral justification of the gluing between Euclidean half-sphere and Lorentzian half-hyperboloid, it remains conceptually unclear whether this logic can lead to exact results (see however~\cite{Iliesiu:2020zld} as well as the concerns raised in \cite{Chaudhuri:2024yau}). On the other hand, the alternative possibility of analytically continuing the Euclidean wave function faces the problem of obtaining an exact result for the latter \cite{Stanford:2020qhm, Chaudhuri:2024yau}. 

Thus, we return to the defining expression \eqref{dexp} and propose taking the unavoidable presence of a second boundary more seriously. In other words, we start with the amplitude
\begin{equation}\label{aPI1a}
     \langle h,\phi|h',\phi'\rangle = \int\limits_{(h',\phi')}^{(h,\phi)}[Dg][D\phi''] \exp{(iS[g,\phi''])}\,,
\end{equation}
assuming that it can eventually be turned into the desired wave function by convolving it with an appropriate distribution of $h',\phi'$.

Let us first evaluate this amplitude following \cite{Halliwell:1990tu}. We write the metric as
\begin{equation}
  ds^2 = -\frac{N^2}{h}dt^2 + h d\theta^2
\end{equation}
and fix the gauge such that both $h$ and the bulk field $\phi$ are spatially constant. We denote these variables by $\bar{h}(t)$ and $\bar{\phi}(t)$. Removing the GHY boundary term in the JT action \eqref{jtL} using
Gauss' theorem, the action in our path integral takes the form
\begin{equation}\label{aIh1}
  S=S[\bar{h},\bar{\phi};N] =  2\pi\int_0^1 dt
  \left(-\frac{1}{2N}\partial_t\bar{\phi}\partial_t\bar{h} -  N\lambda^2\bar{\phi}\right)\ .
\end{equation}
The transition amplitude now reads  \begin{equation}\label{aPI1}
     \langle h,\phi|h',\phi'\rangle = \int dN [D\bar{h}][D\bar{\phi}] \exp{(iS[\bar{h},\bar{\phi};N])}\,,
\end{equation}
where one has to integrate over the real variable $N$ and two functions with boundary conditions $\bar{h} = h'$ and $\bar{\phi} = \phi'$ at $t = 0$ as well as $\bar{h} = h$ and $\bar{\phi} = \phi$ at $t = 1$. 

The integral is easily evaluated by expanding around a saddle point satisfying the equations of motion
\begin{equation}
  \partial^2_t{\bar{h}} - 2\lambda^2 N^2 = 0\ , \quad \partial^2_t{\bar{\phi}} = 0 \ .
\end{equation}
The solutions read
\begin{equation}
  \tilde{h} = h' + (h-h'-\lambda^2 N^2) t + \lambda^2 N^2 t^2\ , \quad
  \tilde{\phi} = \phi'+(\phi-\phi')t \ .
\end{equation}
The total action may be written as a sum, $S=S_0+S_2$, of the saddle point action
\begin{equation}
  S_0(\alpha,\beta;N) = S[\tilde{h},\tilde{\phi};N] = \frac{1}{2}\left(\alpha N -
    \frac{\beta}{N}\right)\ ,
\end{equation}
with
\begin{equation}
  \alpha = -2\pi\lambda^2(\phi+\phi')\ , \quad
  \beta = 2\pi(\phi-\phi')(h-h') \ ,
  \end{equation}
and a term quadratic in the fluctuations,
  \begin{equation}
    S_2[\delta h,\delta \phi;N] = 2\pi
    \int_0^1dt\left(-\frac{1}{2N}\delta \dot{h}\delta \dot{\phi}\right) \ .
  \end{equation}
Path integrals of this type have been studied in detail in \cite{Halliwell:1990tu}. The action \eqref{aIh1} is very similar to the Kantowski-Sachs model considered in \cite{Halliwell:1990tu, Fanaras:2021awm}.

The path integral over the fluctuations is Gaussian and yields a result proportional to $N^{-1}$ so that
the transition amplitude \eqref{aPI1} becomes\footnote{
Note 
that this result depends on the measure chosen in the path integral \eqref{aPI1}, which is related to the factor ordering in the WDW equation \cite{Halliwell:1988wc}. As we shall see, the choice made here corresponds to `canonical'  factor ordering.
}  
\begin{equation}\label{aPI2}
    \langle h,\phi|h',\phi'\rangle = \int \frac{dN}{N} 
    \exp{(i S_0[\alpha,\beta;N])} \ .
\end{equation}
To respect diffeomorphism invariance, including time reversal, we integrate over $N\in (-\infty,0)$ and $N\in(0,\infty)$. Adding these integrals leads to a purely imaginary result.\footnote{
In 
principle, a relative phase could be introduced in this sum, but this would not change our qualitative conclusions.
} 
The oscillatory behaviour of the exponential at $N\to 0$ and $N\to \pm\infty$ ensures that the integral is a priori well defined.
We are interested in the regime $h > h'$, $\phi > \phi'$, such that $\alpha < 0 < \beta$. The explicit evaluation using the steepest descent method then gives \cite{Halliwell:1990tu}\,\footnote{
See
specifically Table I in Sect. III of \cite{Halliwell:1990tu}. Note that our $N$ corresponds to their $iN$.
}
\begin{equation}\label{aRA}
  \langle h,\phi|h',\phi'\rangle = -2\pi i
  J_0((-\alpha\beta)^{1/2}) \quad .
\end{equation}
It can be shown on general grounds that amplitudes defined by integrating $N$ over half-axes are related to Green functions $G$ of the corresponding WDW equations \cite{Halliwell:1988wc}. In our case, the defining equation reads
\begin{equation}\label{aGphi}
(\partial_h\partial_\phi + 2\pi^2\lambda^2\phi) G(h,\phi;h',\phi') =
\delta(\phi-\phi')\delta(h-h') \ .
\end{equation}
Exchanging the variable $\phi$ for $\sigma=\phi^2$ one sees that this is nothing but a Klein-Gordon equation in light-cone coordinates. The relation of the retarded Green function to the amplitude calculated above is then fixed explicitly by 
\begin{equation}
  G(h,\phi;h',\phi') =  J_0((-\alpha\beta)^{1/2})
  \Theta(\phi-\phi')\Theta(h-h') \,,
  \label{gftt}
\end{equation}
cf.~Appendix~\ref{sec:cauchy} for more details. Here by `retarded' we mean retarded w.r.t. the `time' variable $h+\sigma$.

Now we return to our main conceptual point: We want to define the wave function of the universe using the amplitude or, equivalently, the Green function, as just derived. The standard approach for this would be to convolve the Green function with an appropriate source. However, instead of prescribing the source we may also prescribe the `initial' or boundary values of the wave function we are after. Choosing the boundary to be the line $\phi=\phi'$, with $\phi'$ fixed, the full wave function then reads
\begin{equation}\label{apsibulk}
  \begin{split}
  \Psi(h,\phi) &= \frac{1}{2}\int dh' \left(G(h,\phi;h',\phi')\partial_{h'}\Psi(h',\phi')
    - \partial_{h'}G(h,\phi;h',\phi')\Psi(h',\phi') \right)\\
  &= \Theta(\phi-\phi')\left(\Psi(h,\phi')
    - \int_0^h dh' \Psi(h',\phi') \partial_{h'} J_0((-\alpha\beta)^{1/2})\right) \,,
\end{split}
\end{equation}
see Appendix~\ref{sec:cauchy} for calculational details. Note that the last term vanishes as $\phi$ approaches $\phi'$ so that the initial condition is satisfied.

\begin{figure}[t]
 \centering
 \hspace{1cm}
 \includegraphics[width = 0.4 \textwidth]{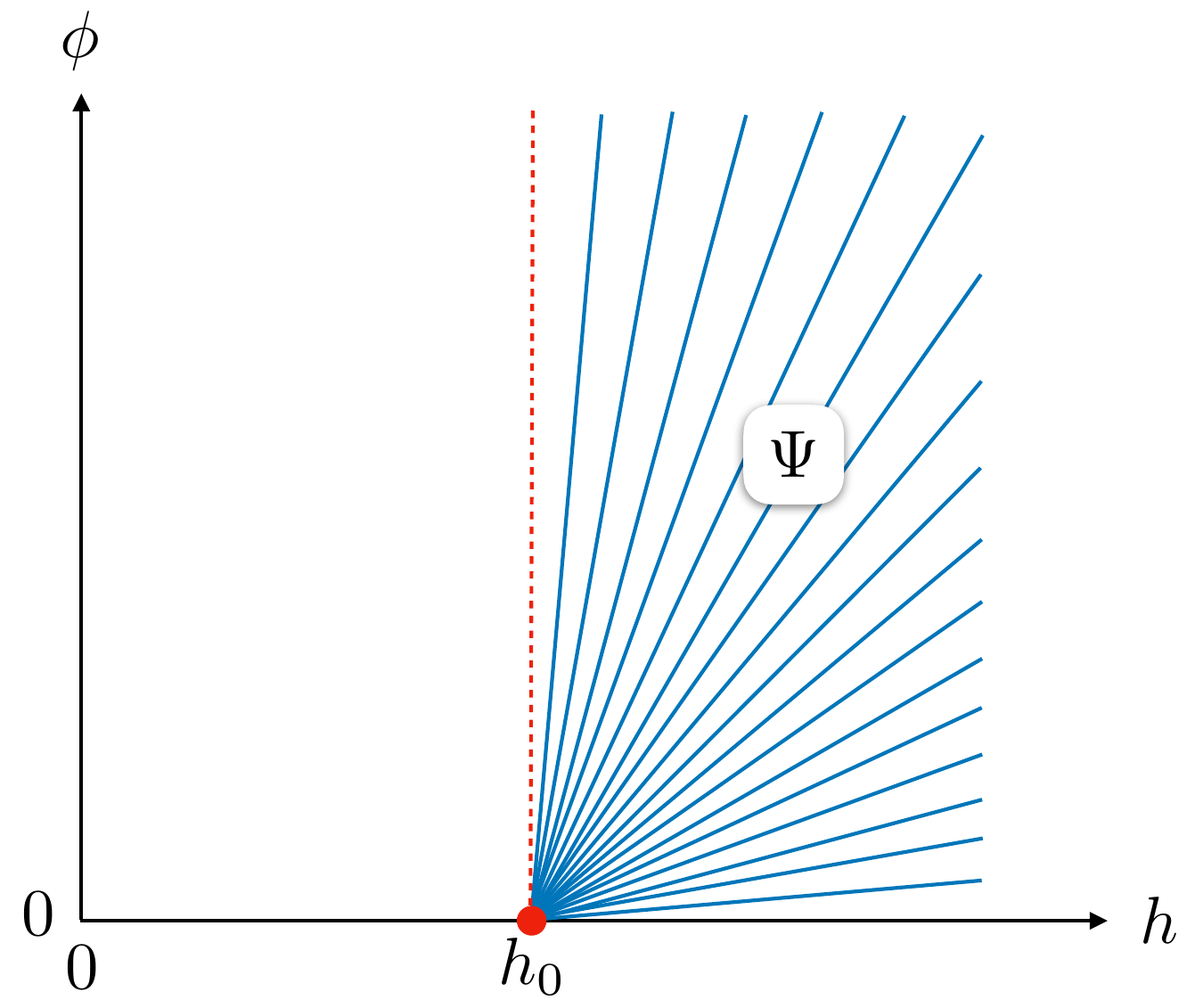}
 \caption{Wave function with non-zero values at $h > h_0$, $\phi > 0$.} \label{fig:ahalfJTbulk}
\end{figure}

Some comments are in order: First, our choice of a constant-$\phi$ boundary is justified by our desire to restrict the model to positive dilaton values by choosing $\phi'=0$. This is in line with interpreting JT gravity as an effective theory for a compactification, with the dilaton corresponding to a volume. It is also in line with the perturbatively controlled regime being at large $\phi$. Second, having made the choice $\phi'=0$, our boundary data is determined by a function $\chi(h')\equiv \Psi(h',0)$. The simple choice 
\begin{equation}\label{eq:aSimpSingSource}
\chi(h')\equiv \chi_{h_0}(h')\equiv\delta(h'-h_0)\,\,,\qquad h_0>0
\end{equation}
then ensures that the wave function vanishes on the axis $h=0$. This is expected since, at this locus, the spatial slice becomes singular in the sense that its volume vanishes. We consider this relevant independently of the fact that the 2d geometry can of course remain smooth. The technical reason for the vanishing of the wave function at this locus is that our Green function or propagator `transports' information only upward and to the right in the $h$-$\phi$-plane, cf.~Fig.~\ref{fig:ahalfJTbulk}. Thus, as advertised in the Introduction, we have obtained (a first version of) a wave function in the upper-right quadrant based on physically motivated boundary data.

We may use \eqref{apsibulk} and \eqref{eq:aSimpSingSource} as well as the relations $\partial_{h'}J_0(\sqrt{-\alpha\beta})=-\partial_{h}J_0(\sqrt{-\alpha\beta})$ and $J_0'=-J_1$ to write the wave function explicitly as
\begin{equation}\label{eq:aBulkWFdeltaSource}
\Psi(h,\phi;h_0)=\delta(h-h_0)- 2\pi^2\lambda^2
\phi^2 x^{-1} J_1(x) \Theta(h-h_0)\ ,\quad x=2\pi\lambda\phi\sqrt{h-h_0}\,.
\end{equation}
Here we have given $\Psi$ an extra argument $h_0$ to emphasize the key role of the $\delta$-function choice for the boundary value on the $h$-axis. We note that this singular choice leads to a persistent singularity of $\Psi$ at $h=h_0$ and positive $\phi$.\footnote{The singular boundary condition \eqref{eq:aSimpSingSource} can also be interpreted as a singular source on the boundary \cite{Morse:1980rh}. The wave function 
$\hat{\Psi}(h,\phi;h_0)) = \Psi(h,\phi;h_0)\Theta(\phi)$ satisfies a WDW equation with singular source: $(\partial_h\partial_\phi + 2\pi^2 \lambda^2\phi) \hat{\Psi} = -\partial_{h_0}\delta(h-h_0)\delta(\phi)$.}

A logical choice for the location of our $\delta$-function is the critical value $h_0=h_c\equiv 1/\lambda^2$. This corresponds to taking the semi-classical Hartle-Hawking picture of gluing the de Sitter hyperboloid to a half-sphere completely seriously. (In the notation of \cite{Cotler:2019nbi} this means enforcing the choice $\alpha=1$ for the bulk metric.)

However, given that the Euclidean half-sphere is only one of the many (quantum) geometries that may connect an $h_0$-sized circle to the singular boundary value $h=0$, it is very reasonable to consider more general wave functions. Indeed, by linearity of the problem we may replace the singular choice $\chi_{h_0}$ in Eq.~\eqref{eq:aSimpSingSource} by a smooth distribution $\rho(h_0)$. This leads to the more general wave function
\begin{equation}
\Psi(h,\phi)=\int dh_0\,\Psi(h,\phi;h_0)\,\rho(h_0)\,.
\label{conv}
\end{equation}

Before closing this section, let us place our findings into the context of other results. First, return to the fixed-$h_0$ wave function of \eqref{eq:aBulkWFdeltaSource} and 
apply the decomposition $2J_1(x)=H_1^{(1)}(x)+H_1^{(2)}(x)$. The `positive frequency part' $H_1^{(2)}(x)$ of this wave function agrees with the outgoing WDW solution obtained by~\cite{Maldacena:2019cbz} for the canonically factor-ordered WDW equation (cf.~footnote 5 of ~\cite{Maldacena:2019cbz} and recall that $H_1^{(2)}(x)=-(2/\pi)K_1(ix)$.)

This, of course, also holds at the asymptotic level: Using the asymptotic form of the Bessel function $J_1$ one obtains from \eqref{eq:aBulkWFdeltaSource} for values $h \gg h_0$:
 \begin{equation}\label{apsiasymp}
\Psi(h,\phi) \sim 
   \frac{1}{\phi}\left(\frac{\phi}{\lambda\sqrt{h}}\right)^{3/2}
   \cos{\left(2\pi\phi\lambda\sqrt{h}\left(1 - \frac{h_0}{2h}\right)\right)}\,.
 \end{equation} 
The positive-frequency part of this,
\begin{equation}\label{apsiasymp2}
 \Psi_+(h,\phi) \sim \frac{1}{\phi}\left(\frac{\phi}{\lambda\sqrt{h}}\right)^{3/2}
   \exp{\left(-i\,2\pi\phi\lambda\sqrt{h}\left(1 - \frac{h_0}{2h}\right)\right)}\,,
\end{equation}
agrees with the result of \cite{Maldacena:2019cbz} (cf.~our Eq.~\eqref{MTY})
to leading order in prefactor and phase. Further agreement in the first subleading piece requires the particular choice $h_0=h_c=1/\lambda^2$. We recall that, by contrast, our proposal for a general wave function is a linear superposition of many such solutions, specified by the distribution $\rho(h_0)$.

\begin{figure}[t]
 \centering
 \includegraphics[width = 0.7 \textwidth]{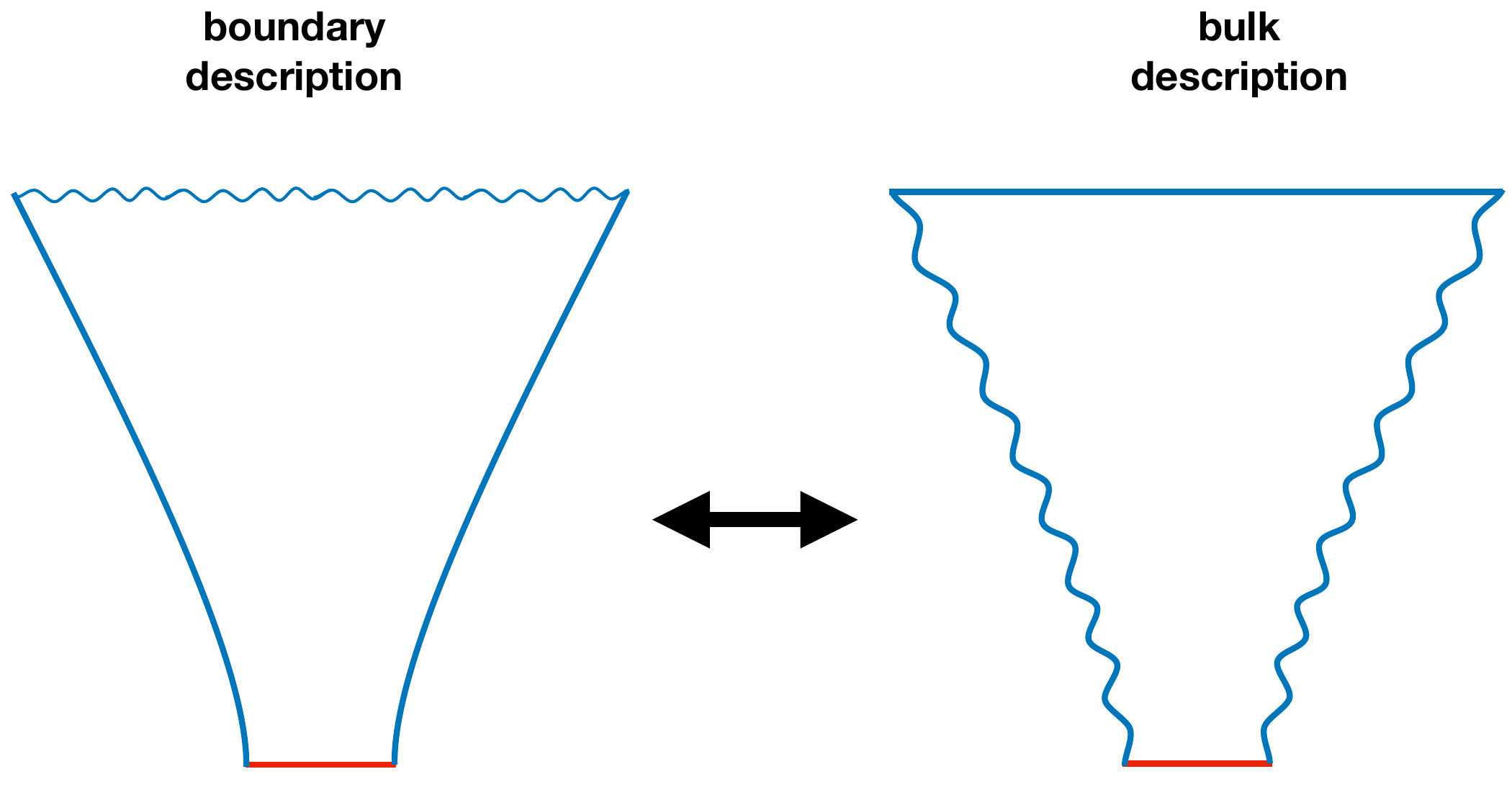}
 \caption{Bulk-boundary duality and wave function for $h > h_0$, $\phi > 0$ from bulk fluctuations.} 
  \label{fig:halfJTbulk}
\end{figure}

The perfect agreement for $\rho(h_0)=\delta(h_0-h_c)$ is a remarkable result as in~\cite{Maldacena:2019cbz} the semiclassical limit of the positive frequency part of the wave function was
determined by integrating out fluctuations of the boundary curve for fixed bulk
geometry. On the contrary, we derived the wave function by integrating over bulk fluctuations of both dilaton and metric, as prescribed by the bulk path integral. In this sense, the agreement is an example of holography, as illustrated in Fig.~\ref{fig:halfJTbulk}.

Finally, we note that our wave function \eqref{apsiasymp} differs from the semiclassical result discussed in Appendix \ref{sec:semiJT} (cf.~\eqref{psiJT}) by a factor $(\phi/\sqrt{h})^{1/2}$. This
difference is an effect of quantum fluctuations that are not included
in the semiclassical approximation.

\section{JT wave functions from the WDW equation}
\label{sec:charivp}
  
In the previous section we have discussed a path integral
representation of the amplitude or propagator to derive a real stationary Hartle-Hawking-type wave function and its complex-valued expanding branch. 
Our analysis focused on a particular (canonical) choice of factor ordering and on $\delta$-function-type boundary conditions.
We now wish to consider more general sources and factor orderings. 
For the latter, the path integral measure is modified (cf.~Appendix~\ref{sec:canonical}) and we do not know how to explicitly evaluate the amplitude and hence the wave function. However, we may take advantage of the fact that all wave functions can be equivalently expressed as solutions of a WDW equation,  supplemented by appropriate boundary conditions. The choice of factor ordering can then be introduced directly in the WDW equation.
In this section we therefore initially solve the WDW equation with two choices of factor ordering known from the literature and implement
boundary conditions consistent with the expected asymptotic behaviour. We will then proceed to discuss more general factor orderings and, correspondingly, more general boundary conditions.

\subsection{Riemann representation of solutions}\label{sec:canon}

Consider the WDW equation \eqref{wdwmty}. Changing variables from
$\phi$ to $\sigma = \phi^2$, one obtains
 \begin{equation}\label{wdwsigma}
  (\partial_h\partial_\sigma + \pi^2\lambda^2 ) \Psi(h,\phi) = 0 \ ,
\end{equation}
which is the Klein-Gordon equation in light-cone coordinates. As argued before, our physical field space is a quarter-plane with lightlike boundaries:
the $h$-axis ($\sigma = 0$, $h \geq 0$) and the $\sigma$-axis
($h = 0$, $\sigma \geq 0$) (see Fig.~\ref{fig:ahalfJTbulk}), which represent characteristics
of the wave equation.
Hence, we are dealing with a
characteristic initial value problem.

The general solution of this problem has been given in \cite{Courant:1965st} in
terms of a ``Riemann function'' $R_>(h,\phi;h',\phi')$\footnote{In
  \cite{Courant:1965st} the solution
  is described in Ch.~V \S~5,  ``Hyperbolic differential equations in two independent
  variables'', for the example of the telegraph equation.}, that
satisfies the WDW equation \eqref{wdwsigma} with $h'$ and $\phi'$ as
parameters. Moreover, $R_>$ is subject to a reciprocity condition,
\begin{equation}\label{riem1}
  R_>(h,\phi; h',\phi') = R_>(h',\phi'; h,\phi)\ , \quad
  R_>(h,\phi; h,\phi) = 1 \ , 
\end{equation}
and, since the lightlike boundaries are characteristics, conditions
on the derivatives,
\begin{equation}\label{riem2}
  \partial_{h'} R_>(h,\phi; h',\phi) =
     \partial_{\phi^{'2}} R_>(h,\phi;h,\phi') = 0 \ .
   \end{equation}
For $h, h',\phi, \phi' \geq 0$ and $(\phi^2-\phi^{'2})(h-h') \geq 0$,
these requirements determine the Riemann function to be given by a regular Bessel function,
\begin{equation}\label{bes}
  R_>(h,\phi; h',\phi') = J_0(2\pi\lambda\sqrt{(\phi^2-\phi^{'2})(h-h')}) \ .
  \end{equation}
One easily verifies that  \eqref{riem1} and
\eqref{riem2} are satisfied. Note the invariance w.r.t.~shifts in $\phi^2$ and $h$. Up to theta-functions, the Riemann
function agrees with the retarded and advanced Green functions of the Klein-Gordon equation (see Appendix~\ref{sec:cauchy}).

A solution of the homogeneous wave equation is uniquely determined by specifying
on a spacelike surface the wave function and its normal derivative. By contrast, on a lightlike boundary surface only the wave function itself
can be freely chosen. In our case the unique solution can be written
in terms of the boundary values along the $h$-axis 
and the $\phi$-axis  \cite{Courant:1965st},
\begin{equation}\label{sol}
  \begin{split}
   \Psi_>(h,\phi) &=
   \frac{1}{2}\left(R_>(h,\phi;0,\phi)\Psi(0,\phi)
     + R_>(h,\phi;h,0)\Psi(h,0)  \right)\\
&+ \frac{1}{2}\int_0^{h} dh'
\left(R_>(h,\phi;h',0)\partial_{h'}\Psi(h',0)
  - \partial_{h'} R_>(h,\phi;h',0)\Psi(h',0)\right)\\
&+ \frac{1}{2}\int_0^{\phi^2} d\phi^{'2}
\left(R_>(h,\phi;0,\phi')\partial_{\phi^{'2}}\Psi(0,\phi')
- \partial_{\phi^{'2}}R_>(h,\phi;0,\phi')\Psi(0,\phi')\right) \ .
\end{split}
\end{equation}
For wave functions that vanish at $h=0$, only the non-zero boundary conditions along the $h$-axis are relevant. In terms of 
\begin{equation}\label{bRlarger}
  \bar{R}_>(h,\phi; h',\phi') \equiv R_>(h,\phi; h',\phi')\Theta(h-h')
\end{equation}
the solution \eqref{sol} can then be expressed as
\begin{equation}\label{solRb}
  \Psi_>(h,\phi)
  =\frac{1}{2}\int_0^{\infty} dh'
\left(\bar{R}_>(h,\phi;h',0)\partial_{h'}\Psi(h',0)
  - \partial_{h'} \bar{R}_>(h,\phi;h',0)\Psi(h',0)\right) \ .
\end{equation}
We now see clearly that, modulo a factor $\theta(\phi-\phi')$, the function $\bar{R}_>$ is the Green function appearing in Eq.s~\eqref{gftt}, \eqref{apsibulk}.

\begin{figure}[t]
 \centering
 \includegraphics[width = 0.4 \textwidth]{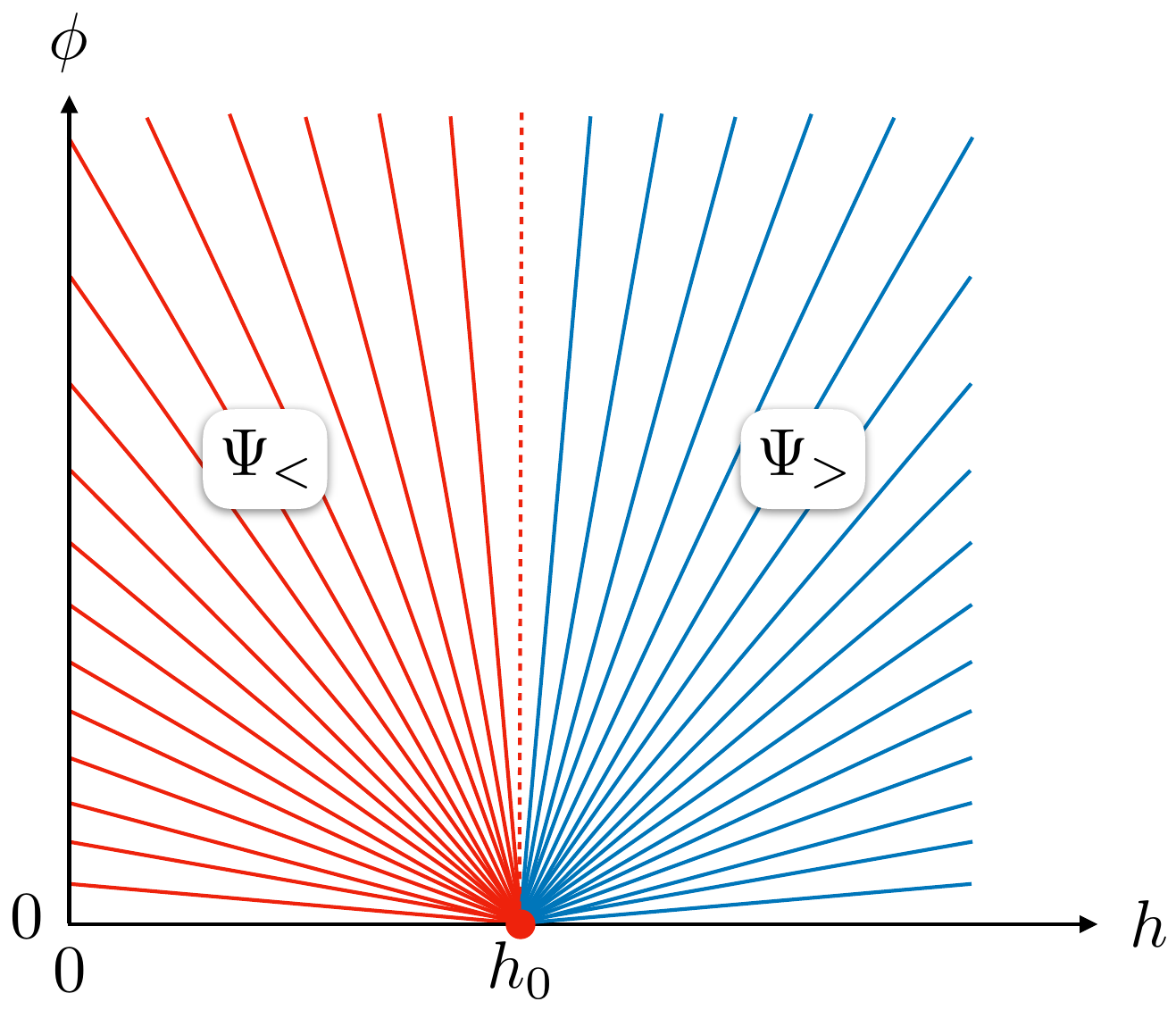}
 \caption{Solution of the WDW equation for $h > h_0$ and $h<h_0$.}
  \label{fig:fullJTbulk}
\end{figure}

So far, we have only rephrased what we already knew. In particular, we prescribed the functional form $\delta(h-h_0)$ on the $h$-axis and propagated this information to larger $\phi,h$ using the Riemann function method. Of course, we can alternatively propagate the information to larger $\phi$, as before, but now to smaller $h$.  We may again use the Riemann function method, adapted for propagation into the upper-left rather than the upper-right quadrant. The two resulting solutions, the first a repetition of \eqref{eq:aBulkWFdeltaSource}, read
\begin{eqnarray}
 \Psi_>(h,\phi;h_0) &=& \delta(h-h_0) 
- 2\pi^2\lambda^2\phi^2 x^{-1}J_1(x)\,\Theta(h-h_0)\,,
\label{psigr}
\\
\nonumber
\\
 \Psi_<(h,\phi;h_0) &=& \delta(h-h_0) 
+ 2\pi^2\lambda^2\phi^2 x^{-1}J_1(x)\,\Theta(h_0-h)
\label{psile}
\\
&= &
\delta(h-h_0) 
+ 2\pi^2\lambda^2\phi^2 (-ix)^{-1}I_1(-ix)\,\Theta(h_0-h)
\ .\phantom{\Bigg|}\label{psile2}
 \end{eqnarray}
See Fig.~\ref{fig:fullJTbulk} for an illustration. In fact, it is straightforward to obtain \eqref{psile} directly from \eqref{psigr}, without repeating the analysis: Propagating to smaller rather than to larger $h$ amounts, at the technical level, to exchanging $h-h_0$ for $h_0-h$. Under such a reflection, our WDW equation \eqref{wdwsigma} remains invariant if we also exchange $\lambda^2$ for $-\lambda^2$. Thus, a solution with propagation to larger $h$ (`$>$') turns into a solution with propagation to smaller $h$ (`$<$') if we perform a reflection in both the parameter $\lambda^2$ and the variable $h-h_0$. Noticing that $x^{-1}J_1(x)$ is an analytic function of $x^2=(2\pi\lambda\phi)^2(h-h_0)$ and hence invariant, we see that \eqref{psigr} turns into \eqref{psile}. Naturally, if one prefers to think in terms of the variable $x$, then this variable is real in \eqref{psigr} and imaginary in
\eqref{psile}, \eqref{psile2}.

In the region $h<h_0$, relevant for $\Psi_<$, the variable $x$ is imaginary. It may hence be useful to also present the result in terms of a modified Bessel function with real argument, cf.~\eqref{psile2}. Comparing \eqref{psile2} with \eqref{psigr} and recalling the asymptotic behaviour of Bessel functions, one observes that the line $h=h_0$ plays the role of a Stokes line, separating regions with oscillatory and exponential behaviour. This is similar to the corresponding behaviour of the minisuperspace Hartle-Hawking wave function in the on-shell and off-shell regimes.

For later use, we record the following concise presentation of our two solutions: 
\begin{equation}
 \Psi_>(h,\phi;h_0) = \partial_h \,[\,J_0(x)\,\Theta(h-h_0)\,]\,\,,
\qquad
\Psi_<(h,\phi;h_0) = -\partial_h \,[\,J_0(x)\,\Theta(h_0-h)\,]\,.
\label{bcom}
 \end{equation}
As an intriguing observation, we now note that the two solutions above may be combined as
\begin{equation}
\Psi_0(h,\phi;h_0) \equiv \frac{1}{2}\left[
\Psi_>(h,\phi;h_0)-\Psi_<(h,\phi;h_0)\right] = - 2\pi^2\lambda^2\phi^2 x^{-1}J_1(x)
\label{psi0d}
\end{equation}
or, even more concisely,
\begin{equation}
   \Psi_0(h,\phi;h_0) = \partial_h
   J_0(2\pi \lambda\phi\sqrt{h-h_0}) \ .
   \label{psi0der}
\end{equation}
This wave function is particularly interesting as, by its very construction, it is identically zero on the $h$-axis. It is also an analytic function of $h$ and $\phi$ within the upper-right quadrant. In fact, using this feature, one could have obtained \eqref{psi0d} from \eqref{psigr} even more straightforwardly: One starts by observing that $\Psi_>$ is analytic in the region $\phi>0$, $h>h_0$. Then one analytically continues this solution to the whole upper-right quadrant, calling the result $\Psi_0$. Now, since $\Psi_>=0$ on the $h$-axis for $h>h_0$, analyticity in $h$ implies that $\Psi_0$ is identically zero on the entire $h$ axis. We note that, as before, more general solutions are obtained by convolving $\Psi_0(h,\phi;h_0)$ with some density function $\rho(h_0)$, cf.~\eqref{conv}.

Of course, our smooth solution $\Psi_0$ has an important shortcoming: It does not vanish on the $\phi$-axis, a condition for which we argued above on physical grounds. This, however, is overcome automatically when we turn to alternative factor orderings below.

\subsection{Henneaux factor ordering}
\label{sec:henn}

Let us now consider the WDW equation with Henneaux factor ordering,
 \begin{equation}\label{wdwsigmaH}
  (\sqrt{h}\partial_h\frac{1}{\sqrt{h}}\partial_\sigma + \pi^2\lambda^2 ) \tilde{\Psi}(h,\phi) = 0 \ .
\end{equation}
Any solution $\Psi$ of the WDW equation \eqref{wdwsigma} with canonical factor ordering gives rise to a solution $\tilde{\Psi}(h,\phi) = \sqrt{h}\Psi(h,\phi)$ of \eqref{wdwsigmaH}. This applies, in particular, to our original solution \eqref{eq:aBulkWFdeltaSource} with $\delta$-function boundary conditions. It also applies to the induced analytic solution \eqref{psi0der}, which hence gives rise to the solution
\begin{equation}
   \tilde{\Psi}_0(h,\phi;h_0) = \sqrt{h}\,\partial_h
   J_0(2\pi \lambda\phi\sqrt{h-h_0})
\end{equation}
of \eqref{wdwsigmaH}. This is a remarkable result since the above solution is analytic in the upper-right quadrant and, thanks to the prefactor $\sqrt{h}$, vanishes not only on the $h$- but also on the $\phi$-axis.

A possible shortcoming is the loss of the `Schwarzian' $h^{-3/4}$ behaviour at asymptotically large-$h$. We emphasised this behaviour in our discussion following \eqref{apsiasymp}, \eqref{apsiasymp2} above. Now it is lost due to the extra prefactor $\sqrt{h}$. It is, however, easily regained by replacing our $\delta$-function boundary condition, on which our whole construction was based, by a more singular choice:
\begin{equation}\label{bddelta}
\chi_{h_0}^{(1)}(h)\equiv\chi_{h_0}(h)=\delta(h-h_0)\qquad\to\qquad
  \chi^{(2)}_{h_0}(h) = \partial_h\delta(h-h_0) \ .
  \end{equation}
By linearity and using the presentation of our canonical solutions in
\eqref{bcom}, the corresponding more singular wave function with Henneaux factor ordering reads
\begin{align}
   \tilde{\Psi}_>(h,\phi;h_0) 
    &=\sqrt{h}\,\partial^2_{h}\,[\, J_0(x)\, \Theta(h-h_0)\,]\phantom{\Big|} \\
 &\hspace*{-1cm}=\sqrt{h}\left[ \partial_h\delta(h-h_0) 
+2\partial_h J_0(x) \delta(h-h_0) + (2\pi^2\lambda^2\phi^2)^2 x^{-2} J_2(x) \Theta(h-h_0)\right]\ .
\nonumber
\end{align}
It does, by construction, satisfy the boundary conditions $ \tilde{\Psi}_>(0,\phi;h_0)
= 0$ and $\tilde{\Psi}_>(h,0;h_0) =  \sqrt{h}\,\chi^{(2)}_{h_0}(h)$. Moreover, it vanishes for $h-h_0 < 0$ and is singular along the characteristic $h=h_0$. For large $h$ and $\phi$ the asymptotic
form of the CJM wave function \cite{Cotler:2019nbi} is now reproduced,
\begin{equation}
 \tilde{\Psi}_>(h,\phi;h_0) \sim
 \left(\frac{\phi}{\sqrt{h}}\right)^{3/2}
\cos{(2\pi\lambda\phi\sqrt{h})} \ .
\end{equation}
Note that, while this has the same `Schwarzian' $h$-dependence as in \eqref{apsiasymp}, \eqref{apsiasymp2}, the $\phi$-dependence has now changed from the one of MTY \cite{Maldacena:2019cbz} to that of CJM \cite{Cotler:2019nbi}.

\begin{figure}[t]
 \centering
 \includegraphics[width = 0.6 \textwidth]{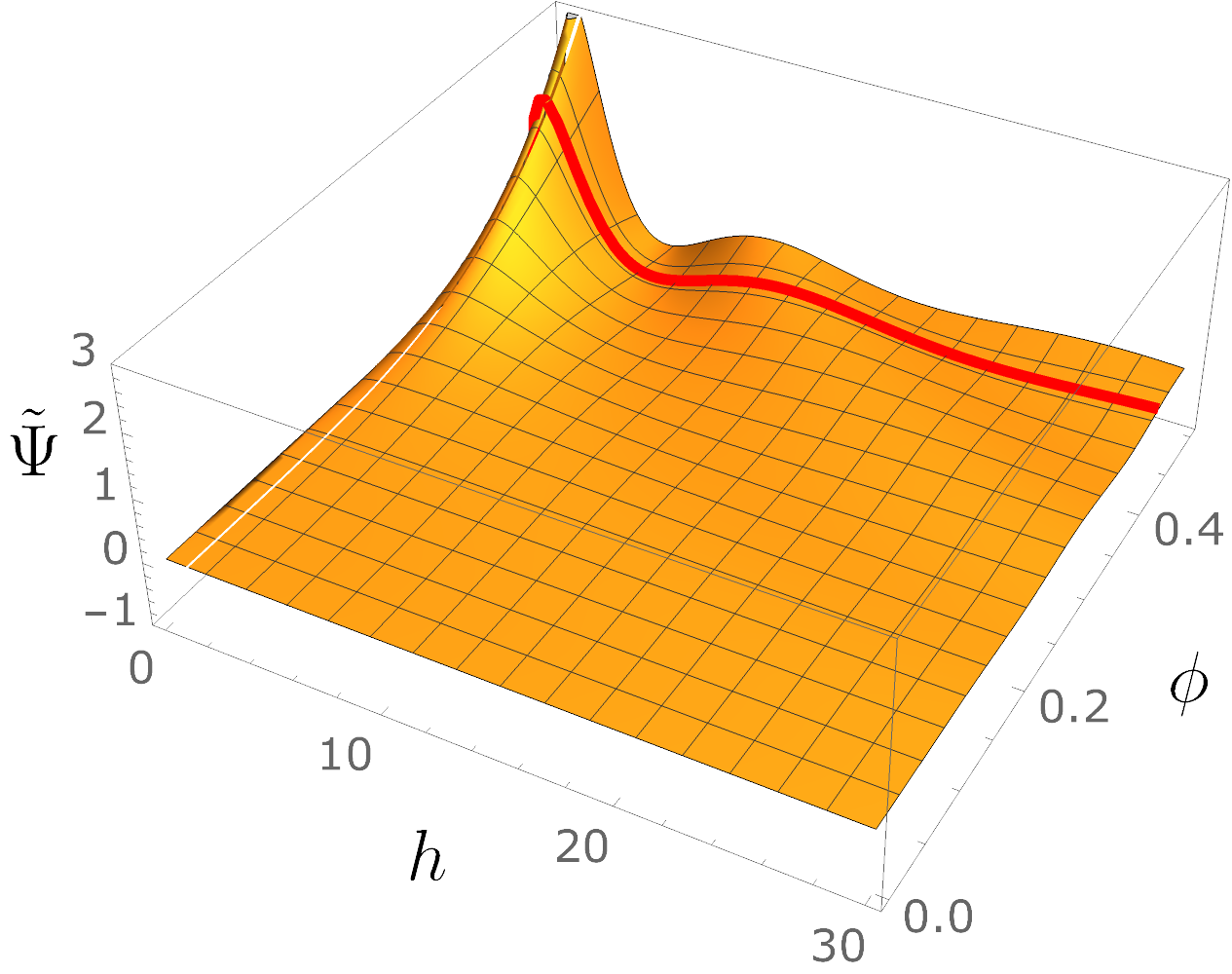}
 \caption{Real solution $\tilde{\Psi}$ of the WDW equation with Henneaux factor ordering and $h_0=1$. The red curve represents $\tilde\Psi|_{\phi=0.45}$.}
  \label{fig:psi3D}
\end{figure}

As in the previous section, a solution that vanishes for $h-h_0 >0$ and is non-zero for
$h-h_0<0$ can also be constructed:
\begin{equation}\label{solH<}
   \tilde{\Psi}_<(h,\phi;h_0) 
    =-\sqrt{h}\,\partial^2_{h}\,[\, J_0(x)\, \Theta(h_0-h)\,]\,.
\end{equation}
In the sum of $\tilde{\Psi}_>$
and $\tilde{\Psi}_<$ the singular terms cancel and one obtains
\begin{equation}
\tilde{\Psi}(h,\phi;h_0) = \frac{1}{2}\left[\tilde{\Psi}_>(h,\phi;h_0) - \tilde{\Psi}_<(h,\phi;h_0)\right]
= \sqrt{h}~\partial^2_h
   J_0(x) \ .
\end{equation}
The wave function $\tilde{\Psi}(h,\phi;h_0=1)$ is shown in Fig.~\ref{fig:psi3D} as function of $\phi$ and $h$ for $h_0=1$.  
Note that $\tilde{\Psi}$ vanishes along the entire boundary, $h=0$ and $\sigma=0$. One clearly sees the increase in $\phi$-direction, the strong rise from zero to $h_0$ in $h$-direction, and the oscillations for $h> h_0$.

For the special choice $h_0 = h_c = 1/\lambda^2$, the result above agrees with that of \cite{Iliesiu:2020zld}. There, it was derived using a perturbative expansion of the
extrinsic curvature, going beyond the Schwarzian limit. More precisely, this expansion was summed up in Euclidean AdS and then analytically continued to Lorentzian dS. Alternatively, this result was also obtained in \cite{Iliesiu:2020zld} by starting
from the general ``HLGK wave function''
\cite{Henneaux:1985nw,Louis-Martinez:1993bge} (cf.~our Appendix~\ref{qth}) and fixing the functional freedom of that solution by demanding that the asymptotic behaviour is Schwarzian. It is remarkable that these findings agree with what we obtain on the basis of a characteristic initial value problem with a $\partial_h\delta(h-h_c)$ boundary behaviour, where $h_c$ singles out the classical 
Euclidean/Lorentzian background geometry.

\subsection{Factor ordering and higher singular sources}
\label{sec:FactOrds}

We can now generalize the results of the previous section to a whole class of factor orderings defined by (recall $\sigma=\phi^2$)
 \begin{equation}\label{wdwsigmaGenFO}
  (h^{p/2}\partial_h h^{-p/2}\partial_\sigma + \pi^2\lambda^2 ) \Psi^{(p)}(h,\phi) = 0\quad,\quad p\in {\mathbb N} \ .
\end{equation}
First, one easily verifies that the Green function $G^{(p)}(h,\phi;h',\phi')$ is given by
  \begin{equation}\label{GreenFGenFO}
  G^{(p)}(h,\phi;h',\phi')=\left(\frac{h}{h'}\right)^{p/2}G(h,\phi;h',\phi') \ ,
\end{equation}
where $G(h,\phi;h',\phi')$ denotes the Green function of the WDW equation in canonical factor ordering, $p=0$. Correspondingly, given a solution $\Psi$ for canonical factor ordering, the wave function
\begin{equation}
\Psi^{(p)}(h,\phi)=h^{p/2}\Psi(h,\phi)
\end{equation}
Moreover, we may allow for more general sources, defined by higher derivatives of a $\delta$-function:
\begin{equation}\label{SingSource}
\chi^{(q)}_{h_0}(h)=\partial_h^{q-1}\delta(h-h_0)\ , \quad q\geq 1 \in {\mathbb N}\ .
\end{equation}

Now, using linearity and the form of our canonical solutions in \eqref{bcom}, one can immediately write down the corresponding solutions with more general factor ordering and boundary conditions:
\begin{eqnarray}\label{psiGret}
 \Psi^{(p,q)}_>(h,\phi;h_0) &=& h^{p/2}\,\partial_h^q \,[\,J_0(x)\,\Theta(h-h_0)\,]\phantom{\Bigg|}
 \\
 &&
 \hspace{-3cm}=\,\,h^{p/2}\,\left[\partial_h^{q-1}\delta(h-h_0)+\sum_{k=1}^{q-1}\left(\!\begin{array}{c}q \\k\end{array}\!\right)\partial_h^{q-1-k}\delta(h-h_0)  \partial_h^kJ_0(x) +\partial_h^qJ_0(x) 
\Theta(h-h_0)\right] \nonumber
\end{eqnarray}
and
\begin{eqnarray}\label{psiGadv}
\Psi^{(p,q)}_<(h,\phi;h_0) &=& -h^{p/2}\,\partial_h^q \,[\,J_0(x)\,\Theta(h_0-h)\,]\,\phantom{\Bigg|}
\\
&&\hspace{-3.3cm}=h^{p/2}\left[\partial_h^{q-1}\delta(h-h_0)+\sum_{k=1}^{q-1}\left(\!\begin{array}{c}q \\k\end{array}\!\right)\partial_h^{q-1-k}\delta(h-h_0) 
\partial_h^kI_0(\tilde x) -\partial_h^qI_0(\tilde x) 
\Theta(h_0-h)\right].\nonumber
\end{eqnarray}
In the last line, we used $J_0(iz)=I_0(z)$ and defined $\tilde{x}=2\pi\lambda\phi\sqrt{h_0-h}$ (recall that $x=2\pi\lambda\phi\sqrt{h-h_0}$).
As before, we can combine the two solutions into a single wave function which is analytic in the upper-right quadrant and vanishes on its boundaries for $p>0$:
\begin{equation}\label{psipq}
\begin{split}
\Psi^{(p,q)}(h,\phi;h_0)&=h^{p/2} \left(\Psi^{(p,q)}_>(h,\phi;h_0) - \Psi^{(p,q)}_<(h,\phi;h_0)\right)\\[5pt]
&=h^{p/2}\, \partial_h^q J_0(2\pi\lambda\phi\sqrt{h-h_0})\,.
\end{split}
\end{equation}

Using the relation $\partial_x(x^{-\nu}J_\nu(x))=-x^{-\nu}J_{\nu+1}(x)$, one straightforwardly obtains
\begin{equation}\label{PsiExact}
\begin{split}
\Psi^{(p,q)}(h,\phi;h_0)&=h^{p/2}(-2\pi^2\lambda^2\phi^2)^q 
x^{-q} J_q(2\pi\lambda\phi\sqrt{h-h_0})\\[5pt]
&=(-\pi\lambda)^q \phi^{p}\left(\frac{\sqrt{h}}{\sqrt{h-h_0}}\right)^p\left(\frac{\phi}{\sqrt{h-h_0}}\right)^{q-p} J_q(2\pi\lambda\phi\sqrt{h-h_0})\ .
\end{split}
\end{equation}
Note that, contrary to appearances, there is no cut in this expression since only even powers of $\sqrt{h-h_0}$ occur after Taylor expanding the Bessel function. Using the asymptotic behaviour 
\begin{equation}
J_{q}(x) \sim \sqrt{\frac{2}{\pi x}}  \cos{\left(x -\frac{q\pi}{2} - \frac{\pi}{4}\right)} \ ,
\end{equation}
the full wave function takes the asymptotic form
\begin{equation}\label{PsiFullAsymp}
\Psi^{(p,q)}(h,\phi;h_0)\sim \lambda^{q-1/2} \phi^{p-1}
\left(\frac{\phi}{\sqrt{h}}\right)^{q-p+1/2}  \cos{\left(2\pi\lambda\phi\sqrt{h}\right)}\ .
\end{equation}

\begin{figure}[t]
 \centering
 \includegraphics[width = 0.65 \textwidth]{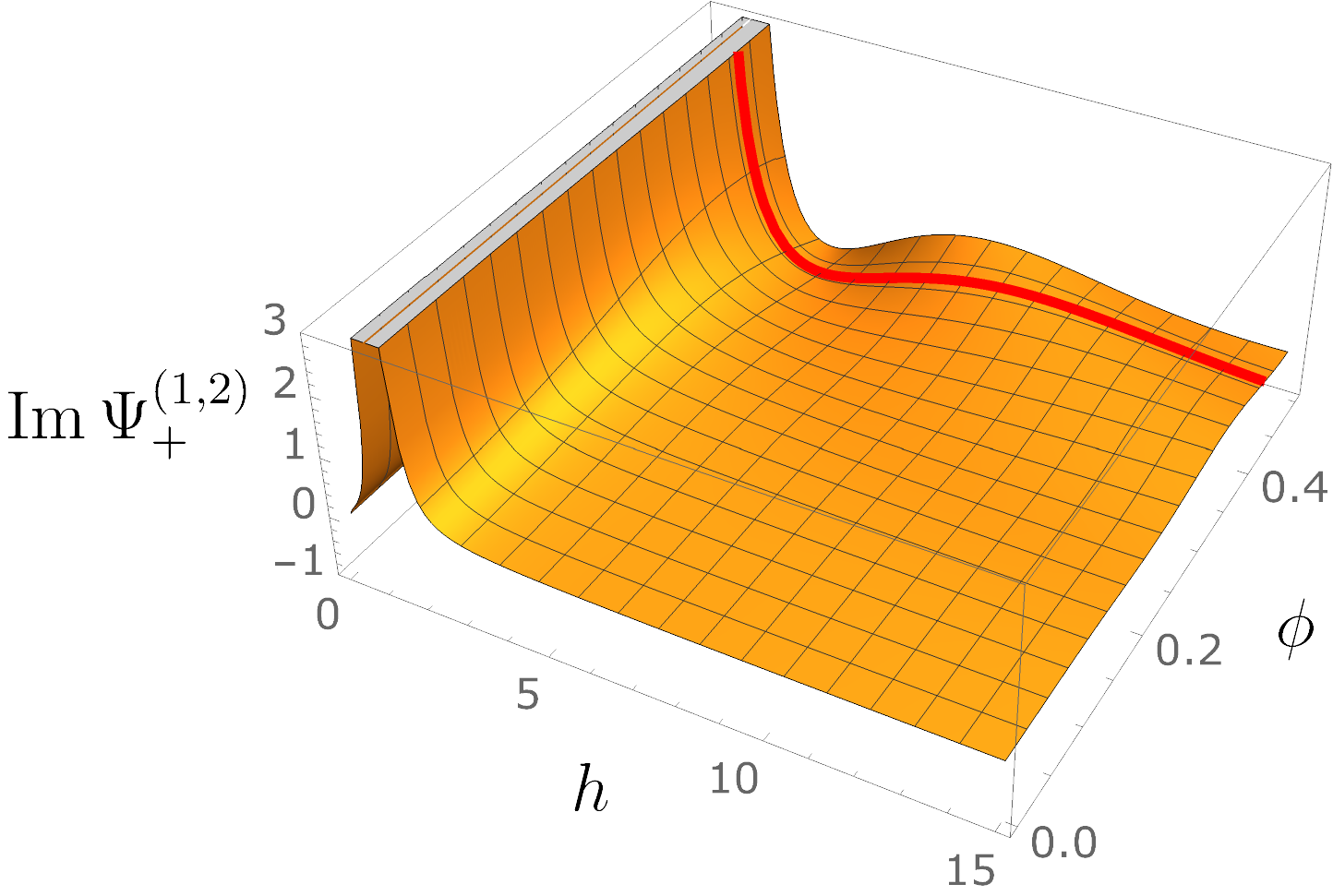}
 \caption{Imaginary part of the complex wave function $\Psi^{(1,2)}_+=\tilde\Psi_+$ (we recall that $\tilde\Psi$ denotes the full real wave function in Henneaux factor ordering in the notation of section~\ref{sec:henn}), representing the outgoing branch for Henneaux factor ordering and $h_0=1$. The red curve represents ${\rm Im}\,\Psi^{(1,2)}_+|_{\phi=0.45}$.)}
  \label{fig:tildepsi+}
\end{figure}

In the Riemann representation solutions of the WDW equation are determined by initial values on the lightlike boundary and by the Riemann function $J_0$. One may then wonder how one can obtain non-zero solutions which vanish on the entire boundary. This can be understood by viewing the
Stokes line $h=h_0$ as boundary instead of the axis $h = 0$. The solutions $\Psi^{(p,q)}$ vanish along $\phi = 0$ but are non-zero along $h=h_0$: 
$\Psi^{(p,q)}|_{h=h_0} = h^{p/2}\partial^q_h J_0|_{h=h_0} \propto h_0^{p/2}\phi^{2q}$. For $h > h_0$ the solution is then given in terms of $J_q$, and for $h < h_0$ in terms of the analytic continuation $I_q$. For canonical factor ordering, $p=0$, the wave function is non-zero also at $h=0$, whereas for nontrivial factor ordering, $p > 0$, it vanishes along
$h=0$.

\subsection{Projection to expanding or contracting universes}\label{sproj}
  
Like the semiclassical JT wave function, the exact wave function \eqref{PsiExact} can be
decomposed into wave functions for expanding and contracting branches in the standard manner. 
For this purpose, we write $J_q$ as linear combination
of the Hankel functions $H^{(2)}_q$ and $H^{(1)}_q$,
\begin{equation}\label{decomp}
  J_q(x) = \frac{1}{2}\left(H^{(2)}_q(x) + H^{(1)}_q(x)\right) \ , \quad h > h_0 \ ,
\end{equation}
which at large $h$ become waves with positive and negative frequencies, respectively. 
Decomposing our wave function from \eqref{decomp} as suggested by \eqref{PsiExact}, we may then write $\Psi^{(p,q)}=\Psi_{+}^{(p,q)}+\Psi_{-}^{(p,q)}$. For $h > h_0$, one finds
\begin{equation} \label{PsiBranchAsymp}
\begin{split}
\Psi_{+}^{(p,q)}(h,\phi;h_0) & =
\frac{1}{2}h^{p/2}\partial_h^q
H^{(2)}_0(2\pi\lambda\phi\sqrt{h-h_0}) \\
&=\frac{1}{2}(-\pi\lambda)^q   \phi^{p}\left(\frac{\sqrt{h}}{\sqrt{h-h_0}}\right)^p\left(\frac{\phi}{\sqrt{h-h_0}}\right)^{q-p}  \\ &\hspace{2.3cm} \times H^{(2)}_q(2\pi\lambda\phi\sqrt{h-h_0})\ , \\[5pt]
&\sim \lambda^q  \phi^{p-1}\left(\frac{\phi}{\sqrt{h}}\right)^{q-p+1/2} \exp{\left(-2\pi i\lambda\phi\sqrt{h}\right)}\\[5pt]
\Psi
_{-}^{(p,q)}(h,\phi;h_0) &= \left(\Psi
_{+}^{(p,q)}(h,\phi;h_0)\right)^* \ .
\end{split}
\end{equation}
This clearly shows that, asymptotically, $\Psi_+^{(p,q)}(h,\phi)$ and $\Psi_-^{(p,q)}(h,\phi)$  describe an outgoing and incoming wave with fast-changing phase. These are the asymptotic wave functions for a semiclassical expanding and contracting universe, respectively.

The Hankel functions $H_q^{(2)}$ have a singularity 
at $h=h_0$. This may be regularized by means of a
Feynman $i\epsilon$-prescription, see Eqs~\eqref{GF2} and \eqref{G+}. In the limit $\epsilon \rightarrow 0$ one then obtains contributions to the wave
function for $h < h_0$ and $h > h_0$
(cf. Appendix~\ref{sec:cauchy}, Eq.s~\eqref{wfG+} and \eqref{wfG+V2}),
\begin{equation}\label{eq:psi+Cont1}
\begin{split}
\Psi_{+}^{(p,q)}(h,\phi;h_0) =
&\frac{1}{2}h^{p/2}\partial_h^q\left(
H^{(2)}_0(2\pi\lambda\phi\sqrt{h-h_0})\Theta(h-h_0)
\phantom{\frac{2i}{\pi}}\right.\\
&\left.+\frac{2i}{\pi}
K_0(2\pi\lambda\phi\sqrt{h_0-h})\cdot\Theta(h_0-h)\right)\ .
\end{split}
\end{equation}
Note that the wave function is purely imaginary on the left of the singularity, which leads to a vanishing
Klein-Gordon current for $h<h_0$.

However, the
definition of $\Psi_{+}^{(p,q)}$ is not unique. 
Choosing the opposite sign in the $i\epsilon$ prescription, one finds (cf.~Eq.s~\eqref{otherwf+} and \eqref{otherwf+V2})
\begin{equation}\label{eq:psi+Cont2}
\begin{split}
\Psi^{(p,q)}_+ (h,\phi;h_0) &=\frac{1}{2}h^{p/2}\partial^q_h\left[
 H^{(2)}_0(2\pi\lambda\phi\sqrt{h-h_0})\Theta(h-h_0) 
\phantom{\frac{2i}{\pi}}\right.\\
& \hspace{2ex}\left. +2\left(I_0(\pi\lambda\phi\sqrt{h_0-h})
+\frac{i}{\pi}K_0(2\pi\lambda\phi\sqrt{h_0-h})\right)\Theta(h_0-h)\right]\ .
\end{split}
\end{equation}
In this case the wave function on the left of the
singularity is complex and the 
Klein-Gordon current there does not vanish.

We can now study the behaviour of the real wave function $\Psi^{(p,q)}(h,\phi;h_0)$ as well as the expanding and contracting branches $\Psi^{(p,q)}_\pm(h,\phi;h_0)$ near $h=h_0$, using the relations \eqref{JY0} and
\eqref{IK0}. As already discussed in Section~\ref{sec:FactOrds},
the real wave function $\Psi^{(p,q)}(h,\phi;h_0)$ is
finite and continuous across $h=h_0$.
On the contrary, when restricting to either the expanding or contracting branch of the wave function, we find near $h=h_0$ a very different behaviour, i.e.
\begin{align} 
\left.\Psi^{(p,q)}_{\pm}(h,\phi;h_0)\right|_{h > h_0}
&\sim \mp\frac{i}{2\pi}h_0^{p/2}\partial^{(q-1)}_h \frac{1}{h-h_0} \\
\left.\Psi^{(p,q)}_{\pm}(h,\phi;h_0)\right|_{h < h_0} 
&\sim \pm\frac{i}{2\pi}h_0^{p/2}\partial^{(q-1)}_h \frac{1}{h_0-h} \ .
\end{align}
Contrary to the full real wave function, its expanding and contracting branch both have a singularity, a pole of order $q$ at $h=h_0$. 
We note that the approach toward the pole is independent of $\phi$.
See Fig.~\ref{fig:tildepsi+} for an illustration.

In the construction of a
Hartle-Hawking wave function for $h > h_c$ in \cite{Maldacena:2019cbz,Cotler:2019nbi}, where the
contributions from the Euclidean and the Lorentzian
sections of the geometry are assumed to factorize, it is
tempting to speculate that the singularity 
arising from the Lorentzian section
corresponds
to a divergence of the sphere partition function \cite{Nanda:2023wne}. The appearance of this
singularity has cast same doubts on the existence of
a physically meaningful Hartle-Hawking wave function
in JT gravity \cite{Iliesiu:2020zld,Fanaras:2021awm,Nanda:2023wne}.
On the other hand, in our approach the singularity
arises as consequence of projecting a real non-singular
wave function onto outgoing and incoming branches,
so that a singularity may be expected, see also Section~\ref{sec:KGmeasure-sing}.

\begin{figure}[t!]
 \centering
 \includegraphics[width = 0.65 \textwidth]{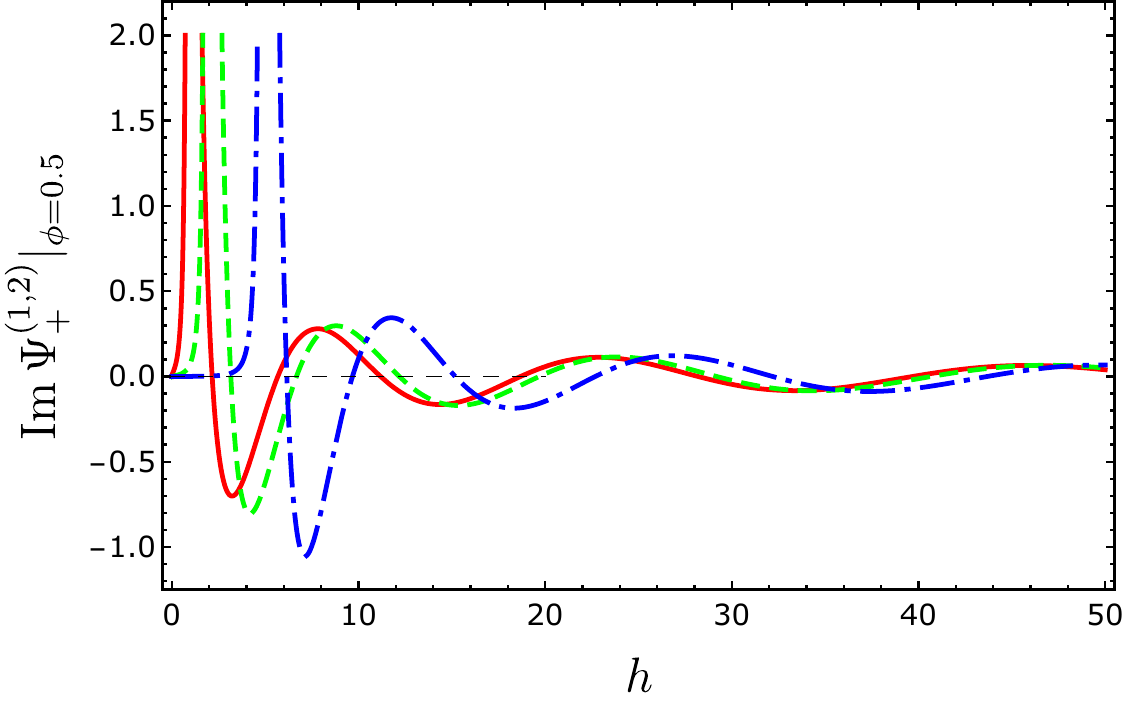}
 \caption{Imaginary part of the complex wave function $\Psi^{(1,2)}_+(h,\phi;h_0)|_{\phi = 0.5}$ with $h_0 =1$ (red solid), $h_0=2$ (green dash) and $h_0=5$ (blue dash-dot), respectively.}
  \label{fig:ImPsi}
\end{figure}

In JT gravity the position of Stokes lines and singularities for real wave functions and individual branches, respectively, are
not determined by a parameter of the theory, like
$h_c$ in the case of 4d de Sitter, but they are
parameters of solutions of the WDW equation.  
The dependence of the singularities of complex wave functions on $h_0$ is illustrated in Fig.~\ref{fig:ImPsi}. For $\Psi^{(1,2)}_+$, the divergent imaginary part has the same sign on both sides of the singularity at $h=h_0$.

The general solution is then a superposition of terms with
different values of $h_0$. Beyond the largest value
$h_0^{\text{max}}$ one approaches a semiclassical regime
where the wave functions coherently oscillate. 
At values below $h_0^{\text{max}}$ one enters a
quantum regime where no semiclassical approximation is possible. 

\begin{figure}[t!]
 \centering
 \includegraphics[width = 0.65 \textwidth]{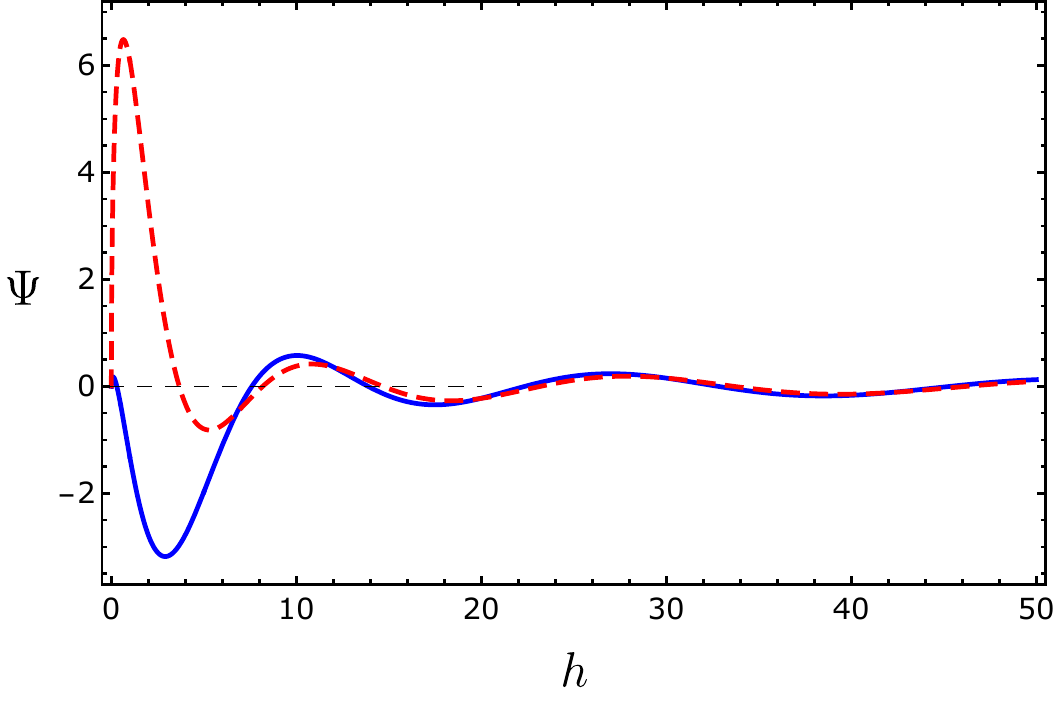}
 \caption{Superposition of three real wave functions $\Psi^{(1,2)}(h,\phi;h_0)$ at $\phi = 0.5$ and $h_0 = 1$, $h_0 = 2$ and $h_0 = 5$, respectively (blue solid); for comparison the wave function of Fig.~\eqref{fig:psi3D} is shown at
 $\phi = 0.5$ (red dash).}
  \label{fig:psisuper}
\end{figure}

This is illustrated in Fig.~\ref{fig:psisuper}.
At small $h$, the behaviour of superposition
and the wave function with $h_0 = 1$ is very different whereas at large $h$, in the semiclassical regime, the two wave functions approach each other.

\subsection{Comparison with semiclassical wave functions}

In this section we have constructed real and complex
solutions to the WDW equation, and it is instructive
to compare the results with semiclassical wave functions that are still frequently used (for recent examples, see \cite{Lehners:2023yrj,Ivo:2024ill}). 

The Hartle-Hawking wave function is real. In the 
unphysical regime it either increases or decreases
exponentially toward the critical scale factor
$\sqrt{h_c}= 1/\lambda$, depending on the chosen
analytic continuation to imaginary time. The Hartle-Hawking choice corresponds to exponential increase, the Linde-Vilenkin choice to exponential
decrease (cf.~Appendix~\ref{sec:semiHH}). In JT gravity the situation is the same (cf.~Appendix~\ref{sec:semiJT}). For the exact real
solutions presented in Section~\ref{sec:FactOrds} there
is no ambiguity. With increasing $h$, the wave functions decrease exponentially down to the oscillatory region, as illustrated by Fig.~\ref{fig:psi3D}.

Semiclassical wave functions that connect the unphysical
region with expanding or contracting universes in the
physical region can be nicely constructed by means of
a complex time path in the complex Lorentzian/Euclidean
geometry \cite{Lehners:2023yrj}. This works in the 
same way for JT gravity (cf.~Appendix~\ref{sec:semiJT}).
Note, however, that such wave functions cannot be the
semiclassical limit of exact solutions of the 
WDW equation, which is a direct consequence of the
conservation of the Klein-Gordon current. 
Semiclassical wave functions also have a different asymptotic
behaviour for large field values compared to exact solutions
of the WDW equation. In particular, they are incompatible with
the Schwarzian behaviour obtained from the functional integral.

Branches of expanding or contracting universes can
emerge from a real wave function by decoherence (see, for example,
\cite{Kiefer:2007ria,Lehners:2023yrj,Halliwell:2011zz}). In JT gravity this would require
to introduce additional matter degrees of freedom.
As shown in the previous paragraph \ref{sproj},
enforcing a projection by hand leads to singularities
for the projected wave functions. We expect this
to be the case also in other models of quantum
gravity, and not a special property of JT gravity.\\

\section{Probabilistic constraints on JT wave functions}\label{prob}

Returning to our set of smooth solutions in \eqref{PsiExact}, we note that the most general wave function for given factor ordering $p$ is constructed as a linear combination of solutions with varying degree of singularity $q$: 
\begin{equation}
\Psi^{(p)}(h,\phi;h_0)
=\sum_{q\geq 1} \psi^{(q)}\,\Psi^{(p,q)}(h,\phi;h_0)\,.
\end{equation}
Explicitly, we have
\begin{equation}
\Psi^{(p)}(h,\phi;h_0)=h^{p/2}\sum_{q\geq 1} \psi^{(q)}\left(\frac{-\pi\lambda\phi}{\sqrt{h-h_0}}\right)^q J_q(2\pi\lambda\phi\sqrt{h-h_0})
\end{equation}
with large-$h$ asymptotics
\begin{equation}\label{PsiSeriesFullAsymp}
\Psi^{(p)}(h,\phi;h_0)\sim  \phi^{p-1}\sum_{q\geq 1}(-\lambda)^q\psi^{(q)}  \left(\frac{\phi}{\sqrt{h}}\right)^{q-p+1/2}  \cos{(2\pi\lambda\phi\sqrt{h-h_0})}\ .
\end{equation}

This asymptotic form shows an encouraging pattern: Solutions constructed on the basis of increasingly singular sources, as quantified by $q$, display an increasingly fast fall-off $\sim h^{-q/2}$ at large $h$. Thus, the large-$h$-behaviour of the full wave function $\Psi^{(p)}$ will be dominated by the smallest contributing $q$ as long as the coefficients $\psi^{(q)}$ stay bounded at $q\to \infty$. From here we see a promise: If there were a way to restrict the singularity strength $q$ to $q\geq p+1$ instead of just $q\geq 1$, then the full wave function would enjoy the Schwarzian limiting behaviour $\sim (\phi/\sqrt{h})^{3/2}$ at large $h$.

\subsection{Factor ordering and normalizability}

In order to obtain a restriction on the range of the singularity strength $q$ we will need to argue from the asymptotic normalizability of the wave function toward large $h$. We will do so using two different proposals for a scalar product on the space of wave functions and hence for a probability measure.

As a first option, we employ the invariant `volume element in field space' times $|\Psi|^2$ to define the probability density. This is motivated by the standard Born rule of quantum mechanics. We will refer to this as the `Hartle-Hawking measure' since it was used in the seminal paper \cite{Hartle:1983ai}.\footnote{In Ref.~\cite{Perry:2021udd} the terms Hartle-Hawking inner product or Hartle-Hawking norm are used.}

To make this explicit, we recall that for canonical factor ordering our wave function satisfies
  \begin{equation}
  (\partial_h\partial_\phi + 2\pi^2\lambda^2\phi ) \Psi(h,\phi) = 0 \ .
  \label{cwdw}
\end{equation}
To allow for different factor orderings, we generalize this using the notion of a Laplacian in field space
\begin{equation}
2\partial_h\partial_\phi \to \Delta\equiv \frac{1}{\sqrt{f}} \partial_\alpha\left(\sqrt{f}f^{\alpha\beta}\partial_\beta\right)\,,
\label{covl}
\end{equation}
where $\alpha,\beta=h,\phi$. In the canonical case, the inverse of the field space metric reads
\begin{equation}
f_c^{\alpha\beta}=\left(\begin{array}{cc}0 & 1 \\ 1 & 0\end{array}\right)\,.
\end{equation}
As discussed in Appendix~\ref{sec:pifo}, the WDW equations with different factor ordering are obtained by rescaling the field space metric according to $f^c_{\alpha\beta}\to f_{\alpha\beta}\equiv \Omega^{-1}f^c_{\alpha\beta}$ with an arbitrary function $\Omega(h)$ and then enforcing invariance of the path integral via a counter-rescaling of the lapse function. This in turn rescales the wave function, taking us from the canonical wave function $\Psi^c$ to $\Psi=\Omega \Psi^c$. Choosing $\Omega(h)=h^{p/2}$ then corresponds to the class of factor orderings described in Section~\ref{sec:FactOrds}.

In full generality, the Hartle-Hawking measure then reads
\begin{equation}
dP_{\rm HH}(h,\phi)=|\Psi|^2\sqrt{f}dhd\phi\sim h^{\frac{p-2q-1}{2}} dh d\phi\,.
\label{hhf}
\end{equation}
To obtain the $h$-dependence in the last expression, we used $|\Psi|^2\sim h^{p-q-1/2}$ (see for example~\eqref{PsiFullAsymp}) and $\sqrt{f}\sim h^{-p/2}$.

Before attempting an interpretation, we want to derive an analogous expression using a second proposal, which we will refer to as the `Klein-Gordon measure'. For this purpose, we recall that solutions of \eqref{cwdw} allow for the definition of a current 
\begin{equation}
  j = (j_h, j_\phi) \ , \quad j_h =
  \frac{i}{2}(\Psi^*\partial_h\Psi - \Psi\partial_h\Psi^*) \ ,\quad
 j_\phi = \frac{i}{2}(\Psi^*\partial_\phi\Psi - \Psi\partial_\phi\Psi^*)\,,
\end{equation}
which is conserved: $\partial_\phi j_h + \partial_h j_\phi = 0$. For general metrics $f_{\alpha\beta}$ the analogous definition and the statement of current conservation read
\begin{equation}
j^\alpha=\frac{i}{2}f^{\alpha\beta}\left(\Psi^*\partial_\beta\Psi-\Psi\partial_\beta\Psi^*\right)
\qquad\mbox{and}\qquad
\nabla_\alpha j^\alpha=0 \, .
\end{equation}
Here $\nabla_\alpha j^\alpha=\partial_\alpha j^\alpha+\Gamma^\alpha_{\alpha\gamma}j^\gamma=f^{-1/2}\partial_\alpha(f^{1/2}j^\alpha)$ is the divergence of the current and the underlying wave function $\Psi$ solves the WDW equation with the Laplacian from \eqref{covl}.

Of course, for a real wave function $\Psi$ the Klein-Gordon current just introduced vanishes. Hence, we can define the Klein-Gordon current only separately for the outgoing and incoming branch of the wave function, $\Psi_\pm$, as discussed in Section~\ref{sproj}.  The overlap between these two branches will become small (a necessary condition for the two branches to decohere~\cite{Halliwell:1989vw}) when the phase becomes fast-changing, that is, at large $h$. The conserved `Klein-Gordon' probability measure~\cite{DeWitt:1967yk, Vilenkin:1988yd}\footnote{The term `DeWitt scalar product' is also widespread.} then reads
\begin{equation}
dP_\pm(h,\phi)=
j_\pm^\alpha(h,\phi)\,\sqrt{f}\,
\epsilon_{\alpha\beta} \,dX^\beta_\pm \ ,
\end{equation}
with
\begin{equation}
j_\pm^\alpha(h,\phi)=f^{\alpha\beta}j^\pm_\beta(h,\phi) \qquad \mbox{and} \qquad
j^\pm_\beta(h,\phi) = \frac{i}{2}
\left(\Psi_\pm^*\partial_\beta\Psi_\pm-\Psi_\pm\partial_\beta\Psi_\pm^*\right)\ .
\end{equation}
This definition depends on the choice of a codimension-one surface $\Sigma$ over which we integrate the Hodge-dual of the current, $P\sim \int_\Sigma *j$. In our case $\Sigma$ is a line and $dX$ is its line element.  We choose the orientation opposite for incoming and outgoing wave function, $dX_- = -dX_+$, such that $dP_- = dP_+$.

Since we are interested in large-$h$ normalizability, we conveniently choose the integration surface by defining
$dX^\alpha=(dh,0)$. This yields
\begin{equation}
 dP_\pm(h,\phi)=j^\pm_h\, f^{h\phi}\,\sqrt{f}\,\epsilon_{\phi h}\,dh = j^\pm_h \, h^{p/2}\, h^{-p/2}\,dh = j^\pm_h \,dh.
\end{equation} 

We now recall that the asymptotic outgoing and incoming branches of our wave function with general factor ordering and singularity strength are given in Eq.~\eqref{PsiBranchAsymp},
\begin{equation}
\Psi^{(p,q)}_\pm(h,\phi)\sim \lambda^q  \phi^{p-1}\left(\frac{\phi}{\sqrt{h}}\right)^{q-p+1/2} e^{\mp i2\pi\lambda\phi\sqrt{h}}\,.
\end{equation}
This implies 
\begin{equation}
|\Psi^{(p,q)}_\pm|^2\sim \frac{1}{h^{q-p+1/2}}\quad,\quad ( j^\pm_h , j^\pm_\phi) \sim \frac{1}{h^{q-p+1/2}}(1/\sqrt{h},\sqrt{h})\sim \left(\frac{1}{h^{q-p+1}},\frac{1}{h^{q-p}}\right)\,.
\end{equation}
Hence, we get
\begin{equation}
dP_\pm(h,\phi) = j^\pm_h\, dh\sim h^{p-q-1}dh\,.
\label{kgf}
\end{equation}
Thus, our results so far are the two probability distributions \eqref{hhf} for the Hartle-Hawking measure together with \eqref{kgf} for the Klein-Gordon measure.

We may now impose the two physical conditions of
\begin{itemize}
\item[i)] Finiteness of the integral over $h$ at $h\to \infty$
\item[ii)] Schwarzian behaviour in $h$ of the large-$h$ wave function: $\Psi \sim 1/\sqrt{h}^{3/2}$.
\end{itemize}
We see that

\begin{itemize}
\item Condition i) implies 
$q> (p+1)/2$ for the Hartle-Hawking and $q>p$ for the Klein-Gordon measure.
\item Condition ii) tells us: $q\geq p+1$.
\end{itemize}
Given integrality of $p$ and $q$, Conditions i) and ii) are hence equivalent in the Klein-Gordon case. By contrast, in the Hartle-Hawking case the two conditions are only equivalent for $p=0$ and $p=1$. For larger $p$, Condition ii) constrains more strongly which of the solutions $\Psi^{(p,q)}$ are allowed to contribute to the wave function $\Psi^{(p)}$.

As a result, we find an intriguing correlation between the requirements of normalizability and Schwarzian behaviour. Moreover, if we exclude the case $p=0$ because the physical requirement $\Psi^{(0,q)}(0,\phi)=0$ is violated, then $p=1$ is singled out as the potentially unique factor ordering prescription: Only in this case is Hartle-Hawking normalizability at large $h$ equivalent to asymptotic Schwarzian behaviour.

\subsection{Klein-Gordon current near the singularity}
\label{sec:KGmeasure-sing}

It is instructive to consider the Klein-Gordon current $j^\phi = f^{\phi h} j_h$ for
all values of $h$, in particular near the singularity at $h_0$. The current of the outgoing branch reads
\begin{align}
\big(j^{(p,q)}_+\big)_h &= \frac{i}{2}\big(\Psi^{(p,q)*}_+\partial_h\Psi^{(p,q)}_+ -
\Psi^{(p,q)}_+\partial_h\Psi^{(p,q)*}_+ \big)\nonumber\\
&= \frac{(\pi\lambda)^{2q}}{4}\frac{h^p\phi^{2q}}{(h-h_0)^{q}}\big(J_q\partial_h Y_q - Y_q\partial_h J_q\big) 
\Theta(h-h_0)
\nonumber \\
&= \frac{(\pi\lambda)^{2q}}{4\pi}\frac{h^p \phi^{2q}}{(h-h_0)^{q+1}} \Theta(h-h_0)\,.
\end{align}
It vanishes for $h<h_0$ due to the choice $\epsilon>0$ in taking the limit $i\epsilon \rightarrow 0$ (cf.~\eqref{eq:psi+Cont1}). By contrast, for $\Psi^{(p,q)}_-$ this
current component is non-zero also for $h < h_0$ and is proportional to
\begin{align}
\text{Im}\big(\Psi^{(p,q)*}_-\partial_h\Psi^{(p,q)}_-\big)
&=-\frac{(\pi\lambda)^{2q}}{\pi}\frac{h^p\phi^{2q}}{(h_0-h)^{q}}\big(I_q\partial_h K_q - K_q\partial_h I_q\big)  \nonumber\\
&= -\frac{(\pi\lambda)^{2q}}{2\pi}\frac{h^p \phi^{2q}}{(h_0-h)^{q+1}} \ . 
\end{align}
The complete current for $\Psi^{(p,q)}_-$ then reads
\begin{align}
\big(j^{(p,q)}_-\big)_h &= \frac{i}{2}\big(\Psi^{(p,q)*}_-\partial_h\Psi^{(p,q)}_- -
\Psi^{(p,q)}_-\partial_h\Psi^{(p,q)*}_- \big) \nonumber\\
&= \frac{(\pi\lambda)^{2q}}{4\pi} \Big(-\frac{h^p\phi^{2q}}{(h-h_0)^{q+1}}\Theta(h-h_0) + \frac{2h^p\phi^{2q}}{(h_0-h)^{q+1}}\Theta(h_0-h)\Big) \ .
\end{align}
The corresponding currents $\big(j^{(p,q)}_\pm\big)_\phi$ are
given by
\begin{align}
\big(j^{(p,q)}_+\big)_\phi 
&=\frac{(\pi\lambda)^{2q}}{2\pi}\frac{h^p \phi^{2q-1}}{(h-h_0)^{q}} 
\Theta(h-h_0)\ ,\\
\big(j^{(p,q)}_-\big)_\phi 
&= \frac{(\pi\lambda)^{2q}}{2\pi} \Big(-\frac{h^p\phi^{2q-1}}{(h-h_0)^{q}}\Theta(h-h_0) - \frac{2h^p\phi^{2q-1}}{(h_0-h)^{q}}\Theta(h_0-h)\Big) \ .
\end{align}
Using the metric $f^{\phi h} = f^{h\phi} = h^{p/2}$ one
finally obtains for the vector currents $j^{(p,q)}_\pm$: 
\begin{align}
\big(j^{(p,q)}_+\big)^\phi 
&= \frac{(\pi\lambda)^{2q}}{4\pi}\frac{h^{3p/2} \phi^{2q}}{(h-h_0)^{q+1}} \Theta(h-h_0)\ ,\label{j+phi}\\
\big(j^{(p,q)}_-\big)^\phi 
&= \frac{(\pi\lambda)^{2q}}{4\pi} \Big(-\frac{h^{3p/2}\phi^{2q}}{(h-h_0)^{q+1}}\Theta(h-h_0) + \frac{2h^{3p/2}\phi^{2q}}{(h_0-h)^{q+1}}\Theta(h_0-h)\Big) \ ,\label{j-phi}\\
\big(j^{(p,q)}_+\big)^h
&=\frac{(\pi\lambda)^{2q}}{2\pi}\frac{h^{3p/2} \phi^{2q-1}}{(h-h_0)^{q}} 
\Theta(h-h_0)\ ,\label{j+h}\\
\big(j^{(p,q)}_-\big)^h 
&= \frac{(\pi\lambda)^{2q}}{2\pi} \Big(-\frac{h^{3p/2}\phi^{2q-1}}{(h-h_0)^{q}}\Theta(h-h_0) - \frac{2h^{3p/2}\phi^{2q-1}}{(h_0-h)^{q}}\Theta(h_0-h)\Big) \ \label{j-h}.
\end{align}

\begin{figure}[t]
 \centering
\includegraphics[width = 0.6 \textwidth]{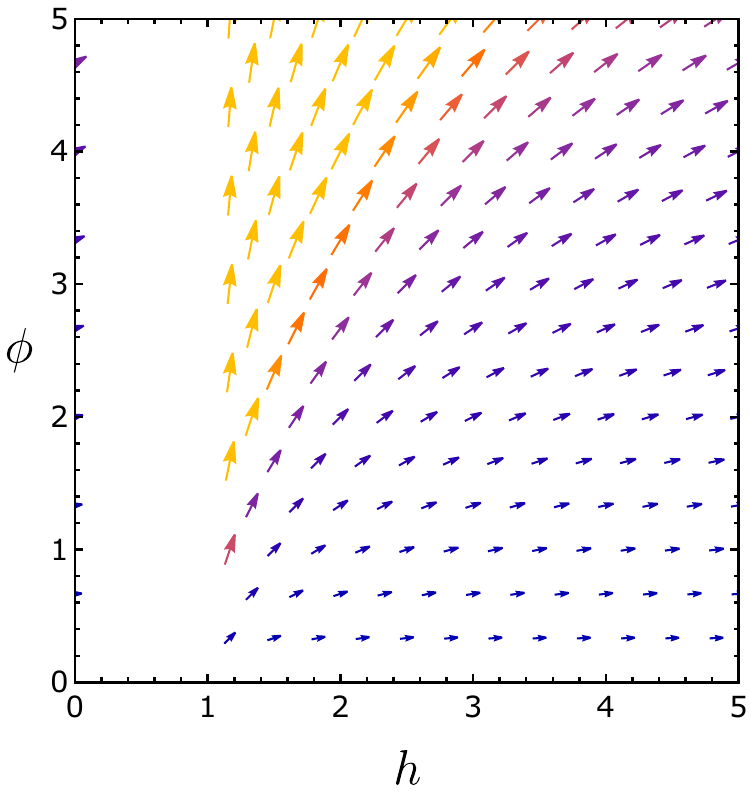}
 \caption{Vector field plot of the KG current $j_+=(j_+^h,j_+^\phi)$ of the outgoing branch $\Psi^{(1,2)}_+$ with the continuation to $h<h_0$ given in Eq.s~\eqref{j+phi}, \eqref{j+h}. The arrows represent the vector field.}
  \label{fig:jh}
\end{figure}

The Klein-Gordon currents for the wave function branches $\Psi_+$, $\Psi_-=\Psi-\Psi_+$ defined with the continuation to $h<h_0$ given by the $i\epsilon$-convention of~Eq.s~\eqref{wfG+} and \eqref{wfG+V2} have a pole along the line $h=h_0$ whose strength depends on $\phi$. The current in $h$-direction is given by $j_\pm^h =  2j_\pm^\phi (h-h_0)/\phi$. For
$h < h_0$ the current $j_+$ vanishes since there the wave function branch $\Psi_+$ is purely
imaginary. This is illustrated in Fig.~\ref{fig:jh}.

The very existence of the non-vanishing currents at $h>h_0$ is an artefact of splitting the real and thus current-free wave function $\Psi=\Psi_++\Psi_-$ into complex branches. Moreover, it is this split due to which the outgoing and incoming branches are each built from derivatives of Bessel functions which at the line $h=h_0$ show a logarithmic singularity vis a vis its derivatives, and which are in turn responsible for possibility of the branch KG current to develop a pole there, despite the associated real DeWitt wave function being completely smooth everywhere. 

However, the logarithmic singularity of the relevant Bessel functions will show up in the resulting outgoing and incoming branch wave functions only if the source $\chi_{h_0}^{(q)}$ is completely localized at one particular value of $h_0$ and thus the source is itself singular. Upon smearing out the source into a smooth function with sufficiently fast fall-off towards large $h$, both the real DeWitt wave function and its outgoing and incoming branch components will be completely smooth.  Hence, the projection onto a branch is a necessary condition for an outgoing and incoming branch complex WDW wave function to show a pole at $h=h_0$. The pole will appear in the branches if the source is chosen to be singular.

The interpretation of the wave functions for individual branches as solutions of a WDW
equation with a singular source has already been conjectured
in \cite{Fanaras:2021awm}. In the construction of a Hartle-Hawking wave function in \cite{Maldacena:2019cbz} this
source could not be seen since
the wave function was not constructed in the range $0 < h < h_c$ but instead solely asymptotically at large-$h$.

\begin{figure}[t]
 \centering
 \includegraphics[width = 0.6 \textwidth]{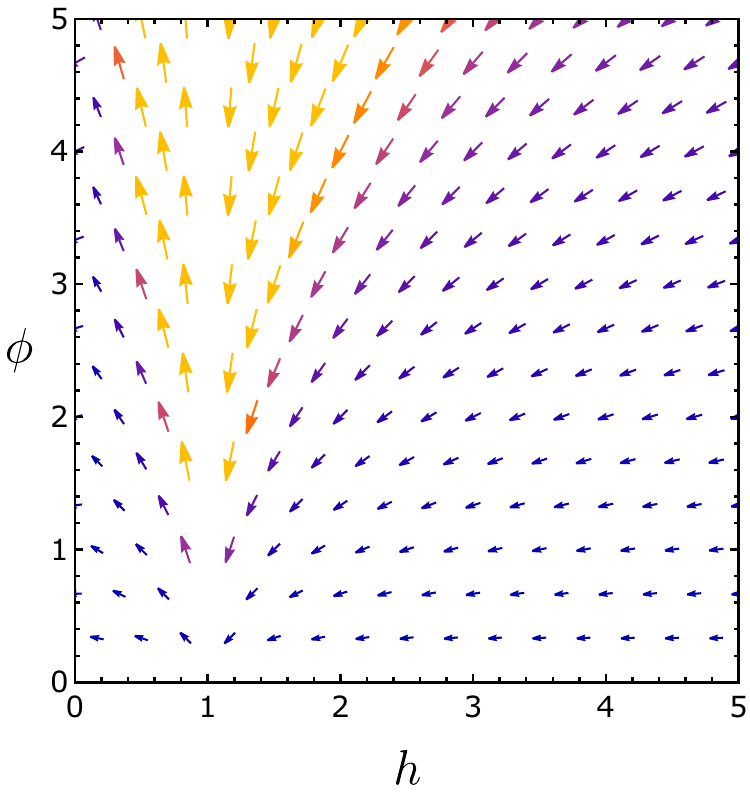}
 \caption{Vector field plot of the KG current $j_-=(j_-^h,j_-^\phi)$ of the incoming branch $\Psi^{(1,2)}_-$ with the continuation to $h<h_0$ given in Eqs.~\eqref{j-phi}, \eqref{j-h}. The arrows represent the vector field.}
  \label{fig:phh}
\end{figure}

We note further the behavior of the KG current for the incoming branch $\Psi_-$. As the outgoing branch $\Psi_+$ defined in Eq.s~\eqref{wfG+} and \eqref{wfG+V2} is purely imaginary for $h<h_0$, we get $\Psi_-$ to be complex there. Hence, the current $j_-$ does not vanish for $h<h_0$, and instead seems to 'bounce off' from $j_-$ at $h>h_0$ in the point $(h=h_0,\phi=0)$ to stream into the $h<h_0$ region, see Fig.~\ref{fig:phh}. The KG currents of both $\Psi_\pm$ vanish on the boundary $\phi=0$ and the KG current of $\Psi_-$ vanish on the boundary $h=0$ as well. This is consistent with the notion of both branches being complex projections of a real DeWitt wave function. 

To see this, we recall that DeWitt wave functions are defined as wave functions which solve the source-free WDW equation and satisfy a quantum-mechanical boundary condition $\Psi(\{{\rm sing.\,loci}\})=0$. Here, the set $\{{\rm sing.\,loci}\}$ of singular loci is to be understood as the joint union of all loci in field space where field energy densities and/or curvature invariants diverge (called the `barrier' by DeWitt). In our case, this clearly includes $h=0$ where the 2d JT geometry becomes singular,  while the locus $\sigma=\phi^2=0$ can be argued to be singular upon embedding 2d JT gravity via compactification of 4d Einstein gravity on a 2-sphere with volume $\phi^2$. Given this embedding, $\phi^2=0$ represents a curvature singularity of the compactification 2-sphere. With this motivation we take $\{h=0\}\cup \{\phi=0\}$ as the singular locus of JT gravity. Our construction of real-valued DeWitt wave functions for JT gravity satisfies the condition of barrier avoidance for all factor ordering choices $p\geq 1$, except for canonical factor ordering.  Hence, the DeWitt wave functions here vanish on the boundaries $\{h=0\}\cup \{\phi=0\}$ and thus their outgoing and incoming branch components as well as their respective KG currents should vanish on the boundaries as well.

\section{Summary and conclusions}
\label{sec:conclusion}

In this work we have constructed Wheeler-DeWitt (WDW) wave functions
for JT quantum gravity in 1+1 dimensional de Sitter (dS) space-time, with configuration space restricted to the domain $0\leq h=a^2 <\infty$ and $0\leq \phi<\infty$. Here $a$ is the scale factor of the spatial $S^1$ and $\phi$ is the dilaton. Our wave functions are exact, real analytic solutions to the JT WDW equation in this full region. We focused on the subset of solutions which, in the semiclassical limit at large scale factor, reproduce the behaviour derived from the path integral of the Schwarzian boundary mode of 2d dS space.

Our solutions follow from the bulk path integral or, equivalently, the Riemann method for solving the initial value problem of the WDW equation. Either way, boundary data at the two axes ($\phi=0, h>0$) and $(\phi>0,h=0$) is required. We found simple solutions by requiring the wave function to vanish everywhere on the boundary except for one point, $(\phi=0,h=h_0)$, at which a singular source is introduced. Remarkably, by analytic continuation each such solution can be turned into a real analytic solution vanishing on all boundaries that is characterized by a transition region containing a Stokes line separating regions with exponential and oscillatory behavior, respectively.

The additional choice $h_0=h_c=1/\lambda^2$ then leads to what, in our approach, is the closest analogue of the semiclassical Hartle-Hawking (HH) proposal whose complex lapse function contour turns from real to imaginary lapse precisely at $h=h_c=1/\lambda^2$.

Matching the boundary mode Schwarzian behaviour asymptotically corresponds to choosing a particular minimal singularity strength for the source. Meanwhile, adding successively stronger singularities to the source corresponds to adding further solutions which display a faster fall-off at large scale factor. 

Hence, the general WDW wave function is sourced by boundary data consisting of a sum of increasingly singular contributions at in general different loci $h_0$. As a consequence, we find this general WDW wave function to consist of a superposition of real `modes', with the asymptotically dominant mode matching the Schwarzian scaling behaviour. It is in this sense that we found a duality between the full WDW wave function based on the bulk path-integral amplitude and the asymptotic description of the wave function using Schwarzian boundary dynamics.

Furthermore, we were able to find such a set of WDW solutions parametrized by singular boundary data and reproducing the Schwarzian limit asymptotically for any choice within a certain class of factor orderings. Hence, it is possible to quantize JT gravity in 2d dS space for different choice of factor ordering while the resulting WDW wave functions show consistent semiclassical behavior independent of the factor ordering choice. 
We find no preference for particular factor orderings. This includes the `Henneaux factor ordering' which just has the feature of providing an explicit functional integral form of the wave function. For all choices of factor ordering other than the `canonical' one, we found the real WDW solutions to actually vanish on both axes $\phi=0$ and $h=0$, thus satisfying a physical requirement for WDW wave functions at the boundary of their configuration space discussed first by DeWitt~\cite{DeWitt:1967yk}. It is for this reason that we called these solutions DeWitt wave functions.

Real WDW wave functions have vanishing Klein-Gordon (KG)
current. However, each such real solution can be decomposed into a
linear superposition of an `outgoing' and `incoming' branch. These
branches can be defined in the semiclassical limit due to inter-branch decoherence. The outgoing and incoming branches have an outward and inward directed KG current, corresponding to an ensemble of semiclassical growing and shrinking universes, respectively.

Crucially, we noted that once supplying a singular source, the outgoing and incoming branches of the associated real and smooth WDW solution possess a power-law pole in $h$ at $h=h_0$. As the outgoing and incoming branches are obtained as projections of the real DeWitt wave functions, we thus found the pole behavior of each of the two branches, which was observed before in literature, to be a projection effect given presence of a singular source.\footnote{We leave a potential interpretation of the source in terms of a spacelike End-of-The-World brane, as suggested by the `Boundary proposal'~\cite{Friedrich:2024aad}, as a suggestive hint for future work.}
Smearing out the source smooths out the pole behaviour for the branches as well.

Finally, there is the question of the measure, that is, how to
calculate probabilities from the wave function of the universe. We
tentatively applied the two main measures discussed in the literature
-- the KG measure based on the KG current and the HH measure. The KG
measure can only be defined on either an outgoing or incoming branch of a real WDW wave function in the semiclassical limit. Interestingly, imposing the condition of normalizability at large-$h$ on candidate wave functions turned out to be equivalent to demanding asymptotically Schwarzian scaling behavior, irrespective of the choice of factor ordering. In contrast, the HH measure gives non-zero probabilities also to the real wave function. Furthermore, the HH measure is less restrictive than KG measure upon imposing the Schwarzian limit.

We may now discuss the implications of these results. The outcome, that the power-law singularities observed at small $h$ for candidate outgoing wave functions obtained from the Schwarzian boundary theory turn out to be projection effects, may suggest `real analyticity' to be a feature of an exact WDW wave function. Real analyticity implies vanishing KG current, and therefore the absence of current sources at small scale factor. We may view this condition as a quantum mechanical `left-over' of the semiclassical no-boundary proposal.

Furthermore, we saw that real analyticity leaves the freedom of a whole class of increasingly singular boundary data. Hence, in contrast to the original no-boundary proposal,  our class of DeWitt wave functions,  being superpositions of many modes, does not select a unique ground state wave function. Choosing a unique ground state among them amounts to choosing a particular compact distribution of boundary data at small $h$ for which the input of a UV-complete theory of quantum gravity seems necessary.

In closing we now wish to make some further comments on the relation between our DeWitt wave functions for JT gravity and the long-standing notion of the semiclassical Hartle-Hawking no-boundary wave function, as well as on the limits of the semiclassical regime. The no-boundary proposal is based on a semiclassical picture: Starting from a classical complex Euclidean/Lorentzian geometry, a
Wheeler-DeWitt wave function is constructed which changes across a region containing a
Stokes point from an exponential behaviour to an oscillatory
behaviour that approaches a semiclassical regime at large scale
factors. The corresponding Hartle-Hawking wave function is real and  interpreted as a superposition of expanding and contracting universes.
The real DeWitt wave functions constructed in this paper are of this
type. They are exact solutions of the WDW equation. 
They do not have just a single Stokes point but are rather
superpositions of wave functions with many Stokes lines. 
For large field values they show the asymptotic behaviour
determined by the Schwarzian degrees of freedom of a boundary
curve. However, in order to describe an expanding or contracting
universe, one has to project on the corresponding branch of the
real wave function. Within JT gravity this leads to a singular
behaviour on the boundary and the various wave functions with
the same asymptotic behaviour can be described by different
singular boundary conditions. As illustrated by Figs.~\ref{fig:jh} and \ref{fig:phh},  singular source choices are correlated with pole-like features in the flux of expanding or contracting
branches. In a higher-dimensional theory, for which JT gravity
is supposed to be an effective low-energy description, the role
of a singular source will be replaced by the dynamics of additional
degrees of freedom. This could be branes as in the Boundary
Proposal \cite{Friedrich:2024aad} or decoherence induced by an
environment of additional degrees of freedom \cite{Halliwell:2011zz}.

Finally, we observe that while our DeWitt wave functions are real like the semiclassical Hartle-Hawking wave function, the latter is known to display an exponentially growing behavior away from the singular locus at $h=0$ towards the semiclassical gluing point $h=h_c$. The DeWitt wave functions we obtained in this work (up to a monomial factor from factor ordering) in contrast show exponentially decaying behavior from $h=0$ towards $h=h_0$. It is interesting to note that unlike the sign choice discriminating the semiclassical notions of the Hartle-Hawking no-boundary and the Linde-Vilenkin tunneling wave function, we had no such freedom. The decaying behavior of the DeWitt wave functions for $h<h_0$ is an outcome of our solutions. We leave it as a problem for future work what this may imply for the long-standing competition between the HH and LV wave function and their respective scaling with the dS cosmological constant.

\subsection*{Acknowledgments}
We thank Klaus Fredenhagen, Bj\"orn Friedrich, and Jean-Luc Lehners for valuable discussions, and Marc Henneaux for clarifying correspondence. AW is partially supported by the Deutsche Forschungsgemeinschaft under Germany’s Excellence Strategy - EXC 2121 “Quantum Universe” - 390833306, by the Deutsche Forschungsgemeinschaft through a German-Israeli Project Cooperation (DIP) grant “Holography and the Swampland”, and by the Deutsche Forschungsgemeinschaft through the Collaborative Research Center SFB1624 ``Higher Structures, Moduli Spaces, and Integrability’'. AH is partially supported by the Deutsche Forschungsgemeinschaft (DFG, German Research Foundation) under Germany’s Excellence Strategy EXC 2181/1 - 390900948 (the Heidelberg STRUCTURES Excellence Cluster).

\appendix

\section{Semiclassical wave functions}
\subsection{Semiclassical Hartle-Hawking wave function}
\label{sec:semiHH}

 Let us briefly recall how the semiclassical Hartle-Hawking wave function
 \cite{Hartle:1983ai} is constructed. For recent reviews, see
 \cite{Kiefer:2007ria,Lehners:2023yrj}. In the simplest version, one considers 4d de
 Sitter space with cosmological constant $\lambda$. The Lorentzian
 action including the Gibbons-Hawking-York (GHY) boundary term reads
 \begin{equation}\label{HHL}
  S[g] = \frac{1}{2} \int_{\mathcal{M}}d^4x \sqrt{g} (R - 2\lambda) +
  \int_{\partial\mathcal{M}} d^3y \sqrt{h} K \ ,
\end{equation}
where $g$, $R$, $\lambda$, $h$ and $K$ denote  metric tensor, 
Ricci scalar, cosmological constant, induced metric on
$\partial\mathcal{M}$ and extrinsic curvature, respectively. 
In the minisuperspace approximation, the metric depends on the lapse
function $N(t)$ and the scale factor $a = \sqrt{h}$,
 \begin{equation}\label{metricL}
   ds^2 = - N^2(t)dt^2 + a^2(t) d\Omega^2_3 \ ,
 \end{equation}
 for which the Lorentzian action becomes
 \begin{equation}\label{IL}
  I_L[a] = 2\pi^2 \int dt N \left(-\frac{3}{N^2}a\dot{a}^2 + 3a-\lambda
    a^3\right)\ .
\end{equation}
With $\pi_a=- 12\pi^2 a\dot{a}/N$, this leads to the Hamiltonian constraint
\begin{equation}\label{HHHC}
  \frac{1}{144\pi^4 a} \pi_a^2 + a -\frac{\lambda}{3} a^3 = 0 \ .
  \end{equation}
For the simplest, `canonical' factor ordering, one obtains the
WDW equation \cite{Hartle:1983ai}  \begin{equation}\label{wdwHH}
    \left(\frac{\hbar^2}{144\pi^4}\frac{\partial^2}{\partial a^2} - a^2 +
    \frac{\lambda}{3} a^4\right)\Psi(a) = 0 \ .
\end{equation}

The classical equation of motion for $a(t)$ has the solution $\bar{a}(t) =
H^{-1}\cosh{(Ht)}$, with $\lambda = 3H^2$, which describes the
de Sitter half-hyperboloid connecting
a circle of minimal radius $H^{-1}$ at $t = 0$ with a circle of radius
$a > H^{-1}$ at $t_a = H^{-1} \operatorname{arcosh}{(Ha)}$. 
Using Eqs.~\eqref{IL} and \eqref{HHHC}, one obtains for the on-shell action
corresponding to the saddle point $\bar{a}(t)$
\begin{equation}
\begin{split}  
I_L^{os}(a) &= 12\pi^2 \int_0^{t_a} dt \left(\bar{a}(t) - H^2\bar{a}^3(t)\right)\\
&= - \frac{4\pi^2}{H^2}\left(H^2a^2-1\right)^{3/2} \ .
\end{split}
\end{equation}
The semiclassical wave function
\begin{equation}
  \Psi_L(a) \simeq \exp{\left(\frac{i}{\hbar}I_L^{os}(a)\right)} = 
\exp{\left(- \frac{4\pi^2i}{\hbar
      H^2}\left(H^2a^2-1\right)^{3/2}\right)} \ , \quad Ha > 1 \ ,
\end{equation}
is a solution of the WDW equation \eqref{wdwHH} to leading order in $\hbar$.
\begin{figure}[t]
 \centering
 \includegraphics[width = 0.4 \textwidth]{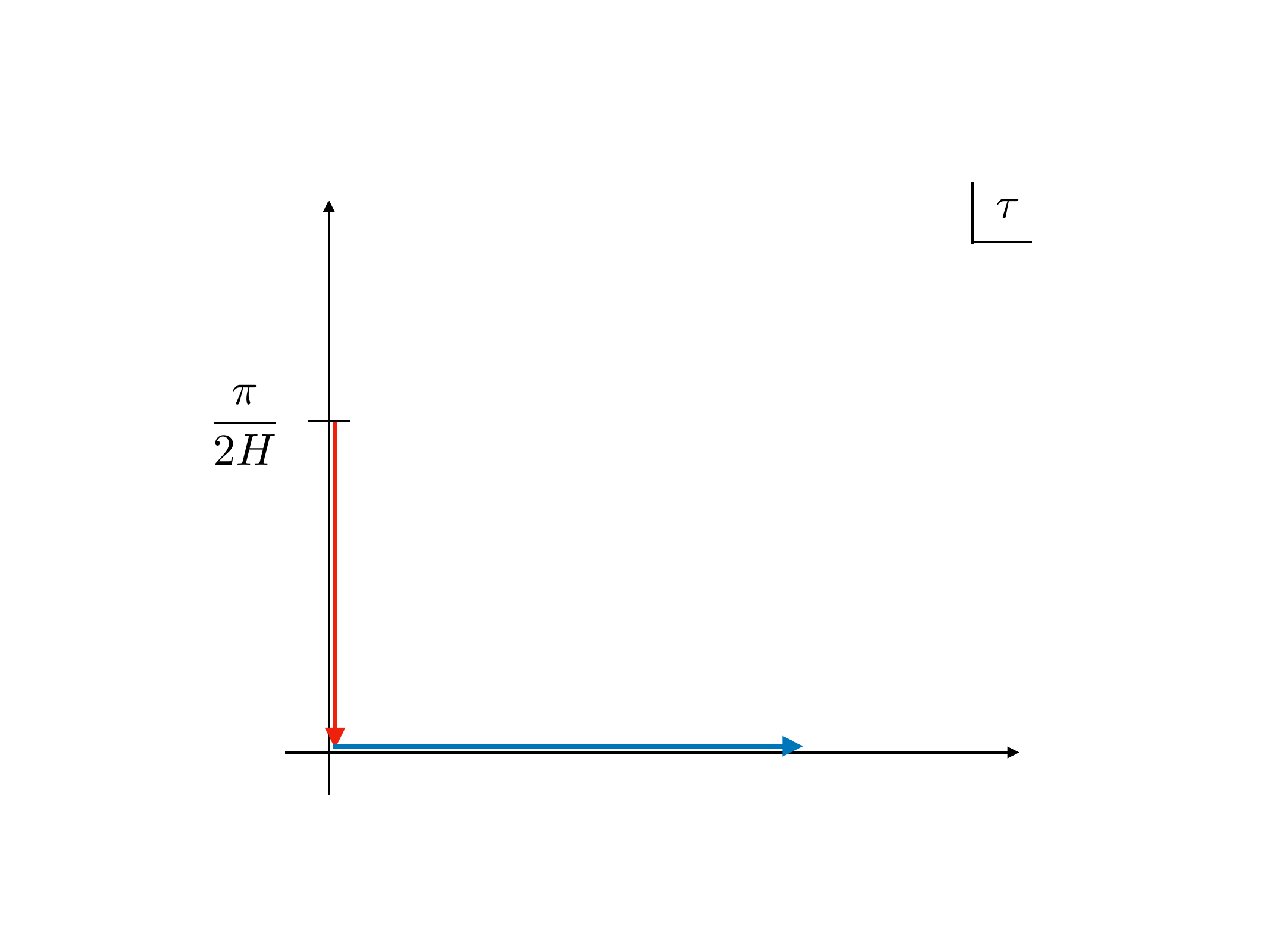}
 \hspace{1cm}
 \includegraphics[width = 0.4 \textwidth]{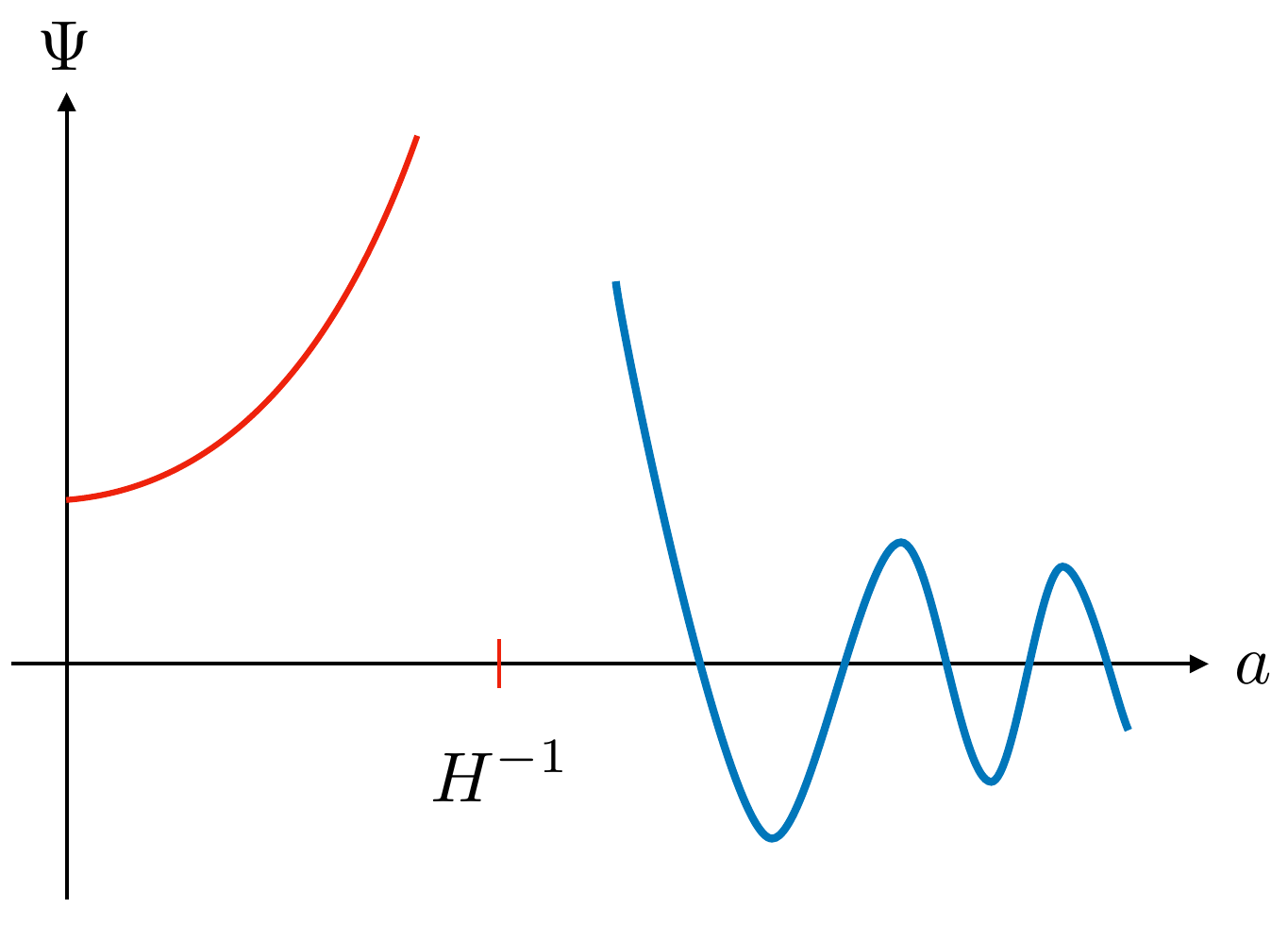}
 \caption{Semiclassical $\text{dS}_4$ wave function. Left: Complex time-path corresponding to the complex Lorentzian/Euclidean geometry \cite{Lehners:2023yrj}. Right: Wave function as function of the scale parameter $a$. In general, the wave function does not vanish at $a=0$.}
  \label{fig:scHH}
\end{figure}
  
The Hartle-Hawking geometry matches the Lorentzian de Sitter
hyperboloid at minimal size $H^{-1}$ to a Euclidean half-sphere at the
equator of size $H^{-1}$, which is described by the Euclidean metric
\begin{equation}\label{metricE}
   ds^2 = N^2(\tau)d\tau^2 + a^2(\tau) d\Omega^2_3 \ ,
 \end{equation}
 where $\tau = it$ is Euclidean time. The Lorentzian action $iI_L$
 becomes $-I_E$, with
 \begin{equation}\label{IE}
  -I_E = 2\pi^2 \int d\tau N \left(\frac{3}{N^2}a\dot{a}^2 + 3a-\lambda
    a^3\right)\ ,
\end{equation}
where the dot now denotes differentiation with respect to $\tau$. The
action has a saddle point $\bar{a}(\tau)=H^{-1}\cos(H\tau)=H^{-1}\cosh(-iH\tau)$ that
interpolates between the ``South Pole'' of the half-sphere at $\tau =
\pi/(2H)$ and a circle of radius $a < H^{-1}$ at $\tau_a =
H^{-1}\arccos{(Ha)}$. At $t=\tau = 0$, the Lorentzian and Euclidean
saddle points match, as well as their first derivatives.
Analogous to the Lorentzian section one finds
for the on-shell action
\begin{equation}
\begin{split}  
  -I_E^{os}(a) &= -12\pi^2 \int_{\pi/(2H)}^{\tau_a} d\tau
  \left(\bar{a}(t) - H^2\bar{a}^3(t)\right)\\
&= - \frac{4\pi^2}{H^2}\left(\left(1-H^2a^2\right)^{3/2}-1\right) \ .
\end{split}
\end{equation}
One can again define a semiclassical wave function 
\begin{equation}\label{psiEH}
  \Psi_E(a) \simeq \exp{\left(-\frac{1}{\hbar}I_E^{os}(a)\right)} = 
  \exp{\left(\frac{4\pi^2}{\hbar H^2}\left(1-\left(1-H^2a^2\right)^{3/2}
      \right)\right)}\ , \quad Ha < 1 \ ,
\end{equation}
which is also a solution of the WDW equation \eqref{wdwHH} to leading
order in $\hbar$.

The Hartle-Hawking geometry corresponds to the Euclidean/Lorentzian metric
\begin{equation}\label{metricEL}
 \begin{split} 
ds^2 &= d\tau^2 + H^{-2}\cos^2(Ht) d\Omega^2_3 \ , \quad 
   \frac{\pi}{2H} \geq \tau \geq 0 \ , \\
  ds^2 &= - dt^2 + H^{-2}\cosh^2(Ht) d\Omega^2_3 \ , \quad  0 \leq t  \ ,
\end{split}
 \end{equation}
 and the complete on-shell action for $Ha > 1$ is given by
 $iI^{os}_{E/L}(a) = - I^{os}_E(H^{-1}) + iI^{os}_L(a)$ (see Fig.~\ref{fig:scHH}). This yields
 the wave function \cite{Lehners:2023yrj}
 \begin{equation}\label{psiELH}
   \begin{split}
  \Psi_{E/L}(a) &\simeq \exp{\left(\frac{1}{\hbar}\left(-I^{os}_E(H^{-1}) +
        iI_L^{os}(a)\right)\right)}  \\
    &\simeq \exp{\left(\frac{4\pi^2}{\hbar H^2}\left(1 -i\left(H^2a^2-1\right)^{3/2}\right)\right)} \ , \quad Ha > 1 \ .
\end{split}
\end{equation}
Including the standard corrections to order $\hbar$ one obtains  for
the semiclassical wave functions of the Euclidean and Lorentzian sections
\begin{align}
\hat{\Psi}_{E}(a) &\sim \frac{C}{\sqrt{a}\left(1 - H^2a^2\right)^{1/4}}
  \exp{\left(-\frac{4\pi^2}{\hbar 
                        H^2}\left(1-H^2a^2\right)^{3/2}\right)} \ ,\label{psihatE}\\
  \hat{\Psi}_{E/L}(a) &\sim \frac{C}{\sqrt{a}\left(H^2a^2-1\right)^{1/4}}
                        \exp{\left(-i\frac{4\pi^2}{\hbar H^2}\left(H^2a^2-1\right)^{3/2}
                        \right)} \ ,\label{psihatL}
\end{align}
where $C = \exp{\big(\frac{4\pi^2}{\hbar
    H^2}\big)}$. The wave function is
singular at $a=0$ and at $a = H^{-1}$. 

The constant
$C$ depends on the cosmological constant as 
\begin{equation}\label{CC}
C = \exp{\left(\frac{12\pi^2}{\hbar\lambda}\right)} \ ,
\end{equation}
which has been used to argue
that a small value of $\lambda$ is preferred for the vacuum of a
quantum theory of gravity\cite{Hawking:1984hk}.
Note, that the sign of the exponent depends on the
analytic continuation. Changing $t \rightarrow -i\tau$
to $t \rightarrow i\tau$ reverses the sign in the
exponent of Eq.~\eqref{CC} \cite{Linde:1983mx}. Now large values of $\lambda$
appear to be preferred, as for the `tunneling wave function'
\cite{Linde:1983mx,Vilenkin:1984wp}.

The wave function $\hat{\Psi}_{E/L}(a)$ of the Lorentzian section is the
analytic continuation of the Euclidean wave function \eqref{psiEH}
from $Ha < 1$ to $Ha > 1$ with $(1 - H^2a^2)^{3/2} \rightarrow
i(H^2a^2-1)^{3/2}$. It describes an `expanding'
universe. The wave function $\hat{\Psi}^*_{E/L}(a)$ describes a
`contracting' universe. 
The sum of expanding and contracting branch of the Lorentzian
wave function gives
the well-known result of Hartle and Hawking \cite{Hartle:1983ai},
\begin{equation}\label{psiHH}
  \Psi_{\text{HH}}(a) = \hat{\Psi}_{E/L}(a) + \hat{\Psi}^*_{E/L}(a)
  \simeq \frac{2C}{\sqrt{a}\left(H^2a^2-1\right)^{1/4}}
 \cos{\left(\frac{4\pi^2}{\hbar H^2}\left(H^2a^2-1\right)^{3/2}\right)} \ .
                  \end{equation}
Note that this wave function is not connected to the wave
function $\hat{\Psi}_E(a)$ of the Euclidean section by analytic
continuation. This may not be surprising due to the singularity at
$H^2a^2 = 1$ (see Fig.~\ref{fig:scHH}). Exact real solutions of the WDW equation are
regular at this point, as discussed in Sections~\ref{sec:canon}
and \ref{sec:D.1}.

Let us finally emphasize key features of the semiclassical
complex wave functions that describe expanding and contracting
branches, respectively.
The matching of the classical Euclidean and Lorentzian geometries is possible at
a critical scale factor $a_c = H^{-1}$ where time of the Lorentzian
section is analytically continued to imaginary values. The
semiclassical wave functions in the two domains $a > a_c$ and $a < a_c$ satisfy the same WDW equation and are related by
analytic continuation. 

The singularity at $a = a_c$ in Eq.~\eqref{psiHH} is an artifact of the expansion in powers of $\hbar$ and not present in the exact real solution of the WDW equation. However, similar to Eq.~\eqref{psihatL}, the projections of the exact solution to expanding and contracting branches do have a singularity at $a=a_c$, as discussed in Section~\ref{sproj}.

\subsection{Semiclassical dS JT wave function}
\label{sec:semiJT}

Jackiw-Teitelboim gravity \cite{Teitelboim:1983ux,Jackiw:1984je}
in de Sitter space is defined by the action \eqref{jtL},
\begin{equation}
  S[g,\phi] = \frac{1}{2} \int_{\mathcal{M}}d^2x \sqrt{g} \phi (R - 2\lambda^2) +
  \int_{\partial\mathcal{M}} d\theta \sqrt{h} \phi K \ .\nonumber
\end{equation}
Compared to Eq.~\eqref{HHL}, the case of pure gravity with cosmological
constant, the action depends linearly on a dilaton field $\phi$.
In minisuperspace, which in 2d is just a gauge choice \cite{Louis-Martinez:1993bge}, the metric
\begin{equation}\label{mLd2}
  ds^2 = -N^2(t)dt^2 + a^2(t) d\theta^2 
\end{equation}
yields the action ($h=a^2)$
\begin{equation}\label{ILjt}
  I_L[h,\phi] = 2\pi \int dt N\left(-\frac{1}{N^2} \dot{a} \dot{\phi} - \lambda^2 a\phi
    \right) \ .
  \end{equation}
This implies the Hamiltonian constraint
    \begin{equation}\label{jtHC}
\dot{a} \dot{\phi} - \lambda^2 a\phi = 0 \ ,
\end{equation}
which leads to the WDW equation\footnote{From now on we set $\hbar=1$.} ($N=1$)
\begin{equation}\label{wdwjt}
  \left(\frac{1}{4\pi^2}\frac{\partial^2}{\partial
      a\partial\phi}
    +\lambda^2 a \phi\right)\Psi(a,\phi) = 0 \ .
\end{equation}
As in 4d de Sitter, the solution of the equations of motion is $\bar{a}(t) =
H^{-1}\cosh{(Ht)}$, with $H = \lambda$, which interpolates between
a circle of minimal radius $H^{-1}$ at $t = 0$ and a circle of radius
$a > H^{-1}$ at $t_a = H^{-1} \operatorname{arcosh}{(Ha)}$ of the
de Sitter hyperboloid. The corresponding solution for the dilaton field, satisfying
the constraint \eqref{jtHC} and the boundary condition
$\bar{\phi}(t_a) = \phi$, reads
\begin{equation}\label{phi0}
  \bar{\phi}(t) = \phi_0 \sinh(Ht)\ , \quad
  \phi_0 = \frac{\phi}{\sinh(Ht_a)} \ .
\end{equation}
$\bar{\phi}(t)$ vanishes at $t=0$. From Eqs.~\eqref{ILjt} and \eqref{jtHC} one
obtains the on-shell action ($h_c = \lambda^{-2})$
\begin{equation}
  I^{os}_L(h,\phi) = -2\pi\lambda \phi\sqrt{h-h_c} \ ,
\end{equation}
which can also be directly read off from Eq.~\eqref{jtL} by using $R = 2\lambda^2$
and inserting the extrinsic curvature $K = -\lambda h^{-1/2}(h-h_c)^{1/2}$.
The semiclassical wave function 
\begin{equation}
  \Psi_L(h,\phi) \simeq
  \exp{\left(iI^{os}_L(h,\phi)\right)}
  = \exp{\left(-2\pi i\lambda \phi\sqrt{h-h_c}\right)} \ ,
  \quad h > h_c \ ,
\end{equation}
is a solution of the WDW equation to leading order in $\hbar$.

\begin{figure}[t]
 \centering
 \includegraphics[width = 0.6 \textwidth]{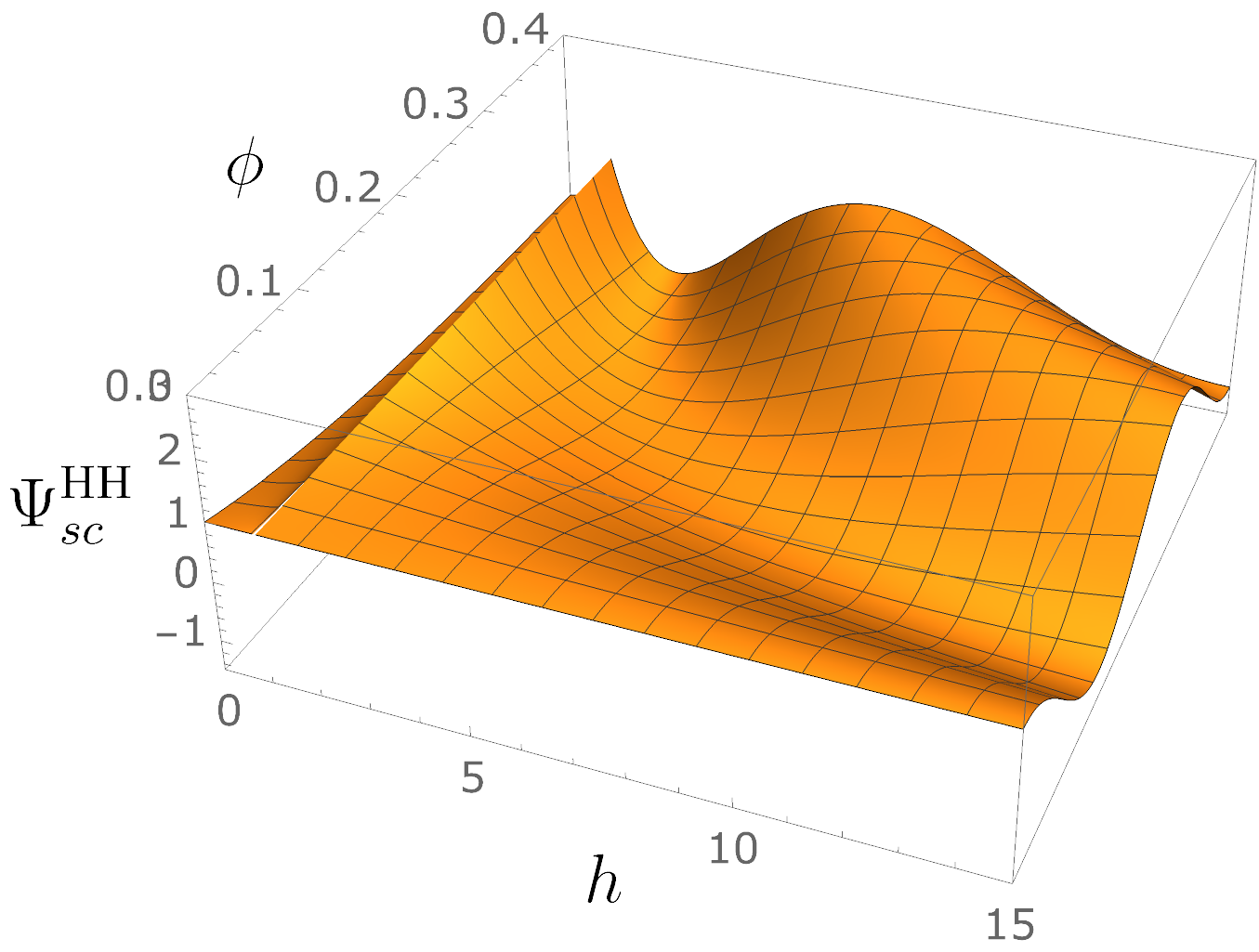}
 \caption{Hartle-Hawking-type semiclassical wave function $C^{-1}\Psi_{sc}$ in the $h\!\!-\!\!\phi$-plane; analytic continuation $\tau = it$ at $h_0=h_c=1$.}
  \label{fig:psiscHH}
\end{figure}

 Continuing analytically to the Euclidean half-sphere with $t=-i\tau$
 and metric
 \begin{equation}\label{mEd2}
   ds^2 = N^2(\tau) d\tau^2 + a^2(\tau) d\theta^2 \ ,
\end{equation}
the Lorentzian action $iI_L$ becomes
\begin{equation}\label{IEjt}
  -I_E[h,\phi] = 2\pi \int d\tau N\left(\frac{1}{N^2} \dot{a}\dot{\phi} - \lambda^2 a\phi\right) \ .
\end{equation}
The saddle point of $I_E$ that matches the Lorentzian saddle point at
$t = \tau = 0$ reads
\begin{equation}
  \bar{a}(\tau) = H^{-1}\cos(H\tau), \quad
  \bar{\phi}(\tau) = \phi_0\sin(H\tau) \ .
\end{equation}
Also the first derivatives of $\bar{a}(\tau)$ and $\bar{\phi}(\tau)$
agree with the first derivatives of the Lorentzian saddle point
$\bar{a}(t)$ and $\bar{\phi}(t)$ at $t = \tau = 0$. $\phi_0$ is the
value of the dilaton field at the ``South Pole'' $\tau = \pi/(2H)$.
It is a function of $a$ and $\phi$, given in
Eq.~\eqref{phi0}.\footnote{This is analogous to the saddle point
  discussed in \cite{Maldacena:2019cbz} where the dilaton is continued
  to imaginary values in the Euclidean region.}
From Eq.~\eqref{IEjt} one obtains the on-shell action
\begin{equation}
  -I_E^{os}(h,\phi) = 2\pi\left(\phi_0 - \lambda\phi\sqrt{h_c-h}\right) \ ,
\end{equation}
which corresponds to the imaginary Euclidean extrinsic curvature
$K = i\lambda h^{-1/2}\sqrt{h_c-h}$,
with $\bar{a}(\tau_a) = a$ and $\bar{\phi}(\tau_a)=\phi_0\sin(H\tau_a)=\phi$.
The semiclassical wave function 
\begin{equation}\label{psiEjt}
  \Psi_E(h,\phi) \simeq \exp{\left(-I_E(h,\phi)\right)}
  = \exp{\left(2\pi\left(\phi_0 - \lambda\phi\sqrt{h_c - h}\right)\right)} \ ,
    \quad h < h_c \ ,
  \end{equation}
  is again a solution of the WDW equation \eqref{wdwjt}.

\begin{figure}
 \centering
 \includegraphics[width = 0.6 \textwidth]{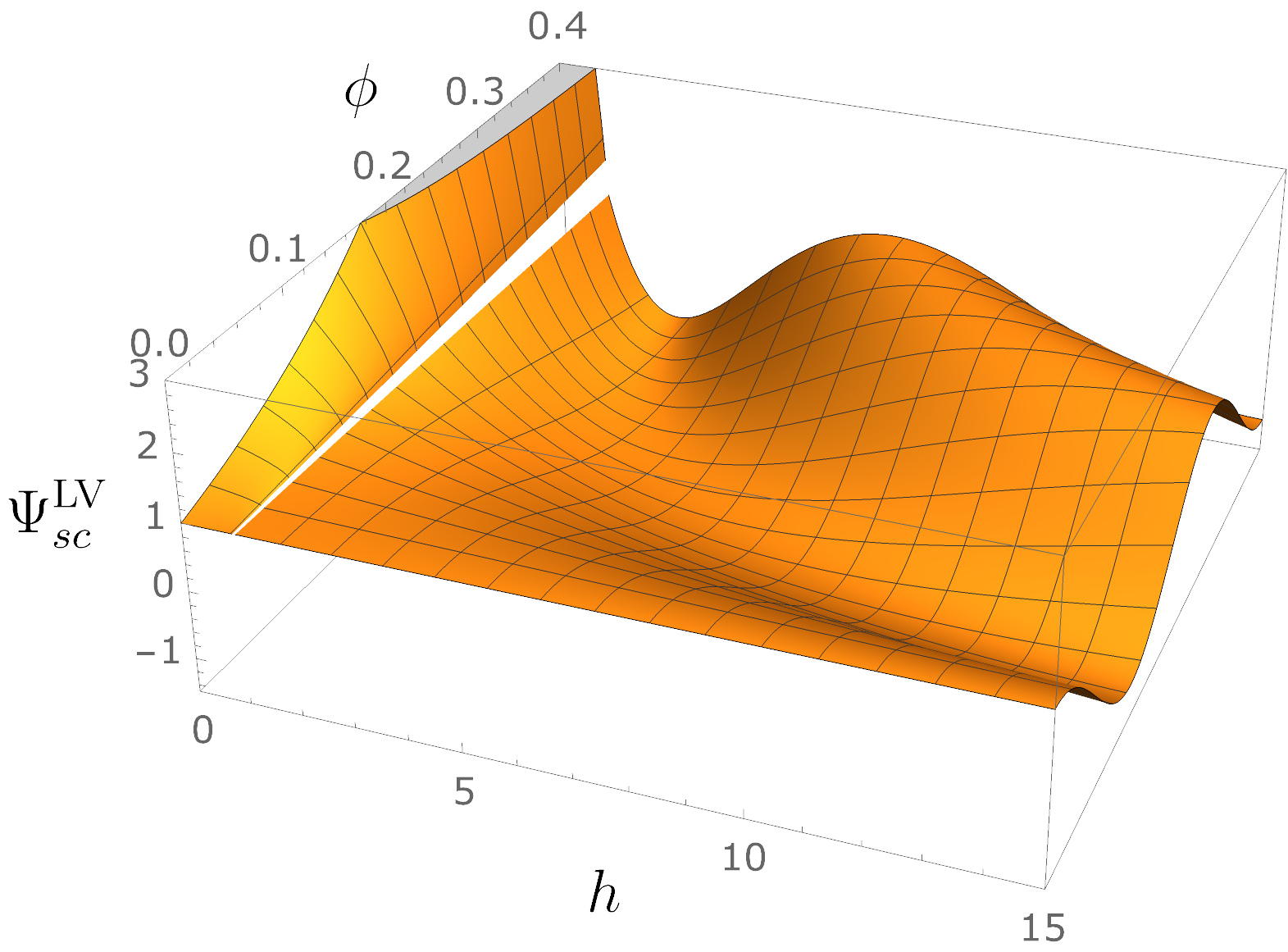}
 \caption{Linde-Vilenkin-type semiclassical wave function $C^{-1}\Psi_{sc}$ in the $h\!\!-\!\!\phi$-plane; analytic continuation $\tau = -it$ at $h_0=h_c=1$.}
  \label{fig:psiscLV}
 \end{figure}

For the Euclidean/Lorentzian metrics \eqref{mLd2} and \eqref{mEd2}
 the complete on-shell action for $h > h_c$ is given by
 $iI^{os}_{E/L}(h,\phi) = - I^{os}_E(\lambda^{-2},0) + iI^{os}_L(h,\phi)$. This yields
 the wave function
 \begin{equation}\label{psiELjt}
   \begin{split}
  \Psi_{E/L}(h,\phi) &\simeq \exp{\left(-I_E^{os}(\lambda^{-2},0) +
        iI_L^{os}(h,\phi)\right)}  \\
    &\simeq \exp{\left(2\pi\left(\phi_0 -
          i\lambda\phi\sqrt{h-h_c} \right)\right)} \ , \quad h > h_c \ .
\end{split}
\end{equation}
The complete real wave function is given by
\begin{equation}\label{psisc}
\begin{split}
  \Psi_{sc}(h,\phi) &= \Psi_{E}(h,\phi)\Theta(h_c-h)
 + \frac{1}{2}\left(\Psi_{E/L}(h,\phi) + \Psi_{E/L}(h,\phi)^*\right)
    \Theta(h-h_c) \\
 &= C \big(\exp{(-\lambda\phi\sqrt{h_c - h})}\Theta(h_c-h)
+\cos{(2\pi\lambda\phi\sqrt{h-h_c})}\Theta(h-h_c)\big) \ ,
\end{split}
\end{equation}
with the normalization constant
\begin{equation}
C = \exp{\left(2\pi\phi_0\right)} \ .
\end{equation}
Note that, contrary to the
Hartle-Hawking wave function, $C$ does not depend on the
cosmological constant but instead on the value of the dilaton at
the South Pole.

Including the standard corrections $\mathcal{O}(\hbar)$ one obtains for
the semi-classical wave functions of the Euclidean and Lorentzian sections
\begin{align}
 \hat{\Psi}_{E}(h,\phi) &\sim
  \frac{C}{\lambda\sqrt{h_c-h}}
  \exp{\left(-2\lambda\pi
  \phi\sqrt{h_c-h}\right)} \ , \label{psihatjtE}\\
 \hat{\Psi}_{E/L}(h,\phi) &\sim
  \frac{C}{\lambda\sqrt{h-h_c}}
  \exp{\left(-i2\pi\lambda\phi\sqrt{h-h_c}\right)}
                            \ , \label{psihatjtL}
\end{align}
Contrary to the
Hartle-Hawking wave function, $C$ does not depend on the
cosmological constant but instead on the value of the dilaton at
the South Pole.
$\hat{\Psi}_{E/L}(a,\phi)$ is the analytic continuation of the Euclidean wave function
\eqref{psiEjt} from $h < h_c$ to $h > h_c$ with $\sqrt{h_c - h)} \rightarrow
i\sqrt{h-h_c}$. $\hat{\Psi}_{E/L}(h,\phi)$ describes an expanding universe and the complex conjugate $\hat{\Psi}^*_{E/L}(h,\phi)$ a contracting universe. 
The sum of expanding and contracting
branch of the Lorentzian wave function yields
\begin{equation}\label{psiJT}
 \hat{\Psi}_{E/L}(h,\phi) + \hat{\Psi}^*_{E/L}(h,\phi)
\sim \frac{C}{\lambda\sqrt{h-h_c}}
\cos{\left(2\pi\lambda\phi\sqrt{h-h_c}\right)}\ , 
\end{equation}
which is the analogue to the Hartle-Hawking wave function \eqref{psiHH}.

The semiclassical wave function $\Psi_{sc}$, Eq.~\eqref{psisc}, with $\tau = it$, is shown
in Fig.~\ref{fig:psiscHH}. With increasing $h$, $\Psi_{sc}$ increases
exponentially, until at $h=h_c$ it starts to oscillate. On the contrary,
with $\tau = -it$, the wave function decreases exponentially toward
the oscillatory regime, see Fig.~\ref{fig:psiscLV}. For both signs,
$\Psi_{sc}$ does not vanish at $h=0$.

\section{Canonical quantization and factor ordering}
\label{sec:canonical}

For completeness, we briefly review canonical quantization of JT gravity in this section. Subsequently, we discuss the
connection between path integral and factor ordering. For an interesting recent discussion of the Hilbert space of dS JT gravity, see \cite{Held:2024rmg}.

\subsection{Classical theory}

Canonical quantization of JT gravity was first studied by
Henneaux \cite{Henneaux:1985nw}, and it has been shown that in this
model the momentum constraints can be solved such that a wave
functional can be explicitly constructed that solves the functional
WDW equation \cite{Henneaux:1985nw,Louis-Martinez:1993bge}.

Starting point is the Lorentzian action \eqref{jtL} of the JT model including the GHY boundary term,
\begin{equation}
  S[g,\phi] = \frac{1}{2} \int_{\mathcal{M}}d^2x \sqrt{g} \phi (R - 2\lambda^2) +
  \int_{\partial\mathcal{M}} d\theta \sqrt{h} \phi K \ .\nonumber
\end{equation}
Following \cite{Halliwell:1988ik}, we use the ADM decomposition of the
metric \cite{Arnowitt:1962hi}
\begin{equation}\label{adm}
  ds^2 = -\frac{N^2}{h}dt^2 + h(d\theta + \tilde{N}dt)^2 \ ,
\end{equation}
where $N$ and $\tilde{N}$ are lapse function and shift vector,
respectively.
The normal vector on slices $\Sigma_t$ of constant $t$ is $n^\alpha =
(\sqrt{h}/N)(\partial_t - \tilde{N} \partial_\theta) x^\alpha, 
\alpha = 0,1$, with $n^\alpha n_\alpha = -1$.
One can now express the action in terms of the scalar field $\phi$ and the metric components introduced 
in the decomposition \eqref{adm}. Removing the GHY boundary term by means of Gauss's theorem,
one obtains after a straightforward calculation,\footnote{The extrensic
  curvature is given by $K=\nabla_\alpha n^\alpha =
  (\partial_t\sqrt{h} - \sqrt{h}D_\theta \tilde{N})/N$, where
  $D_\theta$ is the covariant derivative w.r.t. the induced
  metric $h$ and $\partial^\theta\phi = h^{-1}\partial_\theta \phi$ . The normal derivative of $\phi$ reads
  $\partial_n\phi = (\sqrt{h}/N)(\partial_t - \tilde{N} \partial_\theta)\phi$.}
\begin{equation}\label{SADM}
  S[g,\phi] = \int d^2x\left(-\frac{1}{2N}(\partial_t h - 2 hD_\theta \tilde{N})
    (\partial_t - \tilde{N}\partial_\theta)\phi - N\left(\lambda^2 \phi
    +D_\theta\partial^\theta\phi\right)\right) \ .
    \end{equation}
The classical phase space consists of the fields $h$ and $\phi$, and their conjugate momenta
\begin{equation}
  \begin{split}
    \Pi_\phi &= -\frac{1}{2N}(\partial_t h - 2 hD_\theta \tilde{N}) =
    - \sqrt{h} K \ ,\\
    \Pi_h &= -\frac{1}{2N}(\partial_t - \tilde{N}\partial_\theta)\phi =
    - \frac{1}{2\sqrt{h}} \partial_n\phi \ .
\end{split}
\end{equation}
The Hamiltonian is given by
\begin{equation}
H = \int d\theta (N\mathcal{H} + \tilde{N}\tilde{\mathcal{H}}) \ ,
\end{equation}
with the Hamiltonian and momentum densities
\begin{equation}\label{densities}
  \mathcal{H} = -2 \Pi_h \Pi_\phi + \lambda^2 \phi
  +D_\theta\partial^\theta\phi \ , \quad
  \tilde{\mathcal{H}} = -\sqrt{h} \partial_\theta(\sqrt{h}\Pi_h)
  + \partial_\theta\phi \Pi_\phi \ .
\end{equation}
Variation with respect to the Lagrange multiplier fields
$N$ and $\tilde{N}$ yields the first-class constraints
(for a review, see, for example~\cite{Kiefer:2007ria}),
\begin{equation}\label{constraints}
\mathcal{H} = 0 \ , \quad \tilde{\mathcal{H}} = 0 \ ,
\end{equation}
which are satisfied on a constrained phase space. 
Given two first-class constraints, there are no
propagating modes.
An important role is played by the functional \cite{Louis-Martinez:1993bge}
\begin{equation}
C[h,\Pi_h,\phi,\Pi_\phi] 
= 4 h \Pi_h^2 - h^{-1}(\partial_x\phi)^2 - \lambda^2\phi^2 ,
\end{equation}
whose Poisson brackets with the generators of time and
space translations vanish,
\begin{equation}
\{\mathcal{H},C\} = 0 \ , \quad\ 
\{\tilde{\mathcal{H}},C\} = 0 \ .
\end{equation}
Hence, $C$ is constant on the constrained phase space, i.e., a global variable. This is also the case for 
\begin{equation}\label{momC}
  P_C = - \int_0^{2\pi} d\theta\frac{h \Pi_h}
  {4h\Pi_h^2 - h^{-1}(\partial_x\phi)^2} \ ,
\end{equation}
which is the momentum conjugate to $C$,
\begin{equation}
\{C,P_C\} = 1 \ .
\end{equation}

The invariance with respect to spatial diffeomorphisms
can be used to transform on a slice $\Sigma_t$ the field $h(\theta)$ to a constant value $h$. Moreover, a time slicing 
can be chosen such that on $\Sigma_t$ also the function
$\phi(\theta)$ becomes a constant value $\phi$ 
\cite{Louis-Martinez:1993bge}. After this reduction of an infinite number of degrees of freedom to two, 
the JT model becomes a system with just two variables,
$h$ and $\phi$. Hence, in JT gravity, minisuperspace
is not an approximation but just a particular choice of spacetime coordinates. Note that
different pairs $(h,\phi)$ correspond to
different spatial slices of the manifold. 

The Hamiltonian density $\mathcal{H}$ in 
Eq.~\eqref{densities} is bilinear in the momenta.
As a consequence, the constraints \eqref{constraints}
are equivalent to the two linear constraints \cite{Henneaux:1985nw}
\begin{equation}\label{clin}
\begin{split}
  \Pi_h &= \frac{Q}{2\sqrt{h}} \ , \quad
    Q(h,\phi;C) = (\lambda^2\phi^2+ R^2 +C))^{1/2} , \;
R = \sqrt{h^{-1}(\partial_\theta \phi)^2}\ ,    \\
  \Pi_\phi &= \frac{\sqrt{h}g}{Q} \ , \quad g(h,\phi) = \lambda^2\phi + D_\theta\partial^\theta\phi \ ,
\end{split}
 \end{equation}
where $C$ is a constant. 

\subsection{Quantum theory}
\label{qth}
Quantization of the theory amounts to solving the constraints
$\mathcal{H}\Psi = 0$ and $\tilde{\mathcal{H}}\Psi = 0$ as functional differential equations, where the classical momenta $\Pi_h$ and $\Pi_\phi$ are replaced by
$\hat{\Pi}_h=-i\delta/\delta h$ and $\hat{\Pi}_\phi =
-i\delta/\delta \phi$ (see, for example,
\cite{Kiefer:2007ria}). In general, only approximate solutions to these equation can be found. In JT gravity,
however, it is sufficient to solve the linear differential equations
\begin{equation}\label{qlin}
  \hat{\Pi}_h \Psi = \frac{Q}{2\sqrt{h}} \Psi\ , \quad
  \hat{\Pi}_\phi \Psi  = \frac{\sqrt{h}g}{Q}\Psi \ .
\end{equation}
An exact solution is given by the `HLGK wave function'
\cite{Henneaux:1985nw,Louis-Martinez:1993bge}
\begin{equation}
  \Psi_{\rho}[h,\phi] = \int dC \rho(C) \Psi[h,\phi;C]) \ ,
\end{equation}
where
\begin{equation}\label{psifunc}
\Psi[h,\phi;C] = \exp{\left(-i\int_{0}^{2\pi} d\theta \sqrt{h} \left(Q + \frac{1}{2} R
  \ln{\left|\frac{Q - R}{Q + R}\right|}\right)\right)} \ ,
\end{equation}
and $\rho(C)$ is an arbitrary function. Note that $\Psi[h,\phi;C]$ is
manifestly invariant under spatial diffeomorphisms. By construction,
the wave function $\Psi$ satisfies the WDW equation with a particular
factor ordering
\begin{equation}\label{fwdw}
    \left(2\hat{Q}\frac{\delta}{\delta \phi}
    \hat{Q}^{-1}\frac{\delta}{\delta h} +
    g\right)\Psi(h,\phi;C) = 0 \ ,
\end{equation}
where we have defined $\hat{Q} = Q(\hat{C};h,\phi)$ with
 $\hat{C} \Psi[h,\phi;C] = C \Psi[h,\phi;C]$. 
 The canonically conjugate momentum acts on
 the wave function as
 \begin{equation}
\begin{split}
 -i\frac{\partial}{\partial C}
\Psi[h,\phi;C] 
  &= -\left(\frac{1}{2}\int_0^{2\pi} d\theta \sqrt{h}\frac{Q}{Q^2 - R^2}
\right) \Psi[h,\phi;C] \\
&= 
P_C[h,\phi;C] \Psi[h,\phi;C] \ ,
\end{split}
\end{equation} 
where the function $P_C$ is given by Eq.s~\eqref{momC}, \eqref{clin}.
In
 \cite{Iliesiu:2020zld} this construction of a wave function has been
 applied to Euclidean AdS JT gravity.

Going to minisuperspace, the momentum constraint is automatically fulfilled, and the functional WDW equation
\eqref{fwdw} becomes an ordinary partial differential
equation with a particular factor ordering,
\begin{equation}\label{swdw}
  (Q\partial_\phi Q^{-1}\partial_h + 2\pi^2\lambda^2\phi) \Psi = 0\ .
\end{equation}
 As observed in \cite{Iliesiu:2020zld}, the solutions are simply related to solutions of the WDW equation with canonical factor ordering. Using $Q^{-1}\partial_h\Psi = i\pi/\sqrt{h} \Psi$, one obtains from Eq.~\eqref{swdw} a WDW equation with
 `Henneaux factor ordering',
\begin{equation}\label{HFO}
\left(\sqrt{h}\partial_h \frac{1}{\sqrt{h}} \partial_\phi
+ 2\pi^2\lambda^2\phi\right) \Psi = 0 \ .
\end{equation}
One concludes that $\Psi=\sqrt{h}\tilde{\Psi}$ is a solution ot this equation if $\tilde{\Psi}$ is a solution of the
WDW equation with canonical factor ordering. 

\subsection{Path integral and factor ordering}
\label{sec:pifo}

The appearance of WDW equations with different factor orderings
is related to the ambiguity in the measure of the
path integral, which has been discussed by Halliwell 
in the case of 4d de Sitter gravity \cite{Halliwell:1988wc}. 
The procedure can be directly applied to JT gravity.
We write the action
\eqref{SADM} for minisuperspace in terms of an off-diagonal metric $f_{\alpha\beta}$ in field space,
\begin{equation}
S = \int dt \left(p_h \dot{h} + p_\phi \dot{\phi} - N H\right) \ ,
\end{equation}
with
\begin{equation}
    H = -\frac{1}{2\pi}f^{\alpha\beta}p_{\alpha}p_{\beta} + V(q)\ , \quad f^{h\phi}=f^{\phi h} = 1\ , \quad V(q) = 2\pi\lambda^2\phi \ .
\end{equation}
Demanding invariance w.r.t. field redefinitions, one defines the Hamilton operator in terms of the Laplacian,
\begin{equation}
H = - \frac{1}{2}\Delta + V \ , \quad 
\Delta = \frac{1}{\sqrt{f}}\partial_\alpha\left(\sqrt{f}f^{\alpha\beta}\partial_\beta\right) \ . 
    \end{equation}
The quantum constraint $\mathcal{H}\Psi_c=0$ then yields the WDW equation with canonical factor ordering,
\begin{equation}\label{cFO}
\left(\frac{1}{2}f^{\alpha\beta}\partial_\alpha \partial_\beta
+ 2\pi^2\lambda^2\phi\right) \Psi_c = 0 \ .
\end{equation}

This WDW equation is not unique because of the
ambiguity of the lapse function in the ADM construction.
Define a new lapse function by $N = \tilde{N} \Omega(h)$.
The action now becomes
\begin{equation}
S = \int dt \left(p_h \dot{h} + p_\phi \dot{\phi} - \tilde{N} \tilde{H}\right) \ ,
\end{equation}
with
\begin{equation}
\tilde{H} = -\frac{1}{2}\tilde{\Delta}+\tilde{V}=\Omega H\ , \quad
\tilde{\Delta} = \Omega\Delta\ ,\quad \tilde{V} =\Omega V\ . 
\end{equation}   
Now $\tilde{N}$ is the integration variable in the path integral, corresponding to a different choice of integration measure.
The canonical wave function also satisfies the new quantum constraint, $\tilde{\mathcal{H}}\Psi_c = 0$, with the rescaled Laplacian and 
the rescaled potential. Alternatively, one may use a rescaled wave function, $\Psi^{(\Omega)} = 
\Omega\Psi_c$, which satisfies a WDW equation with
the same potential as in Eq.~\eqref{cFO} but with
different factor ordering,
\begin{equation}\label{OmFO}
\left(\Omega\partial_h \Omega^{-1} \partial_\phi
+ 2\pi^2\lambda^2\phi\right) \Psi^{(\Omega)} = 0 \ .
\end{equation}
The special choice $\Omega = \sqrt{h}$ corresponds to Henneaux factor ordering. Clearly, quantization of JT gravity does not single out a particular factor ordering. Henneaux factor ordering simply allows for an explicit
representation of the wave functional.

There is an analogous relation between the Green functions for
different factor orderings. Defining $\sigma = \phi^2$,
one easily verifies that the solution of
\begin{equation}\label{hfowdw1}
  \left(\Omega(h) \partial_h \Omega(h)^{-1}\partial_\sigma+\pi^2\lambda^2\right)
G^{(\Omega)}(h,\sigma;h',\sigma') = \delta(h-h')\delta(\sigma-\sigma') 
\end{equation}
is given by
\begin{equation}
  G^{(\Omega)}(h,\sigma;h',\sigma') = \frac{\Omega(h)}{\Omega(h')}G(h,\sigma;h',\sigma')\ , 
\end{equation}
where  $G(h,\sigma;h',\sigma')$ is the Green function for
canonical factor ordering.

Rescaling the lapse function by $\Omega(h)$ affects  probability measures whose definition and interpretation
remains an important topic of current research
(for a discussion and references, see, for 
example~\cite{Halliwell:2011zz}).
The Hartle-Hawking measure \cite{Hawking:1987bi}
\begin{equation}
dP_\text{HH}(h,\phi) = |\Psi|^2 \sqrt{f}dhd\phi
\end{equation}
scales as
\begin{equation}
dP^{(\Omega)}_\text{HH} (h,\phi) = \Omega(h)^{-1} d P_\text{HH}(h,\phi) \ .
\end{equation}
In the semiclassical regime one can also use the measure provided by the conserved 
Klein-Gordon current \cite{Vilenkin:1988yd,DeWitt:1967yk}.
Choosing the component of the current in $\phi$-direction
one has for expanding and contracting branches
\begin{equation}
dP_\pm(h,\phi) = j_\pm^{\phi}(h,\phi)\sqrt{f}
\epsilon_{\phi h}dh\ , \quad
j_\pm^{\phi} = \frac{i}{2}f^{\phi h}\left(\Psi^* \partial_{h}\Psi
- \Psi \partial_{h}\Psi^* \right) \ .
\end{equation}
In two dimensions the rescaling of the metric drops out
and the measure is completely determined by the wave function.

\section{Boundary wave function with compact support}
\label{sec:nonSingBWF}

We obtained many results in the main text for reasons of simplicity by specifying the boundary wave function to be of the simple form $\chi(h',0)=\delta(h'-h_0)$. However, this restrictive assumption is not necessary to obtain the asymptotic Schwarzian behaviour. To show this, consider a boundary wave function that is non-zero in the finite interval $[0,h_0]$. As a convenient simplification approximate this more general boundary wave function as a rectangular box function $\Psi(h',0)=\Theta(h')-\Theta(h' - h_0)$. 

We can now directly use Eq.~\eqref{apsibulk}, exploit there again that 
$$\partial_{h'}J_0(\sqrt{-\alpha\beta})=-\partial_{h}J_0(\sqrt{-\alpha\beta}) \ ,$$
and then expand this in $h_0\ll h$. This leaves us with
\begin{equation}\label{psibulk2}
  \begin{split}
  \left.\Psi(h,\phi)\right|_{h>h_0}&=  \int_0^h dh' \Psi(h',\phi') \partial_{h} J_0((-\alpha\beta)^{1/2}) \nonumber\\
    &=\int_0^{h_0} dh' \partial_{h}\left[\left.J_0((-\alpha\beta)^{1/2})\right|_{h'=0}+\left.\partial_{h'}J_0((-\alpha\beta)^{1/2})\right|_{h'=0}\ h'+\ldots\right]
\end{split}
\end{equation}
which at large $h\gg h_0$ becomes
\begin{equation}
\Psi(h,\phi)= \partial_{h}J_0(\underbrace{2\pi\lambda\phi\sqrt{h}}_{x})\ h_0+\partial_{h}^2J_0(2\pi\lambda\phi\sqrt{h})\ h_0^2+\ldots=-\frac{\pi\lambda\phi}{\sqrt{h}}J_1(x)\ h_0+\ldots\quad.
\end{equation}
This is the result obtained before, which tells us that all we require is the boundary wave function to be normalizable and with compact support on $h\in [0,h_0]$. For our analysis this is beneficial because we can keep using a simplifying choice of class of boundary wave functions given by delta functions and their derivatives with the background knowledge that the results obtained this way will carry over when generalizing the boundary wave function choice to a general normalizable function with compact support. In a sense, this is reminiscent of the electric field of a compact charge distribution having a dominant monopole contribution at large distances.

\section{Exact solutions with Airy functions}
\label{sec:airy}

\subsection{Hartle-Hawking wave function}
\label{sec:D.1}

The semiclassical wave functions \eqref{psihatE} and \eqref{psihatL} diverge at $a=0$ and at the Stokes point $a = H^{-1}$ where,
after an exponential increase in $a$, the wave function starts to
oscillate. To understand this behaviour better it is useful to rescale the lapse function and to use instead of
the metric \eqref{metricL} the metric  \cite{Halliwell:1988wc}
\begin{equation}\label{metricLh}
   ds^2 = - \frac{N^2(t)}{h(t)} dt^2 + h(t) d\Omega^2_3 \ .
 \end{equation}
The Lorentzian action now reads
\begin{equation}\label{ILh}
I_L = 6\pi^2\int dtN\left(-\frac{1}{4N^2}\dot{h}^2 + 1 -\frac{\lambda}{3}h\right)
\end{equation}
and the WDW equation becomes
\begin{equation}\label{wdwHHh}
    \left(\frac{\hbar^2}{36\pi^4}\frac{\partial^2}{\partial h^2} - 1 +
    \frac{\lambda}{3} h\right)\Psi(h) = 0 \ ,
\end{equation}
which  corresponds to a stationary Schroedinger equation with  linear potential. 
The general solution is a linear combination of Airy
functions \cite{Halliwell:1988wc,Lehners:2021jmv}
\begin{equation}\label{AiBi}
  \psi(h) = \alpha \text{Ai}(x) + \beta\text{Bi}(x)\ ,
\end{equation}
where $x(h) = (6\pi^2/\hbar H^2)^{2/3}(1-H^2h)$, with $H^2 = \lambda/3$. At small $h$, i.e. large $x$,
the Airy functions behave as 
\begin{align}
\text{Ai}(x(h)) &\sim
               \frac{C'}{\left(1 - H^2h\right)^{1/4}}
  \exp{\left(-\frac{4\pi^2}{\hbar H^2}(1-H^2h)^{3/2}\right)} (1 +
                  \mathcal{O}(x^{-1})) \ , \label{aismall}\\
\text{Bi}(x(h)) &\sim
               \frac{C'}{\left(1 - H^2h\right)^{1/4}}
  \exp{\left(\frac{4\pi^2}{\hbar H^2}(1-H^2h)^{3/2}\right)} (1 +
                  \mathcal{O}(x^{-1})) \ , \label{bismall}
\end{align}
with $C' = \big(\frac{\hbar H^2}{\pi^2}\big)^{1/6}$, whereas
at large $h$, i.e. large $-x$, one has
\begin{align}
\text{Ai}(x(h)) &\sim \frac{C'}{\left(H^2h-1\right)^{1/4}}
  \cos{\left(\left(\frac{4\pi^2}{\hbar H^2}\right)(H^2h-1)^{3/2}\right)} (1 +
                  \mathcal{O}(x^{-3/2})) \ , \label{ailarge}\\
  \text{Bi}(x(h)) &\sim
               \frac{C'}{\left(H^2h-1\right)^{1/4}}
  \sin{\left(\left(\frac{4\pi^2}{\hbar H^2}\right)(H^2h-1)^{3/2} \right)} (1 +
                  \mathcal{O}(x^{-3/2})) \ .                    \label{bilarge}
\end{align}
At $x=0$, i.e. $h=H^{-2}$, both Airy functions are finite,
$\text{Ai}(0) = 1/\big(3^{2/3}\Gamma{\big(\frac{2}{3}\big)}\big)$ and 
$\text{Bi}(0) = 1/\big(3^{1/6}\Gamma{\big(\frac{2}{3}\big)}\big)$. Hence,
contrary to the semiclassical wave functions, the wave functions
$\psi(h)$ are finite at the Stokes point $h=H^{-2}$.
Both Airy functions are non-zero at $h=0$. A wave function
that vanishes at $h=0$ can be obtained as linear combination.
Note that a wave function is real or complex for all
values of $h$. In particular, a real wave function at
small $h$ can not turn smoothly near $h=H^{-1}$ into a complex wave function at large $h$, in agreement with the conservation of the Klein-Gordon current.

It is instructive to compare the asymptotic behaviour of
the exact solutions with the semiclassical approximation.
Starting from the action \eqref{ILh}, and using the same procedure as in Section~\ref{sec:semiHH}, one finds
\begin{align}
  \psi_E(h)  &\sim
               \frac{C}{\left(1 - H^2h\right)^{1/4}}
  \exp{\left(-\frac{4\pi^2}{\hbar H^2}(1-H^2h)^{3/2}\right)} \ , \quad h < H^{-2} \ , \label{psiEh} \\
  \psi_L(h) & \sim \frac{C}{\left(H^2h-1\right)^{1/4}}
  \exp{\left(-i\frac{4\pi^2}{\hbar H^2}(H^2h-1)^{3/2}\right)} \ , \quad h > H^{-2} \ ,\label{psiELh} 
\end{align}
where $C = \exp{\left(\frac{4\pi^2}{\hbar H^2}\right)}$.
Matching \eqref{psiEh} to an Airy-function solution implies
$\beta = 0$ in \eqref{AiBi}. But then the oscillating part of the exact Airy-function solution 
can not be given by a complex wave function. Hence, the semiclassical complex wave function discussed in Section~\ref{sec:semiHH} is inconsistent with an exact Airy-function solution.

The Airy wave functions differ from the semiclassical
wave functions in Section~\ref{sec:semiHH} by a factor $\sqrt{a}$ and hence 
there is no singularity at $h = 0$. This is a consequence of
the rescaled lapse function in the metric \eqref{metricLh},
which affects the factor ordering. In terms of the variable
$a$ the WDW equation \eqref{wdwHHh} reads
\begin{equation}
    \left(\frac{\hbar^2}{144\pi^4}a\frac{\partial}{\partial a}\frac{1}{a}\frac{\partial}{\partial a}- a^2 +
    \frac{\lambda}{3} a^4\right)\Psi(a) = 0 \ ,
\end{equation}
which differs from the WDW equation \eqref{wdwHH} by
a specific factor ordering.

\subsection{JT wave equation}

In the case of JT gravity we start from the WDW equation
\eqref{wdwjt},
\begin{equation}\label{wdwjt2}
  H \Psi = \left(\frac{1}{2\pi^2}\frac{\partial^2}{\partial h\partial\phi}
    +\lambda^2 \phi\right)\Psi(h,\phi) = 0 \ .
\end{equation}
Changing variables to $h_\pm = \lambda^2 h \pm \phi$, the
Hamiltonian becomes a sum, and the wave function $\Psi$ factorizes,
\begin{equation}
H = H_+ - H_- \ , \quad \Psi = \Psi_+ \Psi_- \ , \quad
H_\pm \Psi_\pm = \left(\frac{1}{\pi^2}\frac{\partial^2}{\partial h^2_\pm}
+h_\pm - c\right)\Psi_\pm = 0 \ ,
\end{equation}
where $c$ is an arbitrary constant. As in Section~\ref{sec:D.1}, the solutions $\Psi_\pm$
are linear combinations of Airy functions,
\begin{equation}
\psi_\pm (h_\pm) = \alpha_\pm \text{Ai}(x_\pm) + \beta_\pm\text{Bi}(x_\pm) \ ,
\end{equation}
with $x_\pm = \left(\frac{\pi}{\lambda}\right)^{2/3}(c-h_\pm)$. The Airy functions have two Stokes lines,
at $h_\pm = c$, which start at the Stokes point
$h_0 = c/\lambda^2$ on the boundary $\phi = 0$.

Consider first large field values with $h_\pm \gg \lambda^2 h_0$, i.e. 
$-x_\pm \gg \lambda^2 h_0$. The asymptotic behaviour of the Airy functions can be read off from Eq.s~\eqref{ailarge} and \eqref{bilarge},
\begin{align}
\text{Ai}(x_\pm) &\sim \frac{C'}{\left(h_\pm-1\right)^{1/4}}
  \cos{\left(\left(\frac{2\pi}{3\lambda}\right)(h_\pm-1)^{3/2} \right)}\ , \label{aiJTlarge}\\
  \text{Bi}(x_\pm) &\sim
               \frac{C'}{\left(h_\pm-1\right)^{1/4}}
  \sin{\left(\left(\frac{2\pi}{3\lambda}\right)(h_\pm-1)^{3/2}\right)} \ ,                    \label{biJTlarge}
\end{align}
with $C' = (\frac{\la}{\pi})^{1/3}$. A particular solution of the WDW equation is given by
\begin{equation}\label{onestokes}
\Psi(h,\phi;h_0) = \text{Ai}(x_+)\text{Ai}(x_-) 
- \text{Bi}(x_+)\text{Bi}(x_-) \ .
\end{equation}
Note that in addition to the field variables $h$ and $\phi$ the wave function depends on the Stokes point $h_0$ where an exponential behaviour at small $h$ changes to an oscillatory behaviour at large $h$. 
The general solution of the WDW equation is a superposition of wave functions with different Stokes points,
\begin{equation}
\Psi_\rho(h,\phi) = \int dh_0 \rho(h_0) \Psi(h,\phi;h_0) \ ,
\end{equation}
where the function $\rho(h_0)$ is assumed to vanish beyond some maximum
value $h_{\text{max}}$.
For large field values $h_\pm \gg \lambda^2 h_{\text{max}}$ one finds
\begin{equation}
\Psi_\rho(h,\phi) \sim \frac{C_\rho}{h_+^{1/4} h_-^{1/4}} \cos{\left(\frac{2\pi}{3\lambda}\left(h_+^{3/2}-h_-^{3/2}\right)\right)}\ ,
\end{equation}
where $C_{\rho} = C^{'2} \int dh_0 \rho(h_0)$ .
Expanding in powers of $\phi$ yields the final result
\begin{equation}\label{airyJTL}
\Psi_\rho(h,\phi) \sim \frac{C_\rho}{\lambda\sqrt{h}}  
\left(\cos{\left(2\pi\lambda\phi \sqrt{h}\right)} 
 + \mathcal{O}\left(\frac{\phi}{\lambda\sqrt{h}}\right)^2\right) \ .
\end{equation}
This wave function agrees to leading order with the semiclassical wave function \eqref{psiJT} obtained in Section~\ref{sec:semiJT}. On the contrary, it differs from the asymptotic behaviour of the exact solutions discussed
in Sections~\ref{sec:holo} and \ref{sec:charivp}.
At $h=0$, $\Psi_\rho$ is some function of $\phi$. To realize 
$\Psi(0,\phi) = 0$, one has to consider linear combinations of products
of Airy functions.

\section{Initial-value problems in 2d}
\label{sec:cauchy}

Solutions of the inhomogeneous wave equation are conveniently
expressed in terms of Green functions
\begin{equation}
  (\Box - m^2)G(x-x') = -\delta^2(x'-x)\ ,
  \end{equation}  
satisfying retarded ($G_R$), advanced ($G_A$) or Feynman ($G_F$) boundary
conditions and appropriate boundary terms \cite{Morse:1980rh}. In 
Cauchy's problem initial values of a wave function $\psi(x)$ and its time derivative
$\partial_{t}\psi(x)$, with $x^\mu = (t,\bx)$, are specified at an
initial time $t'$. At some later time $t$ the wave function can be
expressed in terms of these initial values and the retarded Green function,
\begin{equation}\label{psiR}
  \psi(x) = \int d\bx' \big(G_R(x-x')\partial_{t'}\psi(t',\bx')
  -\partial_{t'}G_R(x-x')\big) \psi(t,\bx')) ,
  \; (x-x')^2 < 0 \ .
\end{equation}

The Green functions are conveniently expressed in terms of odd and
even solutions of the homogeneous wave equation (see, for example,
\cite{Bjorken:1965sts,Itzykson:1980rh}),
\begin{align}
\Delta(x) &= \frac{i}{2\pi}\int \frac{d\bk}{2\omega}
            \left(e^{-i(\omega t - \bk\bx)} - e^{i(\omega t - \bk\bx)}\right)
            = i(\Delta_+(x) - \Delta_-(x))\ , \\
  \Delta_1(x) &= \frac{1}{2\pi}\int \frac{d\bk}{2\omega}
 \left(e^{-i(\omega t - \bk\bx)} + e^{i(\omega t - \bk\bx)}\right) =
                  \Delta_+(x) + \Delta_-(x) \ ,
\end{align}
where $\omega =\sqrt{\bk^2+m^2}$. From the integral representations
\begin{align}
  \Delta(x) = \frac{1}{2\pi}\int_{-\infty}^\infty \frac{d\bk}{\omega}
  \sin{(\omega t - \bk\bx)}\ , \quad
\Delta_1(x) = \frac{1}{2\pi}\int_{-\infty}^\infty \frac{d\bk}{\omega}
  \cos{(\omega t - \bk\bx)}
  \end{align}
one easily obtains explicit expressions for timelike, lightlike and
spacelike distances in two dimensions,
\begin{align}
\Delta(x) &= \frac{1}{2}\epsilon(t) 
            J_0(m\sqrt{-x^2})\Theta(-x^2)\ , \\
  \Delta_1(x) &= -\frac{1}{2} Y_0(m\sqrt{-x^2}) \theta(-x^2) + \frac{1}{\pi} K_0(m\sqrt{x^2}) \Theta(x^2)\ ,
\end{align}
with
\begin{equation}\label{t0}
 \partial_t\Delta(x)|_{t=0} = \delta(\bx) \ , \quad \Delta(x)|_{t=0} =
 0 \ , \quad \partial_t\Delta_1(x)|_{t=0} = 0 \ .
  \end{equation}
Contrary to 4d, $\Delta(x)$ has no singularity on the light
cone, whereas $\Delta_1(x)$ has a logarithmic singularity.
 The contributions of waves with positive and negative frequencies are
\begin{equation}     
  \Delta_+(x) = \frac{1}{2}(\Delta_1(x) -i \Delta(x)) = \Delta_-^*(x) \ ,
\end{equation}  
which reads in terms of Bessel functions
\begin{equation}
\begin{split}
  \Delta_+ & = -\frac{i}{4}\left(\big(\Theta(t) H^{(2)}_0(m\sqrt{-x^2}) - \Theta(-t)H_0^{(1)}(m\sqrt{-x^2})\big) \Theta(-x^2)  \right.\\
  & \left.\quad\quad +\frac{2i}{\pi} K_0(m\sqrt{x^2})\Theta(x^2)\right) \ .
  \end{split}
  \end{equation}
Note that $\Delta_{\pm}\Theta(\pm t)$ satisfy the inhomogeneous equations
\begin{equation}
(\Box - m^2)(\Delta_\pm(x)\Theta(\pm t)) = \frac{i}{2}\delta^2(x)\ .
\end{equation}
The propagators $G_R$, $G_A$ and $G_F$ can also be expressed in terms of $\Delta$ and $\Delta_1$
\cite{Bjorken:1965sts,Itzykson:1980rh}, which yields the explicit expressions
\begin{align}
  G_R(x) &= \Theta(t)\Delta(x) = \frac{1}{2}\Theta(t) J_0(m\sqrt{-x^2})\Theta(-x^2) \ , \label{GR}\\
G_A(x) &= -\Theta(-t)\Delta(x) = \frac{1}{2}\Theta(-t)
         J_0(m\sqrt{-x^2}) \Theta(-x^2)\ ,\label{GA}\\
  G_F(x) &= \frac{1}{2}(\epsilon(t)\Delta(x) + i\Delta_1(x)) \nonumber\\
    &= i(\Theta(t)\Delta_+(x) +\Theta(-t)
    \Delta_-(x)) \nonumber\\
 &= \frac{1}{4}H^{(2)}_0 (m\sqrt{-x^2}) \Theta(-x^2)
 + \frac{i}{2\pi}K_0(m\sqrt{x^2})\Theta(x^2) \nonumber\\
 &=\frac{1}{4}H^{(2)}_0(m\sqrt{-x^2 - i\epsilon})
  \ . \label{GF} 
\end{align}
Contrary to $G_R$ and $G_A$, the Feynman propagator $G_F$ does not
vanish for spacelike distances and has a logarithmic singularity on
the light cone.

A detailed discussion of massive Green functions in
  curved space-time can be found in chapters 8 and 9 of \cite{DeWitt:1984sjp}.
  
For the problems discussed in this paper it is convenient
to express the Green functions in terms of light-cone
coordinates $x_\pm = \frac{1}{2}(t\pm \bar{x})$. Using the relations
\begin{equation}
\begin{split}
\Theta(t)\Theta(-x^2) = \Theta(x_+)\Theta(x_-) \ , \quad
\Theta(-t)\Theta(-x^2) = \Theta(-x_+)\Theta(-x_-) 
\end{split}
\end{equation}
one finds
\begin{align}
\Delta(x) &= \frac{1}{2} \left(\Theta(x_+)\Theta(x_-) -
\Theta(-x_+)\Theta(-x_-)\right)J_0(m\sqrt{-x^2})\nonumber\\
&\equiv D_+(x)\Theta(x_-) + D_-(x)\Theta(-x_-) \ ,
\end{align}
which has the analytic continuation
\begin{align}
\bar{\Delta}(x) &= \frac{1}{2} \left(\Theta(-x_+)\Theta(x_-) -
\Theta(x_+)\Theta(-x_-)\right)I_0(m\sqrt{x^2})\nonumber \\
&\equiv \bar{D}_+(x)\Theta(x_-) + \bar{D}_-(x)\Theta(-x_-) \ .
\end{align}

For the retarded Green function one has $G_R(x) = D_+(x)\Theta(x_-)$, and the Feynman propagator 
can be decomposed as
\begin{align}\label{GF2}
G_F(x) 
\equiv G_+(x)\Theta(x_-) + G_-(x)\Theta(-x_-) \ ,
\end{align}
with
\begin{align}
G_+(x) &= \frac{1}{4}H^{(2)}_0 (2m\sqrt{x_+x_-}) \Theta(x_+)
 + \frac{i}{2\pi}K_0(m\sqrt{-x_+x_-})\Theta(-x_+) \nonumber\\
 &=\frac{1}{4}H^{(2)}_0(2m\sqrt{x_-(x_+ - i\epsilon)})
  \ , \label{G+}  \\
G_-(x) &= \frac{1}{4}H^{(2)}_0 (2m\sqrt{x_+x_-}) \Theta(-x_+)
 + \frac{i}{2\pi}K_0(m\sqrt{-x_+x_-})\Theta(x_+) \nonumber\\
 &=\frac{1}{4}H^{(2)}_0(2m\sqrt{-x_-(-x_+ - i\epsilon)})
  \ . \label{G-} 
\end{align}

From Eqs~\eqref{t0} and \eqref{GR}
one obtains for the wave function \eqref{psiR} propagated with the
retarded Green function
\begin{equation}
  \begin{split}
 \psi(x) = \Theta(t-t')\int d\bx'&\left(-\partial_{t'}
   \Delta(x-x')\Theta(-(x-x')^2)\psi(x')
   \right.\\
  &\left.+ \Delta(x-x')\Theta(-(x-x')^2) \partial_{t'}\psi(x')\right) \ .
\end{split}
\end{equation}
Using Eq.~\eqref{t0} one recovers the initial conditions
$\psi(t,\bx)$ and $\partial_t\psi(t,\bx)$ in the limit $t \rightarrow t'$.

Consider now a wave function that at time $t'$ is localized at
$\bx' = t' = c$, i.e., at $x_+'=c$, $x_-'=0$,
\begin{equation}\label{wf2}
  \psi(x'_+,0) = \delta(x'_+-c) \ ,
\end{equation}
with time derivative
\begin{equation}
\partial_{t'}\psi(x_+',0) = \frac{1}{2}    \partial_{x_+'}\psi(x_+',0) \ .
\end{equation}
Using $G_R(x) = D_+(x)\Theta(x_-)$,
one obtains from Eqs.~\eqref{psiR},
with $\psi(x) \equiv \psi(x_+,x_-)\Theta(x_-)$,
\begin{equation}\label{psiR2}
  \begin{split}
  \psi(x_+,x_-) = \int dx'_+ &\big(D_+(x_+-x'_+,x_-)\partial_{x'_+}\psi(x'_+,0)\\
  &-\partial_{x'_+}D_+(x_+-x'_+,x_-) \psi(x'_+,0)\big)\ ,
\end{split}
\end{equation}
where we have used that at fixed $t'$ one has $d\bx' = dx_+'$.
Equation~\eqref{psiR2} also holds if one adds sources along the line
$x_- = 0$. The corresponding waves are then generated at different times $t'$ and propagated into the forward light cone by the
retarded Green function. Summation over all waves corresponds
to integration over $dx_+'$.
Note that now integration and derivative both correspond to the lightlike variable $x_+'$, contrary to Eq.~(189) where the integration is carried out in a spacelike direction while the derivative acts in a timelike direction.
After partial integration and inserting \eqref{wf2} one finds the final result
\begin{equation}\label{psir}
\begin{split}
  \psi_>(x_+,x_-) &= \partial_{x_+}D_+(x_+-c,x_-) \\
&= \frac{1}{2}\left(\delta(x_+-c) + \partial_{x_+}
  J_0(2m\sqrt{x_-(x_+-c)})\right) \ .
\end{split}
\end{equation}
This wave function is a singular solution of the
homogeneous wave equation. Adding the analytic
continuation $\psi_<(x_+,x_-) = \partial_{x_+}\bar{D}_+(x_+-c,x_-)$ , one obtains an analytic solution of the homogeneous wave equation,
\begin{align}
\psi(x_+,x_-) &= \psi_>(x_+,x_-) + \psi_<(x_+,x_-) \nonumber\\
&= \partial_{x_+}({D}_+(x_+-c,x_-) + \bar{D}_+(x_+-c,x_-)) 
\nonumber \\
&= \frac{1}{2}\partial_{x_+}J_0(2m\sqrt{x_-(x_+ -c -i\epsilon)})\ .
\label{DbarD}
\end{align}

A complex solution of the homogeneous wave equation in the
half-plane $x_->0$ is obtained by replacing in 
Eq.~\eqref{DbarD} $D_+ + \bar{D}_+$ by $G_+$,
\begin{align}\label{wfG+}
  \psi_+(x_+,x_-) &= \partial_{x_+}G_+(x_+-c,x_-) \nonumber\\
&= \frac{1}{4}\partial_{x_+}H^{(2)}_0(2m\sqrt{x_-(x_+ - c - i\epsilon}) \ .
\end{align}
The complex wave function is singular at $x_+ = c$ and has
to be interpreted as distribution, similar to the 
Feynman propagator. 
Unlike $\Delta_\pm \Theta(t_\pm)$, the projections of
$G_F$ to positive and negative $x_-$ are not Green functions. They satisfy the inhomogeneous equations
\begin{equation}\label{G+-WDW}
(\Box - m^2)(G_\pm \Theta(\pm x_-)) 
= \pm \frac{i}{4\pi} \left(\frac{1}{x_+ \mp i\epsilon}
\right) \delta(x_-) \ .
\end{equation}
Hence, the complex wave function $\psi_+(x_+,x_-)\Theta(x_-)$ has a singular
source along the entire axis $x_- = 0$. However, since the singularity of $H^{(2)}_0$ is only logarithmic, non-singular
solutions of the inhomogeneous wave equation can be found by convolution with a smooth source function.

We note that the limit $\epsilon \to 0$ in \eqref{wfG+} can be taken explicitly, giving the complex wave function
\begin{equation}\label{wfG+V2}
\begin{split}
\psi_+(x_+,x_-) &= \frac{1}{2}
\partial_{x_+}
\left(H^{(2)}_0(2m\sqrt{x_- (x_+-c)}) \Theta(x_+-c)\phantom{\frac{2i}{\pi}}\right. \\
&\qquad\left. +\frac{2i}{\pi}
K_0(2m\sqrt{x_- (c-x_+)})\Theta(c - x_+)\right) \ .
\end{split}
\end{equation}

An alternative form of the complex wave function is obtained
by using a different $i\epsilon$-prescription,
\begin{align}\label{otherwf+}
  \tilde{\psi}_+(x_+,x_-) 
= \frac{1}{4}\partial_{x_+}H^{(2)}_0(2m\sqrt{x_-(x_+ - c + i\epsilon)}) \ ,
\end{align}
which yields
\begin{align}\label{otherwf+V2}
 \tilde{\psi}_+(x_+,x_-) &= \frac{1}{2}\partial_{x_+}\Big(
 H^{(2)}_0 (2m\sqrt{x_-(x_+-c)}) \Theta(x_+-c) \nonumber\\
 &\quad + 2\Big( 
I_0(2m\sqrt{x_-(c-x_+)}) + \frac{i}{\pi}K_0(m\sqrt{x_-(c-x_+)})\Big)\Theta(c-x_+)\Big) \ .
\end{align}
\\

\section{Properties of Bessel functions}
\label{sec:bessel}

In this section we list, for convenience, some properties of Bessel functions\footnote{See https://dlmf.nist.gov/10 and 
Ref.~\cite{Abramowitz:1968ams}} that are used in this paper.

Hankel functions as linear combinations of Bessel functions of first and second kind, $J_\alpha$ and $Y_\alpha$, respectively ($\alpha > 0$):
\begin{equation}
  H^{(1)}_\alpha (z) = J_\alpha(z) + i Y_\alpha(z)\ , \quad
 H^{(2)}_\alpha(z) = J_\alpha(z) - i Y_\alpha(z)
\end{equation}

Analytic continuation ($m \in \mathbb{Z}$):
\begin{equation}
J_\alpha(z e^{m\pi i}) = e^{m\alpha \pi i} J_\alpha(z)
\end{equation}

Derivatives and Wronskians of Bessel functions
($C_\alpha = J_\alpha,Y_\alpha; \bar{C}_\alpha = I_\alpha,K_\alpha; m \in \mathbb{N}$):
\begin{align}
\left(\frac{1}{z}\frac{d}{dz}\right)^m \left(\frac{C_\alpha(z)}{z^\alpha}\right) &= 
(-)^m \frac{C_{\alpha + m}(z)}{z^{\alpha + m}} \\
\left(\frac{1}{z}\frac{d}{dz}\right)^m \left(\frac{\bar{C}_\alpha(z)}{z^\alpha}\right) &= 
\frac{\bar{C}_{\alpha + m}(z)}{z^{\alpha + m}} \\
J_\alpha(z)\frac{d}{dz}Y_\alpha (z) -
Y_\alpha(z)\frac{d}{dz}J_\alpha (z) &= \frac{2}{\pi z} \\
I_\alpha(z)\frac{d}{dz}K_\alpha (z) -
K_\alpha(z)\frac{d}{dz}I_\alpha (z) &= -\frac{1}{z} 
\end{align}

Relation to modified Bessel functions:
\begin{align}
    J_\alpha(\pm iz) &= (\pm i)^\alpha I_\alpha(z)\  \\
    \pi i J_\alpha(z) &= (-i)^\alpha K_\alpha(-iz) - i^\alpha K_\alpha(iz) \ , \quad
    |\text{arg} z| \leq \frac{\pi}{2} \ 
\label{JK} \\      
Y_\alpha(z) & = (\pm i)^{(\alpha + 1)} I_\alpha(\mp iz) 
- \frac{2}{\pi}(\mp i)^\alpha K_\alpha(\mp iz) \ , \quad 
    -\frac{\pi}{2} \leq \text{arg} z \leq \pi
\label{YIK}\\
2 (\pm i)^\alpha I_\alpha(z) & = H^{(2)}_\alpha(\pm iz)  
+ H^{(1)}_\alpha(\pm iz) \ , \quad
    |\text{arg} z| \leq \frac{\pi}{2} \label{IH}\\     
    \frac{2}{\pi}K_\alpha(z) &= 
    \begin{cases}
    &i^{\alpha+1}H^{(1)}_\alpha(iz)\ , \quad 
    -\pi \leq \text{arg} z \leq \frac{\pi}{2} \\
    &(-i)^{\alpha+1}H^{(2)}_\alpha(-iz)\ , \quad 
    -\frac{\pi}{2} \leq \text{arg} z \leq \pi    
    \end{cases} \label{KH}
   \end{align}

Asymptotic behaviour for large arguments ($|z| \gg 1$):
\begin{align}
  H^{(1)}_\alpha (z) &\sim \sqrt{\frac{2}{\pi z}}\
  e^{i\left(z-\frac{\alpha \pi}{2} - \frac{\pi}{4}\right)}\ ,
  \quad -\pi < \text{arg} z < 2\pi\  \\
  H^{(2)}_\alpha (z) &\sim \sqrt{\frac{2}{\pi z}}\
  e^{-i\left(z-\frac{\alpha \pi}{2} - \frac{\pi}{4}\right)}\ ,
  \quad -2\pi < \text{arg} z < \pi \\
I_\alpha (z) &\sim \frac{1}{\sqrt{\pi z}}\ e^z \ ,
  \quad -\frac{\pi}{2} < \text{arg} z < \frac{\pi}{2} \\
K_\alpha (z) &\sim \frac{\pi}{\sqrt{2 z}}\ e^{-z} \ ,
  \quad -\frac{3\pi}{2} < \text{arg} z < \frac{3\pi}{2}
\end{align}

Asymptotic behaviour for small arguments
($ 0 < z < \sqrt{\alpha + 1}$):
\begin{align}
  J_\alpha (z) &\sim \frac{1}{\Gamma(\alpha+1)}\left(\frac{z}{2}\right)^\alpha\ ,\quad 
  Y_\alpha (z) \sim
  -\frac{\Gamma(\alpha)}{\pi}\left(\frac{2}{z}\right)^\alpha \label{JY0}\\
  I_\alpha (z) &\sim \frac{1}{\Gamma(\alpha+1)}\left(\frac{z}{2}\right)^\alpha\ ,\quad 
  K_\alpha (z) \sim
  \frac{\Gamma(\alpha)}{2}\left(\frac{2}{z}\right)^\alpha  \label{IK0}
\end{align} 

Asymptotic behaviour for small arguments
and $\alpha = 0$:
\begin{align}
Y_0 (z) \sim
  \frac{2}{\pi}\left(\ln{\frac{z}{2}} + \gamma \right)\ , \quad
K_0 \sim 
  -\ln{\frac{z}{2}} - \gamma \  \label{YK0}
\end{align}


\bibliographystyle{utphys}
\bibliography{JT}

\end{document}